\documentclass[modern]{aastex63}
\usepackage{graphicx}
\usepackage{color}
\usepackage{amsmath} 
\usepackage{amssymb}

\newcommand{\cmfgen}{{\sc cmfgen}\ }

\renewcommand\ion[2]{#1\,\,{\small\MakeUppercase{\romannumeral #2}}}

\newcommand{\civopt}{C\,{\sc iv} $\lambda \lambda$5801,12}
\newcommand{\civres}{C\,{\sc iv} $\lambda \lambda$1548,52}

\newcommand{\osixdoub}{O\,{\sc vi} $\lambda\lambda$3811,34}   

\newcommand{\Mdot}{\hbox{$\dot M$}}

\newcommand{\mum}{\hbox{$\micron$}}

\shorttitle{WO-Type Wolf-Rayet Stars: the Last Hurrah}
\shortauthors{Aadland et al.}

\graphicspath{{./}{figures/}}

\begin{document}

\title{WO-Type Wolf-Rayet Stars: the Last Hurrah of Massive Star Evolution\footnote{This paper includes data gathered with the 6.5 meter Magellan Telescopes located at Las Campanas Observatory, Chile}}


\author[0000-0002-2894-6418]{Erin Aadland}
\affiliation{Department of Astronomy and Planetary Science, Northern Arizona University, Flagstaff, AZ, 86011-6010, USA}
\affiliation{Lowell Observatory, 1400 W Mars Hill Road, Flagstaff, AZ 86001, USA}
\email{aadlander@lowell.edu}

\author[0000-0001-6563-7828]{Philip Massey}
\affiliation{Lowell Observatory, 1400 W Mars Hill Road, Flagstaff, AZ 86001, USA}
\affiliation{Department of Astronomy and Planetary Science, Northern Arizona University, Flagstaff, AZ, 86011-6010, USA}

\author[0000-0001-5094-8017]{D. John Hillier}
\affiliation{Department of Physics and Astronomy \& Pittsburgh Particle Physics,
Astrophysics and Cosmology Center (PITT PACC), \\University of Pittsburgh,
3941 O'Hara Street, Pittsburgh, PA 15260, USA}

\author[0000-0003-2535-3091]{Nidia I. Morrell}
\affiliation{Las Campanas Observatory, Carnegie Observatories, Casilla 601, La Serena, Chile}

\author[0000-0002-5787-138X]{Kathryn F. Neugent}
\affiliation{Dunlap Institute for Astronomy \& Astrophysics, University of Toronto, 50 St. George Street, Toronto, ON M5S 3H4, Canada}
\affiliation{Lowell Observatory, 1400 W Mars Hill Road, Flagstaff, AZ 86001, USA}

\author[0000-0002-1722-6343]{J. J. Eldridge}
\affiliation{Department of Physics, University of Auckland, Private Bag 92019, Auckland, New Zealand}

\begin{abstract}
Are WO-type Wolf Rayet (WR) stars in the final stage of massive star evolution before core-collapse?  Although WC- and WO-type WRs have very similar spectra, WOs show a much stronger O\,{\sc vi} $\lambda \lambda$3811,34 emission-line feature. This has usually been interpreted to mean that WOs are more oxygen rich than WCs, and thus further evolved.  However, previous studies have failed to model this line, leaving the relative abundances uncertain, and the relationship between the two types unresolved.  To answer this fundamental question, we modeled six WCs and two WOs in the LMC using UV, optical, and NIR spectra with the radiative transfer code \cmfgen in order to determine their physical properties.  We find that WOs are not richer in oxygen; rather, the O\,{\sc vi} feature is insensitive to the abundance. However, the WOs have a significantly higher carbon and lower helium content than the WCs, and hence are further evolved.  A comparison of our results with single-star Geneva and binary BPASS evolutionary models show that while many properties match, there is more carbon and less oxygen in the WOs than either set of evolutionary model predicts.  This discrepancy may be due to the large uncertainty in the $^{12}$C+$^4$He$\rightarrow^{16}$O nuclear reaction rate;  we show that if the Kunz et al.\ rate is decreased by a factor of 25-50\%, then there would be a good match with the observations. It would also help explain the LIGO/VIRGO detection of black holes whose masses are in the theoretical upper mass gap.

\end{abstract}

\section{Introduction} \label{sec:intro}

Wolf-Rayet (WR) stars are massive stars whose spectra consist of a hot continuum and strong, broad emission lines.  The majority of them are classified into two subtypes. The spectra of the WN-type WRs show helium and nitrogen;  a few also show traces of hydrogen. The spectra of the WC-type WRs show helium, carbon, and oxygen, and no hydrogen.  \citet{Gamow43}  proposed that the WNs were showing the equilibrium products of the H-burning CNO cycle at their surface.  The CNO cycle is the dominant hydrogen fusion process in massive stars; as hydrogen fuses into helium, nitrogen increases at the expense of carbon.  By contrast, WCs show the products of the He-burning (carbon and oxygen) at their surfaces.  In order for WRs to form, some mechanism--either stellar winds or binary stripping--must first remove the hydrogen-rich outer layers, revealing these evolved products \citep{P1967,Conti1976}.  The WN- and WC-types are further divided into excitation classes, similar to subtypes of main-sequence stars: WN2-WN11, and WC4-WC9, with the WN2 and WC4 classes showing the highest ionization states \citep{2001NewAR..45..135V}.  

A few percent of the WRs are classified as WO-type. These stars have similar spectra to the WC4-type WRs, except that the O\,{\sc vi} $\lambda \lambda$3811,34 doublet is very strong. (WC4s show O\,{\sc iv} and O\,{\sc v} lines but only very weak O\,{\sc vi}.)\footnote{Although there does not appear to be a quantitative definition in the O\,{\sc vi} $\lambda\lambda$3811,34 line strength to distinguish WC4s from WOs, we will note that the WC4s in our sample have equivalent widths of -20\,\AA\ to -35\,\AA\ for this doublet, while the WOs have equivalent widths of -200~\AA\ to -450\,\AA. In our experience, this is not a continuum; i.e., the feature is about 10$\times$ stronger in WOs than in WCs, at least for the sample we've examined. However, this lack of continuity may simply due to small number statistics.} \citet{Sand1971} called attention to five such stars, one in the Small Magellanic Cloud (SMC), one in the Large Magellanic Cloud (LMC), and three in the Milky Way.   \cite{Barlow1982}
analyzed the optical spectrum of one of the Galactic WOs using a simple line recombination approximation, and concluded that it had a higher oxygen abundance than zero-age main-sequence stars, and suggested that this would be the result of the  $^{12}$C+$^{4}$He$\rightarrow^{16}$O reaction. As helium burning proceeds, the carbon content of the core increases, and this secondary reaction will create oxygen.  They introduced the WO designation and proposed subtypes WO1-WO4 based upon the relative strengths of O\,{\sc iv}, O\,{\sc v}, and O\,{\sc vi}.  The classification scheme was subsequently refined by \citet{Crowther1998}, based primarily on the relative strengths of O\,{\sc vi} $\lambda\lambda$3811,34 and O\,{\sc v} $\lambda$5590.  

We emphasize that the WO-type WRs are {\it rare}.  In the LMC there are 13 WC stars known, and only 3 WOs, out of a total of 154 WRs, a census that is thought to be complete \citep{2018ApJ...863..181N}. The rarity of WOs suggests that either the phase is very short-lived, or that only a small fraction of massive stars go through a WO stage.

Are WO stars the ``last hurrah" of the most massive stars before they die in a core-collapse supernova event? \citet{2019A&A...621A..92S} have recently argued as such, based upon the luminosities and temperatures of Galactic WO stars compared to He-burning stars.  Here we compare the surface chemical abundances of WCs and WOs to provide a more fundamental assessment of their evolutionary status. If the oxygen abundance of WOs is actually higher than that of WCs, then this would indeed support an evolutionary progression, with some massive stars evolving first through a WN stage, then a WC stage, and then finally
a WO stage.  

However, it is not entirely clear that the WOs have stronger O\,{\sc vi} $\lambda \lambda$3811,34 than the WCs because of a higher oxygen abundance.  The strength of this high excitation doublet could simply be due to other
physical properties compared to WCs, i.e., stellar wind where the temperature/optical depth profile favors the formation of O\,{\sc vi} over O\,{\sc v}. \citet{1999IAUS..193..116C} has argued that strong O\,{\sc vi} emission may be connected primarily to wind density rather than chemical abundance, although \citet{Kingsburgh1995} concluded that the WOs were more chemically evolved than WCs based on the derived (C+O)/He ratios. However, both the \citet{Barlow1982} and \citet{Kingsburgh1995} studies used Case B recombination theory for their analyses.  This approximation assumes a uniform temperature and density for the line-formation regions, and is applicable when most emission lines are optically thin.  Neither of these assumptions are valid for line formation in WR stars, where the lines are formed in an accelerating stellar wind.  

Several more advanced studies have been conducted so far, but also failed to achieve adequate fits to the optical O\,{\sc vi} doublet that defines the WO class.   \cite{Crowther2000} used the non-LTE radiative transfer code \cmfgen \citep{Hillier1998} to analyze the UV and optical spectrum of Sanduleak~2, the only known WO star in the LMC at the time.  They found that they could not obtain consistent fits for both the O\,{\sc vi} $\lambda \lambda$1032,38 resonance doublet and the O\,{\sc vi} $\lambda \lambda$3811,34 feature.  The optical O\,{\sc vi}  $\lambda \lambda$3811,34 doublet and the C\,{\sc iv} $\lambda$7700 line indicated a very high effective temperature (170,000~K) while the UV resonance doublet O\,{\sc vi} $\lambda \lambda$1032,38 suggested a much lower temperature (120,000 K). Similarly, 
\citet{2013A&A...559A..72T,Tramper2015} also used \cmfgen to analyze optical spectra of DR1 (a WO star found in the nearby galaxy IC~1613) and Sanduleak~2, but were again unable to fit the optical O\,{\sc vi} doublet, attributing the problem to possible excitation of the upper level by X-rays.  \citet{Sander2012} modeled two single Galactic WO2 stars and found a very high temperature (200,000~K) and enhanced oxygen was needed to fit the O\,{\sc vi} $\lambda\lambda$3811,34 doublet, but at this high temperature their models had much weaker O\,{\sc v} $\lambda$5596 flux than the spectra of the stars.

Here we tackle the question afresh: are WO stars more evolved than WC stars? To answer this, we have obtained data for a set of WCs and WOs in the Large Magellanic Cloud (LMC), and applied the same analysis tools to determine these stars' physical properties, including their surface abundances.    The first part of this study was described by \citet{Aadland2022}, where we analyzed high quality UV, optical, and near-IR (NIR) data on four WC4 stars, comparing our results to those predicted by single-star and binary evolutionary models.   Here we complete our study by providing an analysis of two WO-type stars, as well as two additional WC4s. 

In Section \ref{sec:observations}, we describe the new sample and the data we obtained. In Section~\ref{sec:modeling}, we discuss how we modeled these stars. In Section~\ref{sec:results}, we
present our results, and 
compare these to the predictions of the evolutionary models. We present our interpretation of what we have found in Section~\ref{sec:PM}. Lastly, in Section~\ref{sec:conclusions} we summarize our conclusions.

\section{Observations}
\label{sec:observations}

\subsection{The Sample}

The four new stars in our sample (two WCs and two WOs) are listed in Table~\ref{tab:stars}.  Like the four WC4 stars discussed in our earlier study \citep{Aadland2022}, they are located in the LMC, an ideal laboratory for evaluating the relation between WCs and WOs. The LMC has the advantage that the distances are well known, allowing accurate luminosities to be found; in addition, the reddening is tractable, making it practical to obtain UV spectra.  
 
BAT99-61 and BAT99-90 were added to our WC4 sample when an observing opportunity
presented itself, allowing us to increase the sample size for the WC stars. All six of our WCs (the four discussed in \citealt{Aadland2022} and the two here) have been examined for radial velocity variations by \citet{Bartzakos2001} and found to be likely single, along with the WO3 star Sanduleak 2 (Br93)\footnote{Of course, as our late friend and colleague Dr.\ Virpi Niemla would point out, no star other than the sun can be proven to be single.}.

In addition to Sanduleak~2, there are two other WO stars known in the LMC.  \citet{2012AJ....144..162N} reported the discovery of a WO3 star in the Lucke-Hodge 41 association in the LMC, designated LH41-1042 in the photometry list of \citet{2000AJ....119.2214M}.  However, UV {\it HST} images (obtained through the F225W filter under Program ID 12940, PI: Massey) reveal that this star has a close visual hot companion 0\farcs12 to the north\footnote{{This is in addition to the 0\farcs7 separated companion star to the NNW noted by \citet{Tramper2015} from the same UV image.}}. This companion will cause contamination in our optical and NIR spectra, and so we decided not to include this in our analysis.  Finding LH41-1042, with its strong emission lines, motivated a subsequent multi-year survey of the LMC for WRs (see \citealt{2018ApJ...863..181N} for a summary), and in the first year found another WO star also in the Lucke-Hodge 41 cluster, designated LMC195-1 \citep{Massey2014}.  This star shows a slightly higher excitation state than either Sanduleak~2 or LH41-1042, and is classified as a WO2.  We have only two spectra of this star (the discovery spectrum, and one with higher signal-to-noise that we took for this project); they reveal no radial velocity variation at the 10 km s$^{-1}$ level, comparable to our measuring uncertainty.

\begin{deluxetable}{l l c c c c l}
\tablecaption{\label{tab:stars} Our Sample}
\tablewidth{0pt}
\tablehead{
\colhead{Star}
&\colhead{Subtype}
&\colhead{$\alpha_{\rm 2000}$}
&\colhead{$\delta_{\rm 2000}$}
&\colhead{$V$}
&\colhead{$M_V$\tablenotemark{a}}
&\colhead{Alternative IDs}
}
\startdata
BAT99-61 & WC4 & 05:34:19.24  & -69:45:10.3 & 13.12 & -5.9 & Brey 50, HD 37680\\
BAT99-90 & WC4 & 05:37:44.64  & -69 14 25.7  & 14.63 & -5.7 & Brey 74, HD 269888\\
Sanduleak~2 & WO3 & 05:39:34.29  & -68:44:09.2  & 15.20 & -3.8 & BAT99-123, Brey 93\\
LMC195-1   & WO2\tablenotemark{b} & 05:18:10.33  & -69:13:02.5  & 15.15 & -3.8 & \nodata \\
\enddata
\tablecomments{Data in this table comes from the Fifth Catalog of LMC Wolf-Rayet Stars \citep{2018ApJ...863..181N} and references therein, except as noted.}
\tablenotemark{a}{$M_V$ has been computed using the small color excess given in Section~\ref{sec:results}, assuming $A_V=3.1E(B-V)$ and adopting a true distance modulus to the LMC of 18.50 (50 kpc, \citealt{vandenBergh2000,Pietrzynski2019}).} 
\tablenotetext{b}{Subtype WO2 is from \citet{Massey2014} and confirmed here.  Note that it is incorrectly listed as ``WO3" in The Fifth Catalog through an oversight on the part of P.M.}
\end{deluxetable}

\subsection{Observational Data}

All four stars in our current sample were observed in the UV with {\it HST}, either with the Faint Object Spectrograph (FOS) or the Space Telescope Imaging Telescope (STIS). In addition, the two WO stars were also observed with the Cosmic Origins Spectrograph (COS) in order to extend wavelength coverage to shorter wavelengths in order to include the O\,{\sc vi} $\lambda \lambda$1032,38 resonance doublet. The four stars were also observed on the 6.5-m Magellan telescopes at Las Campanas Observatory in Chile, 
in the optical with the Magellan Echellette (MagE; \citealt{MagE}), and in the NIR with the Folded port InfraRed Echellette (FIRE; \citealt{Simcoe2013}). 
 Specifics of the observations are given in Table~\ref{tab:obs}, with the instrument specifications given below.  FOS, MagE, and FIRE data were also used in our study of the other four WCs described in \citet{Aadland2022}; i.e., the data are of similar quality and with the same wavelength coverage.

\begin{deluxetable}{l l  c c c c}
\tablecaption{\label{tab:obs} Spectroscopic Data Used in This Study}
\tablewidth{0pt}
\tabletypesize{\footnotesize}
\tablehead{
\multicolumn{2}{c}{Parameter}
&\colhead{BAT99-61}
&\colhead{BAT99-90}
&\colhead{Sand 2}
&\colhead{LMC195-1}
}
\startdata
FUV \\
&Instrument & \nodata & \nodata &COS & COS \\
&Date       & \nodata & \nodata &2015-07-16 & 2015-07-17 \\
&Program ID & \nodata & \nodata &13781 & 13781 \\
&Datasets   & \nodata & \nodata &LCJ002010-20    & LCJ001020 \\
UV \\
&Instrument & FOS & FOS & FOS & STIS \\
&Date       & 1994-11-16 & 1995-06-30 &1996-03-26 &  2015-07-04 \\
&Program ID & 5723 & 5460 & 5460 & 13781 \\ 
&Datasets    &Y2JE0305T-8T&Y2A10304T-7T&Y2A10404T-8T&OCJ003030-40 \\
Optical (MagE) \\
&Telescope  &Baade & Baade & Clay       & Clay \\
&Date       &2020-12-01      & 2020-12-01      & 2015-01-09 & 2015-01-09 \\
&Seeing     & 1\farcs3     &  0\farcs9     &     0\farcs9       &  0\farcs7 \\
NIR (FIRE) \\
& Telescope &Baade & Baade & Baade & Baade \\
& Date      &2020-11-29      &  2020-11-29     & 2017-02-09 & 2017-02-09 \\
&Seeing     &  0\farcs9      &  0\farcs6      &    0\farcs7             & 0\farcs7 \\
\enddata
\end{deluxetable}

\paragraph{COS} Observations with COS were taken with the G130M/1096 setting, providing wavelength coverage from 940-1080\,\AA\ (Segment B) and 1096-1236\,\AA\ (Segment A) with a resolving power $R=\lambda/\Delta \lambda$ of 11,000.  The pipeline-reduced data were smoothed by 31 pixels to a resolution of 0.3\,\AA\ ($R\sim4000$) to match the other data.  We used the Primary Science Aperture (PSA), a circular aperture with a diameter of $2\farcs5$.  The data were taken under Program ID 13781 (PI: Massey) specifically for this project.  

\paragraph{FOS} Data with the FOS were taken (post CoStar) with gratings G130H, G190H, and G270H, providing overlapping coverage from 1140\,\AA\ to 3300\,\AA.  The resolving power was $R=1100-1600$ using the 0\farcs26 circular aperture.  The stars were observed under Program IDs 5460 and 5723 (PI: Hillier) with similar goals as the present study. 

\paragraph{STIS} As a consequence of safety concerns due to the richness of hot, blue stars in the Lucke-Hodge 41 cluster, COS could not be used for the longer UV wavelengths for LMC195-1, and so we used STIS with the 0\farcs2$\times$0\farcs2 entrance aperture.  Gratings G140L and G230L were used to provide continue wavelength coverage from 1150\,\AA\ to 3180\,\AA\ with resolving powers of 500-1000.
The observations were obtained Program ID 13781 (PI: Massey) specifically for this project.

\paragraph{MagE} The optical MagE spectra cover the wavelength range of 3150 to 9300\,\AA.  To obtain a resolving power of $R=4100$, a 1\farcs0 slit was used.  The data were taken with the slit oriented to the parallactic angle.  Exposure times varied from 300~s (BAT99-61) to 3600~s (Sanduleak~2), depending upon the brightness of the star and how good the seeing was.  Spectrophotometric standards were observed at the start and end of each night, and wavelength calibrations were obtained by taking a ThAr arc exposure at the end of each observation. Daytime dome flats were used to remove fringing in the red part of the spectra; the chip is sufficiently flat that additional flat-fielding was not needed; see \citet{2012ApJ...748...96M}, which also describes our reduction procedure. Time for this project was kindly provided by the Carnegie Observatories and the Arizona Telescope Allocation Committee.

\paragraph{FIRE} The NIR data for all of the stars were taken with FIRE in echelle mode, providing wavelength coverage from 8300 to 25000\,\AA.  The use of a 0\farcs75 slit resulted in a spectral resolving power of $R\sim5000$.  The data were all taken in the standard A-B-B-A scheme using the Sample-Up-the-Ramp readout mode, with integration times of 4$\times$600 s or 8$\times$600s.  Telluric standards (A-type) were observed after each science target in order both to remove the telluric bands and to provide flux calibration. Those exposures were taken in ``Fowler-1" mode also in an A-B-B-A scheme, and were typically $4\times$60s. A ThAr arc exposure was taken after each target.  The data reduction was performed using the {\sc idl} FIRE pipeline \citep{Simcoe2013}.  Time for these observations was kindly provided by the Carnegie Observatories and the Arizona Telescope Allocation Committee. \\

For convenience, we merged the UV, optical, and NIR data, but left the FUV spectra separate.  Although all three sets of data were flux calibrated, small differences in the absolute flux level are expected for the ground-based optical and NIR data as the seeing is never exactly the same between the science targets and the standards used to determine the sensitivity functions. Since the UV data from HST should be accurate to 3-4\% in an absolute sense\footnote{See Table 32.1 in Version 3.1 of the FOS handbook, currently available from https://www.stsci.edu/hst/documentation/handbook-archive.}, our plan was to first scale the optical to the UV, and then scale the NIR to the optical, as there is overlap between adjacent bands.  However, the poorest calibration for the optical data is at the shortest wavelengths where the data overlap with the UV data from HST, and we found that a better procedure was to compute synthetic $V$ band photometry from our spectrophotometry, and then used the measured $V$ to scale the optical data. (For LMC195-1 no accurate $V$ band photometry is available, so we instead used the Gaia G band data for that star.) The NIR data were scaled to the optical.  This process worked well, resulting in good agreement with the optical and UV without the need for further adjustment. We do note that small imprecisions in the process do not affect our model fitting; as described below, we rescaled the models slightly (by a few percent) as needed when changing wavelength regions.  This process differs slightly from that described in \citet{Aadland2022}, but in fact the method described here was actually used there as well. 

Although we are using the most advanced modeling techniques and the best available atomic data to determine the physical properties of these stars, as time goes on, models will become even more sophisticated and atomic data bases will improve. We are therefore making our combined spectra available to others through this publication.

\section{Modeling} 
\label{sec:modeling}

\subsection{CMFGEN}

Accurately modeling WR spectra requires a stellar atmosphere code which includes the effects of non-LTE, spherical expansion, and line blanketing, as all the spectral lines are formed in an extended, outflowing stellar wind.  Here, we use the open-source radiative transfer code ``Co-Moving Frame GENeral," or \cmfgen \citep{Hillier1998,Hillier2003,Hillier2012}.  Over the years, many improvements have been incorporated into the code, such as more complete blanketing, implementing a two-component velocity law, better treatment of clumping, and incorporating updated atomic data. 
For more information on {\sc cmfgen}, and, in particular its use in modeling WR stars, the interested reader is referred to the appendix in \citet{Aadland2022}.

For this study, we used a tailored analysis, as we have done in the past (see e.g., \citealt{Hillier1989, Hillier1999,Neugent2017,Aadland2022}). The alternative, using a grid of pre-computed models, is not practical, given the large number of adjustable parameters.  Uncertainties in the atomic data and the physics of WR atmospheres also make a grid approach more problematical, in the sense that it would invariably have built-in biases.  Our tailored approach allows us to explore some of these biases.  In modeling the spectra of the WC/WOs, we input trial values of the luminosity, radius, mass-loss rate, the terminal velocities of the inner and outer part of the winds, clumping parameters for the wind, and chemical abundances. The effective temperature $T_{2/3}$, which is defined as the temperature at a Rosseland optical depth of two-thirds, is generally not an input parameter, but is one of the outputs of the radiative transfer code.  Adjustments are made until the model spectrum and observed spectrum match. We explored parameter space to ascertain that we have not reached a false local minimum.  Uncertainties in the parameters are determined by computing additional models to see at what point we no longer have an acceptable fit. As emphasized in earlier papers, the interaction between parameters is complicated, and the strengths of some lines are much more sensitive to some parameters than other lines of the same species are (see, e.g., \citealt{Neugent2017}).  A good example in WN stars is the He\,{\sc ii} $\lambda$4686 line, which is extremely sensitive to the mass-loss rate, unlike most other He\,{\sc ii} lines, which are mostly sensitive to the abundance and derived temperature structure. Table~\ref{tab:guidelines} shows the lines that were most used as diagnostics for each parameter. Nevertheless, the final models are always compromises to some degree due to the complicated interactions between parameters.

\begin{deluxetable}{l l c c c c l}
\tablecaption{\label{tab:guidelines} Diagnostic Lines}
\tablewidth{0pt}
\tablehead{
\colhead{C/He\tablenotemark{a}}
&\colhead{O\tablenotemark{b}}
&\colhead{$\dot{M}$\tablenotemark{c}}
&\colhead{L\tablenotemark{d}}
&\colhead{$V_{\infty,1}$\tablenotemark{e}}
&\colhead{$V_{\infty,2}$\tablenotemark{f}}
}
\startdata
C\,{\sc iv} $\lambda$2698/He\,{\sc ii} $\lambda$2733 & O\,{\sc iv} $\lambda$3400 & He\,{\sc ii} $\lambda$4686 & C\,{\sc iii}/C\,{\sc iv} & C\,{\sc iv} $\lambda$5801 & C\,{\sc iii} $\lambda$2289\\
C\,{\sc iv} $\lambda$5473/He\,{\sc ii} $\lambda$5411 & O\,{\sc vi} $\lambda$5291 & He\,{\sc ii} $\lambda 6560$ & He\,{\sc i}/He\,{\sc ii} & & C\,{\sc iii} $\lambda$9710\\
C\,{\sc iv} $\lambda$11905/He\,{\sc ii} $\lambda$11626 & O\,{\sc v} $\lambda$5596 & He\,{\sc i} $\lambda$10830 & & & He\,{\sc i} $\lambda$10830\\
\enddata
\tablecomments{Lines and guidelines for fitting the spectra of WCs and WOs to determine the physical parameters of these stars.}
\tablenotemark{a}{The C/He ratio is based on the line pairs, such that the model's line pair strengths is proportional to the spectrum pair strengths.} 
\tablenotetext{b}{The strongest oxygen lines in the spectra (excluding O\,{\sc vi} $\lambda \lambda$3811,34 as it is not sensitive to the oxygen abundance).}
\tablenotetext{c}{The mass-loss rate will affect the continuum of the spectrum along with these sensitive lines.}
\tablenotetext{d}{The luminosity was determined based both on line-ratio measurements and also the observed fluxes. Increasing the luminosity increases the strength of high-excitation lines (C\,{\sc iv} and He\,{\sc ii}) but will decrease the strength of low-excitation lines (C\,{\sc iii} and He\,{\sc i}).}
\tablenotetext{e}{$V_{\infty,1}$ is based on the line width of high-excitation lines.}
\tablenotetext{f}{$V_{\infty,2}$ is based on the line width of low-excitation lines.}
\end{deluxetable}

In general, the stellar wind velocity law $v(r)$ has the form $$v(r)=v_\infty(1-R_*/r)^\beta,$$
as first described \citet{CAK1975}, where $v_\infty$ is the terminal velocity of the wind (i.e., the velocity at $r=\infty$), and $R_*$ is the assumed core radius at which the expansion velocity of the wind in negligible. Typically the velocity law is modified at low expansion velocities to
yield an exponential density variation at depth, or to join smoothly onto a hydrostatic solution. However, in practice the inner radius of the model cannot be well determined from observations; see, e.g., \citet{1991IAUS..143...59H,Najarro1997} and \citet{Hamann2004}. Fortunately, the spectrum and parameters are not sensitive to the adopted value; its choice affects only the velocity law; see \citet{Aadland2022}.  The value of $\beta$ is a measure of how steeply the wind is accelerating.  \citet{Hillier1989} found that the stellar winds in WCs undergo a large increase at large radii ($r>10R_*$) and thus adopting a two-component approximation results in better fits to the line widths.  We adopt this two-component approach here,  with $\beta=1.0$ assumed for the inner regions, and $\beta=20$ assumed for the outer regions.  Thus our modeling requires fitting both $v_{\infty,1}$ and $v_{\infty,2}$, the terminal velocities of the inner and outer regions, respectively.  These values were determined for each star by fitting the line widths of higher and lower excitation states, respectively. 

As described in \citet{Aadland2022}, \cmfgen generally assumes a clumping law of the form
$$f(r)=f_\infty + (1-f_\infty) e^{(-v(r)/v_{\rm cl})},$$
where $f_\infty$ is the clumping filling factor at large radii, and $v_{\rm cl}$ is the onset velocity.
In accord with other treatments, we will refer to $f_\infty$ simply as $f$.  For all of our stars,
we assumed a value of $f = 0.05$, as \citet{Aadland2022} found this resulted in somewhat better fits.  As discussed in the Appendix of \citet{Aadland2022} values of 
$f_\infty > 0.2$ can be ruled out based on the red side of the wings of many emission line profiles.
As described in the Appendix, we used a modified clumping law to address the long-standing issue with fitting the O\,{\sc vi} $\lambda \lambda$3811,34 doublet.

As we did in \citet{Aadland2022}, our dereddening consisted of two components: (a) We adopted a foreground component of $E(B-V)=0.08$~mags based on the median dust emission \citep{Schlegel}, and corrected
for it using the \citet{Cardelli1989} (``CMM") extinction law. (b) We added in an LMC component using the reddening law found by \citet{Howarth1983}. The use of the latter is important for getting a good match to the UV. In both cases a ratio of total-to-selective extinction $R_V$ of 3.1 was assumed.  The LMC $E(B-V)$ values were adjusted until a suitable agreement with the measured flux was found, after correcting the observed fluxes for the of 50~kpc distance to the LMC (\citealt{vandenBergh2000}, \citealt{Pietrzynski2019}).

Since models with the same effective temperatures, composition, and values for $\frac{\dot{M}^2}{R_*^3}$ will have similar spectra \citep{Schmutz1989}, once we obtained a good fit with some scaling, we then matched the flux level by adjusting the luminosity and radius so that the temperature remains fixed ($L\propto T_*^4 R_*^2 = T_{2/3}^4 R_{2/3}^2$) and adjusted $\dot{M}$ to keep $\frac{\dot{M}^2}{R_*^3}$ constant.

For all of our models, we assumed one-half solar abundance for most elements (Na, Mg, Al, Si, P, S, Cl, Ar, K, Ca, Ti, Cr, Mn, Fe, and Ni)\footnote{Some minor abundances, such as Na, are also expected to change 
             in the core due to other minor nuclear reactions chains (e.g, \citealt{1986ApJ...304..695P}). However the neglect of these  changes will not affect the conclusions reached in this paper.}.  The 0.5 $Z_\odot$ abundances are usually taken as typical for the current generation of stars in the LMC, and is based primarily from the oxygen abundances measured in H\,{\sc ii} regions (\citealt{1990ApJS...74...93R,2017MNRAS.467.3759T}; for further discussion, see \citealt{2017MNRAS.466.4403N} and \citealt{2021ApJ...922..177M}). We set the abundances of hydrogen and nitrogen to 0.0, and fit the values for helium, carbon, and oxygen.  The neon value was set to a mass fraction of 0.011 for consistency with \citet{Aadland2022}. In retrospect, the neon value should probably have been about half that value for some of the stars at least, as the models over-predict some Ne lines in the UV.  However, lowering the neon abundance for a few test cases showed that this had no affect on the fits for the other lines, and we have retained the models computed with these higher values. A slightly lower sulpher abundance might also be warranted, but this conclusion is very uncertain since it is based on a single sulpher line in the UV.

\subsection{Modeling the WC Spectra}
The two WC4 stars BAT99-61 and BAT99-90 were modeled following the guidelines described in \cite{Aadland2022}. This process was made easier as we were able to begin with the models used in that work, and simply adjusted the parameters to obtain good fits for BAT99-61 and BAT99-90.

\subsubsection{BAT99-61}
The initial model for BAT99-61 was based upon the best fit model for BAT99-11 given in \citet{Aadland2022}.  The two stars' continuum flux differs by about a factor of 0.75, so the BAT99-11 model was scaled as described above. The strength and shape of the emission lines in BAT99-61's spectrum matched that of the model fairly well.  This meant that only minor changes needed to be made.

The terminal velocity of the outer component ($v_{\infty,2}$) was decreased to help fit the width of the low-excitation lines, specifically C\,{\sc iii} $\lambda$2289, C\,{\sc iii} $\lambda$9710, and He\,{\sc i} $\lambda$10830.

The mass-loss rate and luminosity were increased to adjust the continuum and the line strengths of C\,{\sc iii} to C\,{\sc iv} and He\,{\sc i} to He\,{\sc ii}.  Increasing the luminosity would increase the strength of C\,{\sc iv} and He\,{\sc ii}, but decrease the strength of C\,{\sc iii} $\lambda$9710 and He\,{\sc i} $\lambda$10830.  Increasing the mass-loss rate would increase the C\,{\sc iii} $\lambda$9710 and C\,{\sc iv} $\lambda$4650 in particular since they were more sensitive to the mass-loss rate than other lines in the spectrum.  The final adopted luminosity and mass-loss rate was a compromise that worked well for all the lines.

The chemical abundances of carbon and helium did not need to be changed substantially, as the C/He line pairs were a good fit to the spectrum.  However, the oxygen lines were too strong by about 40\% for the O\,{\sc v} $\lambda$5596 and by over a factor of 2 for O\,{\sc iv} $\lambda$3400.  Since O\,{\sc iv} $\lambda$3400 has been high in our previous WC models as well, we based the oxygen abundance on the O\,{\sc vi} $\lambda$5291 and O\,{\sc v} $\lambda$5596 lines.

\begin{figure}[ht!]
\plotone{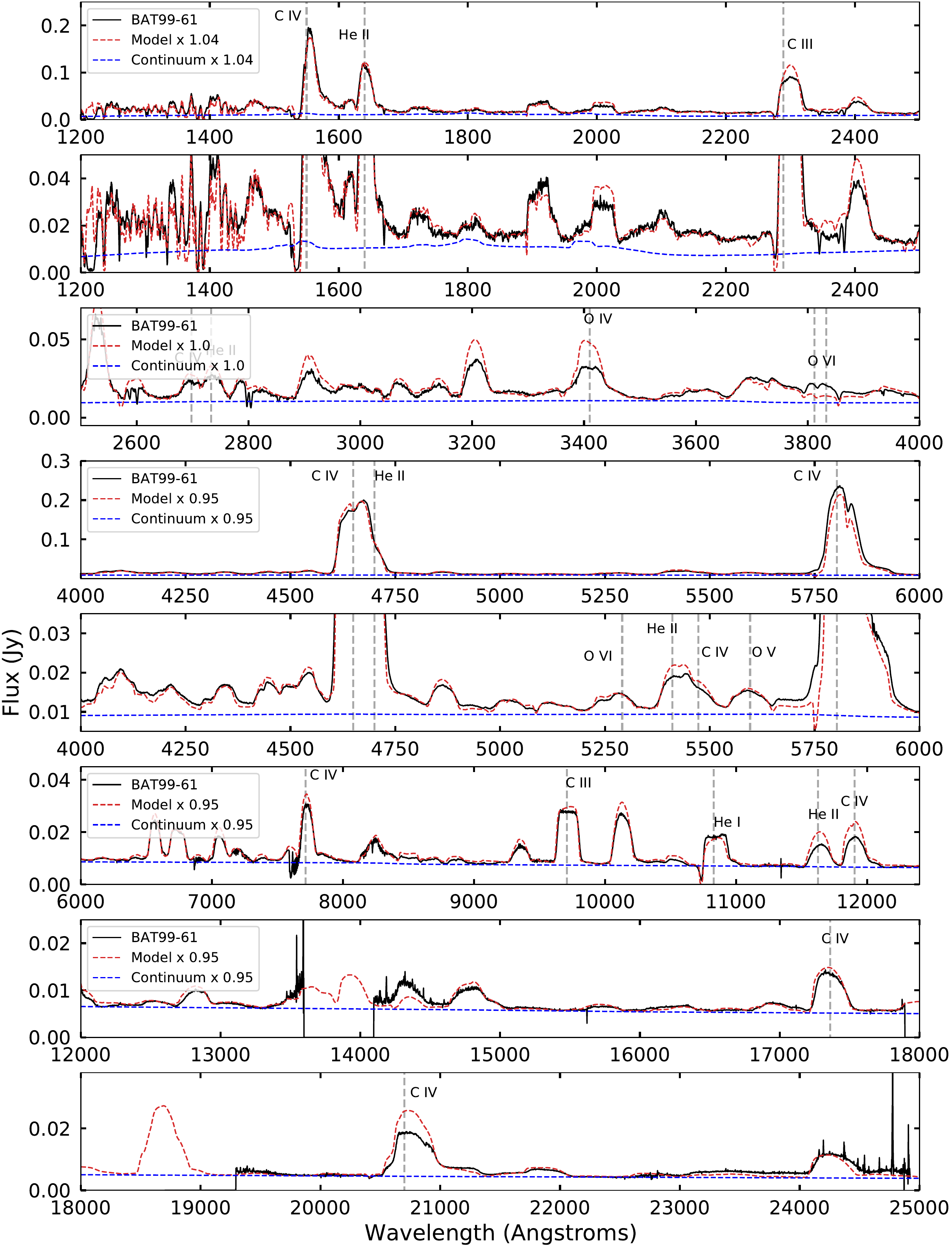}
\caption{Best fit model to the WC4 star BAT99-61. The observed spectrum of BAT99-61 (black) is shown along with the (reddened) best fit \cmfgen model (red dashed) and the model's continuum (blue dashed).  The model was scaled to the stellar continuum. The spectrum of this star, along with those of the others analyzed in this paper and in \citet{Aadland2022}, are available as the data behind the figure.
\label{Fig-BAT99-61}}
\end{figure}

Overall, these slight adjustments resulted in our final model for BAT99-61, shown in Figure~\ref{Fig-BAT99-61}, with its parameters given in Table~\ref{tab:PP}.

\subsubsection{BAT99-90}

The initial model for BAT99-90 was based upon the adopted BAT99-8 model in \citet{Aadland2022}.  The continuum flux needed to be scaled by a factor of 0.5, which was done by adjusting the luminosity, mass-loss rate, and radius.  Again, there was already a good agreement between the model and the observed spectrum so only minor changes needed to be made.

The mass-loss rate was increased to match the strength of the He\,{\sc i} $\lambda$10830,  C\,{\sc iii} $\lambda$9710, and C\,{\sc iv} $\lambda$4650 lines.  By increasing the mass-loss rate, the C\,{\sc iv} $\lambda\lambda$5801,12 lowers in strength slightly, so the luminosity was increased to mitigate this.  The outer component ($v_{\infty,2}$) of the velocity law was decreased to narrow the line width of the low-excitation lines (C\,{\sc iii} and He\,{\sc ii}).  

The only other change needed was increasing the oxygen abundance to better match the O\,{\sc iv}, O\,{\sc v}, and O\,{\sc vi} lines, particularly O\,{\sc vi} $\lambda$5291 and O\,{\sc v} $\lambda$5596.   A side effect of changing the oxygen abundance was that a slight increase in helium and a slight decrease in carbon was needed.  

BAT99-90 is significantly more reddened than the other five WCs: the LMC reddening component $E(B-V)_{\rm LMC}$=0.27~mag, while the other stars are all much more lightly reddened ($E(B-V)_{\rm LMC}$=0.03-0.08). \citet{Grafener1998} found a similar reddening value at $E(B-V)_{\rm LMC}$=0.28~mag.  This higher value is easily explained, as BAT99-90 is located on the southern edge of the 30 Dor H\,{\sc ii} complex, where the reddening is known to be higher than in the rest of the LMC \citep{2011AJ....141..158H}.

\begin{figure}[ht!]
\plotone{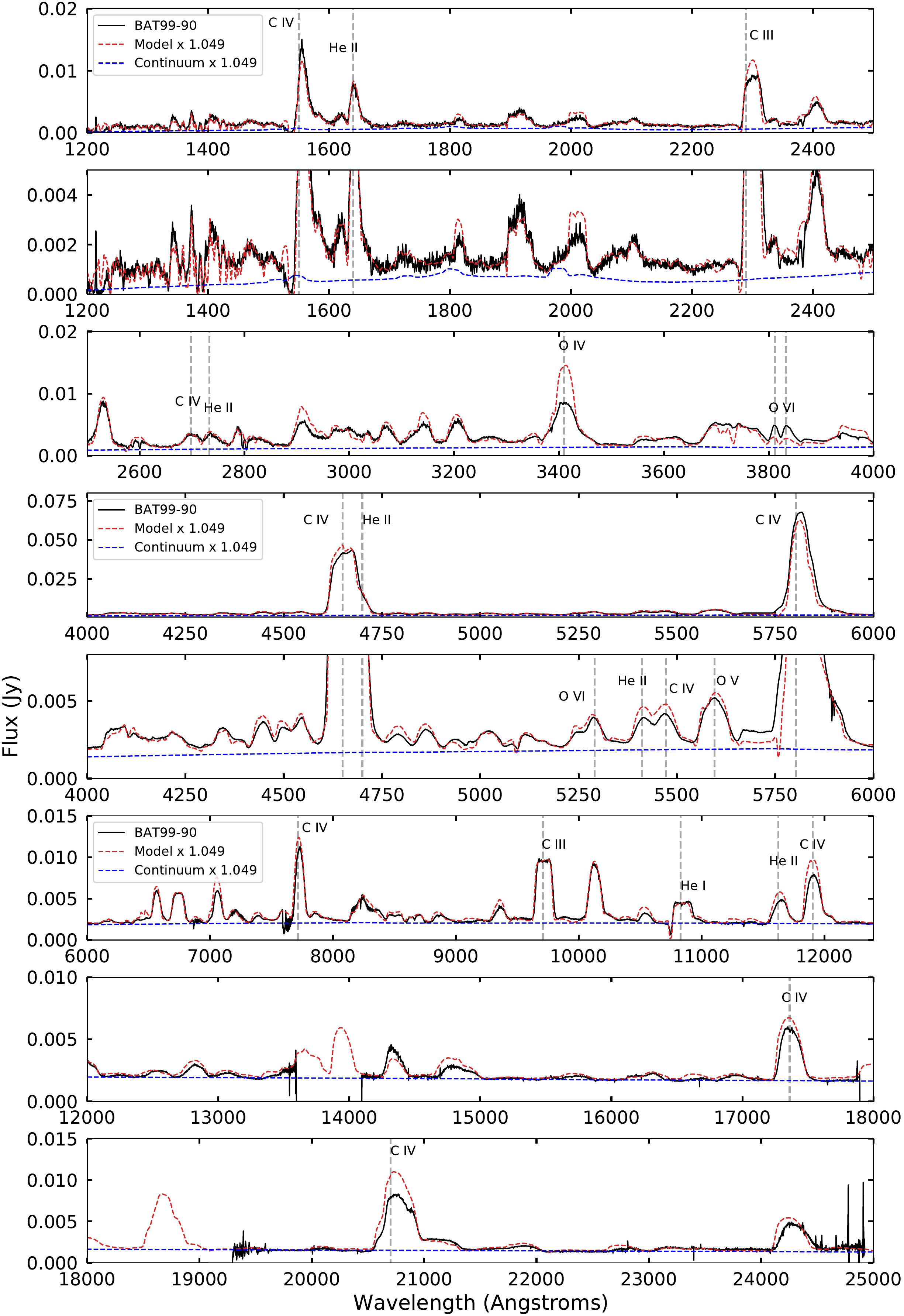}
\caption{Best fit model to the WC4 star BAT99-90. The observed spectrum of BAT99-90 (black) is shown along with the best fit  (reddened) \cmfgen model (red dashed) and the model's continuum (blue dashed).  The model was scaled to the stellar continuum.
The spectrum of this star, along with those of the others analyzed in this paper and in \citet{Aadland2022}, are available as the data behind the figure.
\label{Fig-BAT99-90}}
\end{figure}

The changes resulted in our best fit model for BAT99-90, shown in Figure~\ref{Fig-BAT99-90}.  The parameters are given in Table~\ref{tab:PP}.

\subsection{Modeling WO spectra}

The \cmfgen models described above and in \citet{Aadland2022} were then utilized as a starting basis for our modeling Sanduleak~2 (WO3) and LMC195-1 (WO2), since our WC4 and WO spectra are extremely similar. The spectra of Sanduleak~2 and LMC195-1 are nearly identical, with the difference in spectral subtype due only to the stronger O\,{\sc vi} $\lambda \lambda$3811,34 line in LMC195-1. (The C\,{\sc iv} $\lambda\lambda$5801,12 line is also stronger in Sanduleak~2 than in LMC195-1.)  The WOs were initially modeled together (using one model to represent both WOs) as the changes that needed to be made to the WC spectrum were similar for both WO stars. 

The continuum flux of the WC-type stars is about a factor of 3 higher than that of the WO-type stars. Thus, we began by adjusting the luminosities, mass-loss rates, and radii in order to preserve the strengths of the spectral features.

\begin{figure}[ht!]
\plotone{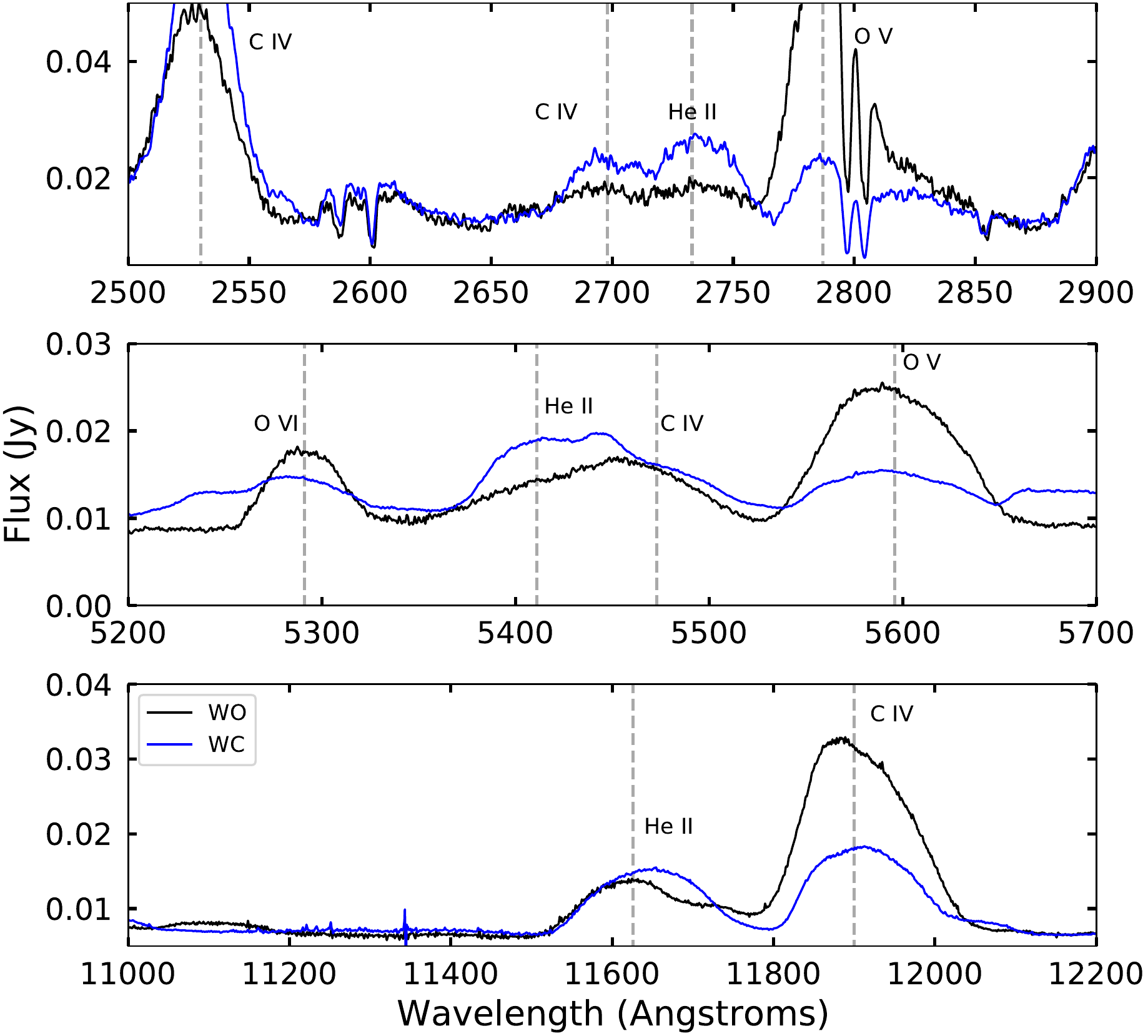}
\caption{Comparison in the C\,{\sc iv}/He\,{\sc ii} line pairs in a WC (BAT99-61) and a WO (Sanduleak~2).  The WO (black) requires a higher carbon abundance and a lower helium abundance than the WC (blue) in order for the model to match the spectrum.  The WO has been scaled by 7.75 in the first panel and by 7 in the last two in order for its continuum to be at a similar flux to the WC.
\label{Fig-WCvsWO}}
\end{figure}

The next change was the adjustment of the C/He abundance ratios in order to get the He\,{\sc ii}/C\,{\sc iv} line pairs to match those of the WO spectra. For instance, the He\,{\sc ii} $\lambda$5411/C\,{\sc iv} $\lambda$5473 line pair is blended to a greater degree than in the WC stars with the He\,{\sc ii} line being a factor of 2 lower in strength than in the WCs.  For the line pair in the NIR, the C\,{\sc iv} $\lambda$11905 was almost a factor of 2 stronger in the WOs and the He\,{\sc ii} $\lambda$11626 line is weaker by about 25\%.  Figure~\ref{Fig-WCvsWO} illustrates these differences in the line pairs for a WC and a WO.  The carbon lines in these  pairs are stronger than the helium, so the helium was decreased and the carbon was increased.  These changes were also needed to match the other He and C lines in the WO spectrum.  

The mass-loss rates and the luminosities were also decreased since the lines sensitive to these parameters were much weaker in the WOs; the He\,{\sc i} $\lambda$10830 line has disappeared\footnote{An emission feature near 10830\AA\ is due to O\,{\sc vi} (10z $^2$Z - 9z $^2$Z), where z is used to denote high angular momentum states.} in the WO spectra, the He\,{\sc ii} $\lambda 6560$ line was a third weaker, and the He\,{\sc ii} $\lambda$4686, C\,{\sc iv} $\lambda$4650 blended feature was about half the flux of that in the WC.  

Although the oxygen lines, O\,{\sc iv} $\lambda$3400, O\,{\sc vi} $\lambda$5291 and O\,{\sc v} $\lambda$5596, are all stronger in the WOs than the WCs (by as much as a factor of 2), once the carbon and helium lines were well fit, the mass fraction of oxygen had to be decreased from that used in the WC models in order to get a good fit.

The equivalent width of the O\,{\sc vi} $\lambda\lambda$3811,34 doublet that defines the WO class is a factor $\sim$5 times stronger in Sanduleak~2, and $\sim$13 times stronger in LMC195-1, than for the WCs with the strongest line. The oxygen abundance was increased as an attempt to fit the doublet.  However, the doublet remained unchanged, similarly to what was found by \cite{Sander2012} and \citet{Aadland2022} in their modeling efforts.  In other words, \textit{this feature is insensitive to the oxygen abundance.} Rather, a good fit was achieved thanks to the higher effective temperature we required for the other lines.  An adjustment in  the clumping (as described in the Appendix) then resulted in an excellent fit.  The only cost of this change was that the fit for the C\,{\sc iv} $\lambda \lambda$1548,41 doublet became worse. Note that this improved approach in clumping resulted in a better fit of the O\,{\sc vi} $\lambda \lambda$3811,34 feature in the WOs than we achieved for the WC4s using our previous approach.

\begin{figure}[ht!]
\plotone{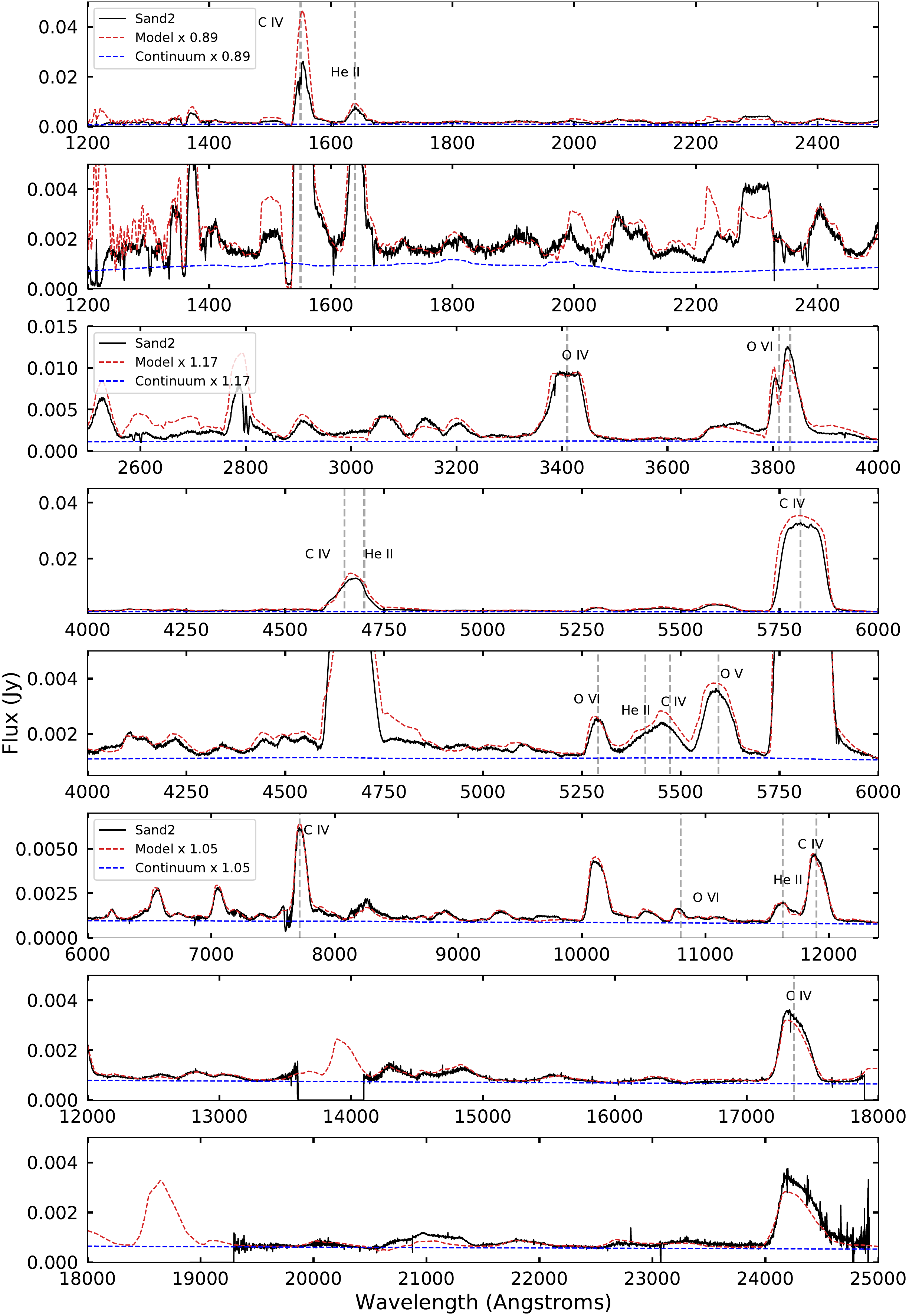}
\caption{Best fit model to the WO3 star Sanduleak~2. The observed spectrum of Sanduleak~2 (black) is shown along with the best fit (reddened) \cmfgen model (red dashed) and the model's continuum (blue dashed).  The model was scaled to the stellar continuum.  Note that in the WOs, the line at $\lambda$10800 is an O\,{\sc vi} line and not the He\,{\sc i} that is in the WCs. The spectrum of this star, along with those of the others analyzed in this paper and in \citet{Aadland2022}, are available as the data behind the figure.
\label{Fig-Sand2}}
\end{figure}

After the majority of the lines were fit for both Sanduleak~2 and LMC195-1, we took the shared model and modeled the differences separately.  Recall that the primary classification criterion between WO3 and WO2 is the ratio of O\,{\sc vi} $\lambda\lambda$3811,34 to O\,{\sc v} $\lambda$5590 blend \citep{Crowther1998}.  The two WO stars differed primarily in the strength of the O\,{\sc vi} $\lambda\lambda$3811,34 doublet as noted above, although
the C\,{\sc iv} $\lambda\lambda$5801,12 line is stronger in Sanduleak~2. The initial model used Sanduleak~2 as a reference for the changes, so nothing was adjusted. Our final fit for Sanduleak~2 is shown in Figure~\ref{Fig-Sand2}.

\begin{figure}[ht!]
\plotone{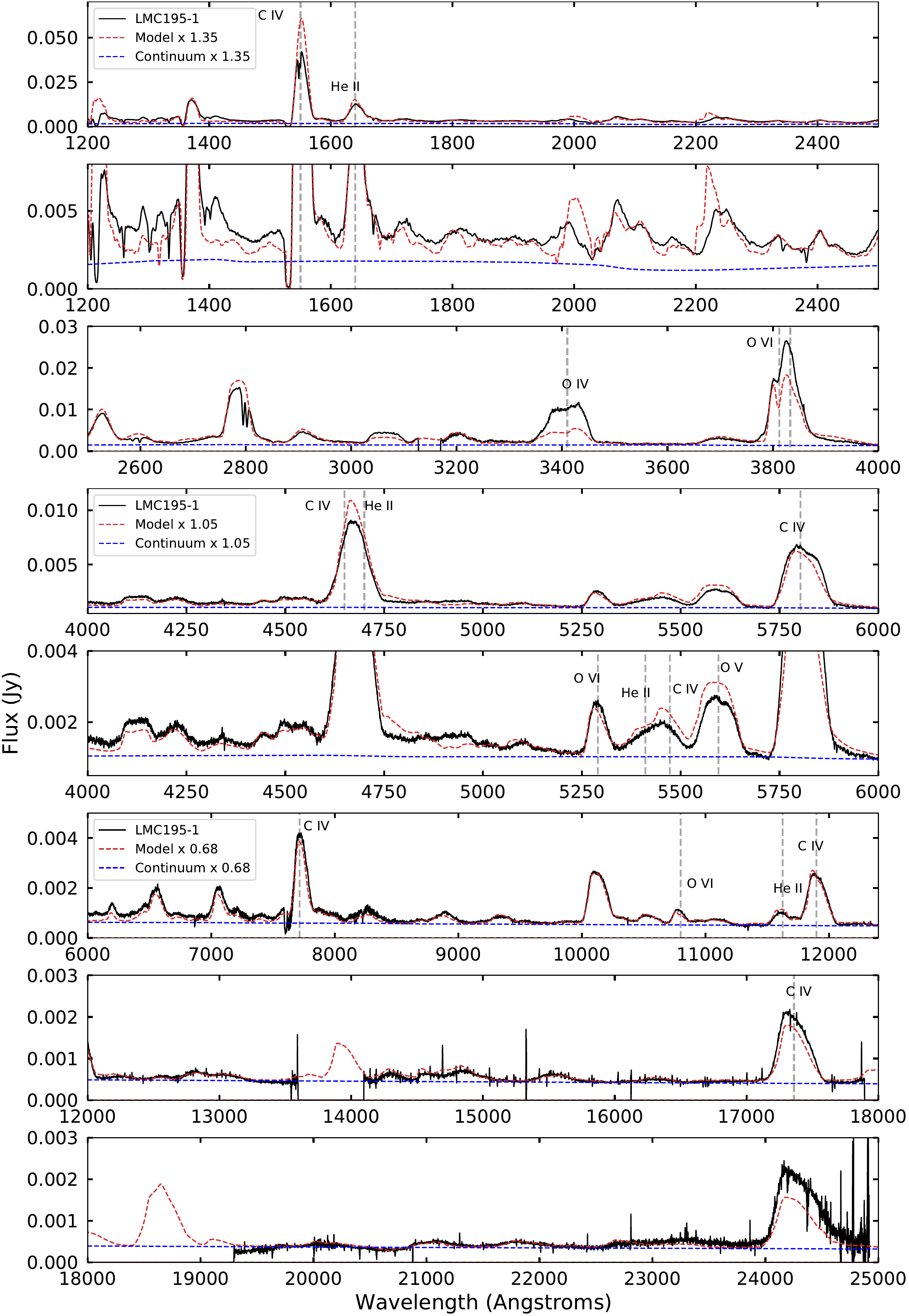}
\caption{Best fit model to the WO2 star LMC195-1. The observed spectrum of LMC195-1 (black) is shown along with the best fit (reddened) \cmfgen model (red dashed) and the model's continuum (blue dashed).  The model was scaled to the stellar continuum. Note that in the WOs, the line at $\lambda$10800 is an O\,{\sc vi} line and not the He\,{\sc i} that is in the WCs. The spectrum of this star, along with those of the others analyzed in this paper and in \citet{Aadland2022}, are available as the data behind the figure.
\label{Fig-LMC195}}
\end{figure}

For LMC195-1, we decreased the mass-loss rate as the C\,{\sc iv} $\lambda\lambda$5801,12 line was extremely sensitive to this parameter after the velocity law and clumping changes\footnote{This can be understood in terms of the branching
ratios from the upper level leading to the C\,{\sc iv} $\lambda \lambda$5801,12 line.
The C\,{\sc iv} $\lambda \lambda$5801,12 doublet comes from the 3p to 3s transition.  However, depopulating
the 3p level to 2s (a line at 312~\AA) is heavily favored, as the branching ratio is $\sim$140:1.  The presence of a strong C\,{\sc iv} $\lambda \lambda$5801,12 doublet in the optical thus requires the 312~\AA\ doublet to be optical thick. In WO stars, the large effective temperature triggers an increase in carbon ionization. As the ionization increases, the ground state population of C\,{\sc iv}  becomes increasingly sensitive to the radiation field, and a small increase can trigger a large reduction in its population causing a large reduction in the optical depth of the  312~\AA\  doublet, and hence a large change in the strength of C\,{\sc iv}$\lambda \lambda$ 5801,12 doublet.}
This line proved to be so sensitive to the mass-loss rate that a difference of less than 0.05 dex was required to decrease the strength of the line by roughly a factor of 5.  The adopted fit for LMC195-1 is shown in Figure~\ref{Fig-LMC195} and the parameters for both WO models are listed in Table~\ref{tab:PP}.

\begin{figure}[ht!]
\plotone{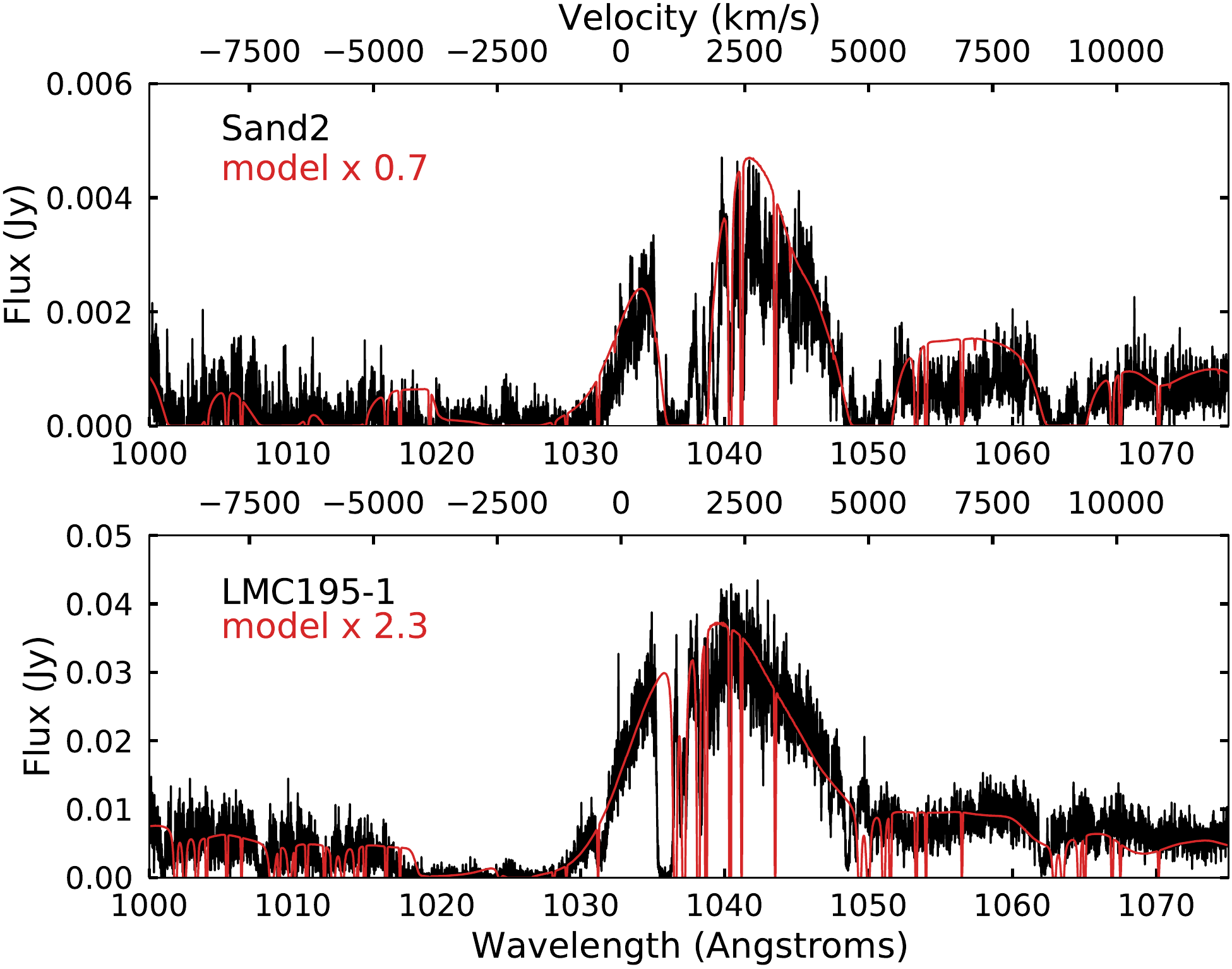}
\caption{Best fit model for the WO stars compared to the O\,{\sc vi} $\lambda \lambda 1032,38$ resonance doublet. A small portion of the FUV COS spectra is shown for Sanduleak~2 (top) and LMC195-1 (bottom) in black. Although we did not use this O\,{\sc vi} line in our fitting, in each case it is well matched by our adopted models.  We also show the interstellar absorption lines from H\,{\sc i} and molecular H$_2$, using column densities of $10^{22}$ and $10^{21.2}$ cm$^{-2}$, respectively, for Sanduleak~2.  For LMC195-1 we used column densities of $10^{21.3}$ and $10^{19.5}$ cm$^{-2}$. However, these were not broadened to the instrumental resolution, and are there to illustrate their locations.} The models have been reddened and were scaled to the stellar continua.  The velocity scale has been corrected for the radial velocity of the star, and is relative to the blue component of the doublet.
\label{Fig-cos}
\end{figure}

We did not use the O\,{\sc vi} $\lambda \lambda$1032,38 resonance doublet when developing our models, but Figure~\ref{Fig-cos} shows how well they work at fitting these lines.  This far in the UV the spectra are strongly affected by interstellar H\,{\sc i} and molecular H$_2$ absorption; in addition, the LMC195-1 profile also has a strong C\,{\sc ii} $\lambda$1036 feature (see, e.g., Figure 13 in \citealt{1996Ap&SS.239..315J}) that we did not attempt to include in our correction. Futhermore, the shape of the extinction law is poorly known.  We have corrected the model as best we can for line absorption from H\,{\sc i} and molecular H$_2$ by adjusting the assumed column densities\footnote{A comparison of our Sanduleak~2 COS spectrum with the FUSE spectrum used by \citet{Crowther2000} shows good agreement in structure but the peak flux at the O\,{\sc vi} $\lambda \lambda$1032,38 feature is about 30\% lower in the COS spectrum.}.

The match between the models and the O\,{\sc vi} feature itself is excellent.  Note that although the line is shifted to the red (compared to the bluer component) due to electron scattering (the ``Auer--van Blerkom Effect", see \citealt{1972ApJ...178..175A} and \citealt{1984ApJ...280..744H}), the models reproduce this shift well.

\subsection{Uncertainties}

The uncertainties for the chemical abundances were found using the same technique as described in \cite{Aadland2022}.  The abundances were adjusted until the model spectrum no longer fit the observed spectrum; in this way, the uncertainties act as the limits for what is an acceptable abundance. These values are extremes; i.e., they should be treated as if they were 3$\sigma$ uncertainties, not 1$\sigma$. 

For the C/He ratio uncertainty, the abundances were adjusted until the He\,{\sc ii} and C\,{\sc iv} line pairs were clearly no longer representative of the spectrum.  The carbon and helium abundance uncertainties are found as a ratio, as these make up the vast majority of the composition, and increasing one requires decreasing the other.  The line pair He\,{\sc ii} $\lambda$5411/C\,{\sc vi} $\lambda$5473 is more sensitive to changes in the carbon and helium abundances than are the line pairs at He\,{\sc ii} $\lambda$2733/C\,{\sc iv} $\lambda$2698 and He\,{\sc ii} $\lambda$11626/C\,{\sc iv} $\lambda$11905.  When all three line pairs no longer represent the spectrum, we use that as a maximum or minimum value for the uncertainty in the C/He ratio.  This process gave a C/He uncertainty for the WCs of 30\% and an uncertainty for the WOs of 50\%.  Note that the uncertainty in the lower-limit for the C/He ratio in the WCs is more generous than what we adopted in \citet{Aadland2022} ($^{+30\%}_{-10\%}$), which emphasizes the somewhat subjective nature of the determinations.

As the oxygen abundance is much smaller than both the carbon and the helium abundance, changing it does not affect the other abundances significantly.  Therefore, the oxygen abundance can be found independently of the carbon and helium abundance.  The oxygen abundance relies primarily on the strengths of the O\,{\sc vi} $\lambda$5291 and O\,{\sc v} $\lambda$5596 lines. (The latter is a blend with O\,{\sc iii} $\lambda$5592, but at these temperatures its contribution will be negligible.)     

\begin{figure}[ht!]
\plotone{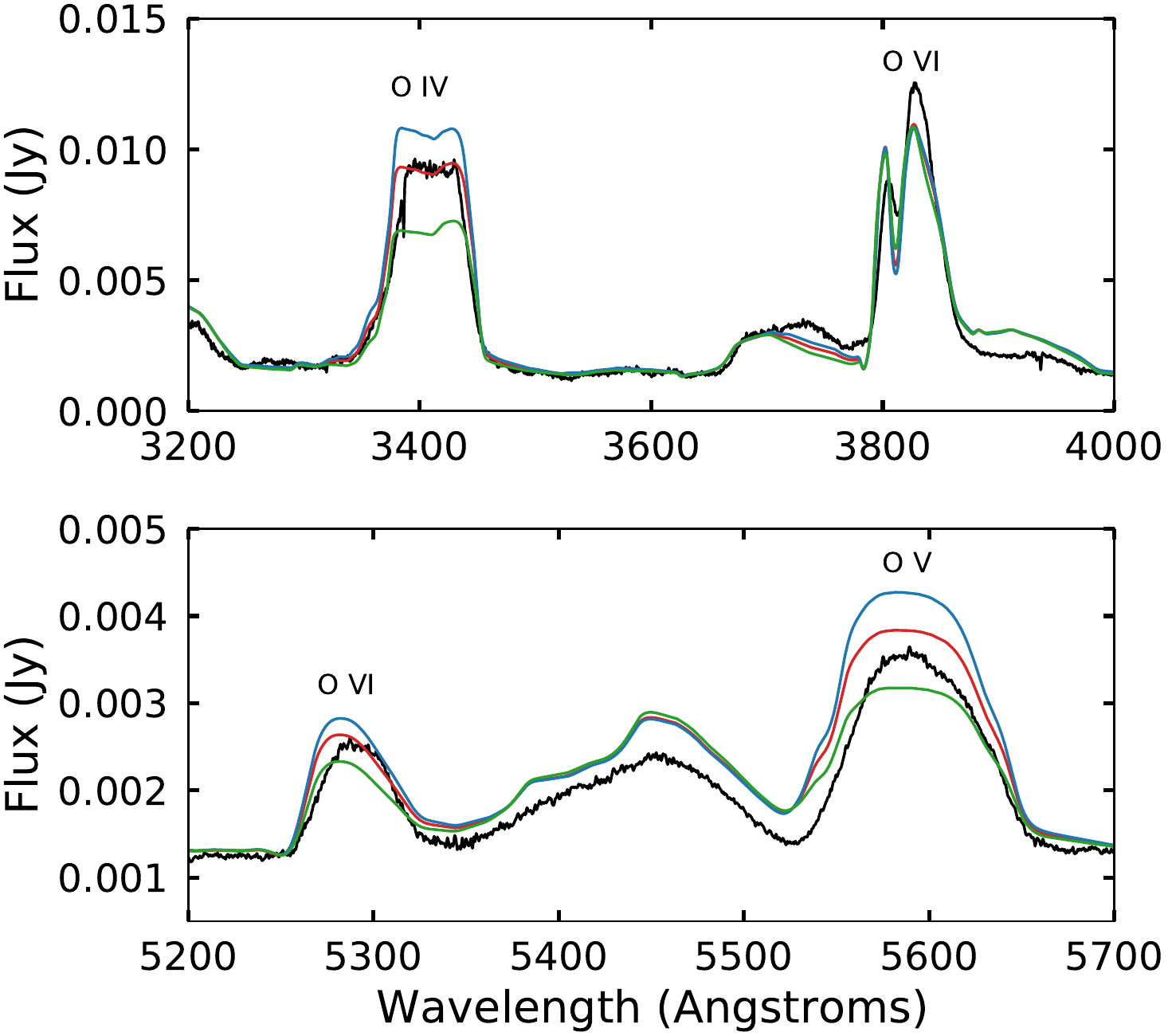}
\caption{Uncertainty determination for the oxygen abundance in the WO3 star Sandueak 2. We show the observed spectrum of Sanduleak~2 (black) with the previously described best fit model (red), the model's upper oxygen limit (blue), and the model's lower oxygen limit (green).  The models were scaled to the stellar continum by 1.17$\times$ for best model, 1.2$\times$ for lower limit, and 1.16$\times$ for upper limit.  The oxygen lines used to determine the oxygen uncertainties on Sanduleak~2 are labeled, as well as the O\,{\sc vi} $\lambda \lambda$3811,34 doublet for comparison.
\label{Fig-S2-oxy-unc}}
\end{figure}

For the WOs, the O\,{\sc iv} $\lambda$3400 line was also used as a reference for the oxygen uncertainty.  In Figure~\ref{Fig-S2-oxy-unc}, the oxygen uncertainties for Sanduleak~2 are shown.  The upper limit on the oxygen was determined by setting the oxygen abundance such that all three of these strong oxygen lines (O\,{\sc iv} $\lambda$3400, O\,{\sc vi} $\lambda$5291, and O\,{\sc v} $\lambda$5596) were high.  The O\,{\sc v} $\lambda$5596 line was already a bit high in our best fit model, so the limit was set when this line was also unreasonably high for the spectrum. The lower limit was set such that the three oxygen lines were all weaker than the spectrum's.  This was the practice followed for the other stars as well. 

These uncertainties are included in Table~\ref{tab:PP} along with the adopted values.

\begin{deluxetable}{l l c c c c c c c c c c c c c l} 
\tablecaption{\label{tab:PP} Physical Parameters}
\tablewidth{0pt}
\tablehead{
\colhead{ID}
&\colhead{log $\dot{M}$} 
&\colhead{log(L/$L_\odot$)}
&\colhead{R$_{2/3}$\tablenotemark{a}}
&\colhead{$T_{\rm eff}$\tablenotemark{a}}
&\colhead{$f$}
&\colhead{$\log \dot{M}/\sqrt{f}$}
&\colhead{X(He)\tablenotemark{b}}
&\colhead{X(C)\tablenotemark{c}}
&\colhead{C/He\tablenotemark{d}}
&\colhead{X(O)}
&\colhead{(C+O)/He\tablenotemark{d}}
&\colhead{E(B-V)$_{\rm LMC}$\tablenotemark{e}}
&\colhead{$V_{\infty,1}, V_{\infty,2}$\tablenotemark{f}} 
&\colhead{Ref} \\
& & & R$_\odot$ & kK & & & & & & & & & mags & km s$^{-1}$
}
\rotate
\tabletypesize{\scriptsize}
\startdata
BAT99-8  & -4.84 & 5.48 & 2.4 & 87 & 0.05 & -4.2 & 0.50 & 0.41 & 0.83$^{+0.25}_{-0.08}$ & 0.077$^{+0.063}_{-0.017}$ & 0.97 &0.05 &  1600, 2600 & A22\\
  BAT99-9 & -4.85 & 5.48 & 2.6 & 84 & 0.05 & -4.2 & 0.66 & 0.29 & 0.43$^{+0.13}_{-0.04}$& 0.041$^{+0.037}_{-0.008}$ & 0.50 &0.08 & 1600, 2300 & A22\\
BAT99-11 & -4.42 & 5.86 & 5.5 & 72 & 0.05 & -3.8 & 0.72 & 0.25  & 0.34$^{+0.11}_{-0.03}$ & 0.025$^{+0.023}_{-0.005}$ & 0.38 &0.06 & 2100, 3000 & A22\\
BAT99-52 & -4.70 & 5.65 & 3.3 & 83 & 0.05 & -4.0 & 0.47 & 0.42  & 0.90$^{+0.27}_{-0.09}$ & 0.094$^{+0.076}_{-0.017}$ & 1.09 &0.03 & 1800, 2600 & A22\\ \hline
BAT99-61 & -4.41 & 5.88 & 5.7 & 72 & 0.05 & -3.8 & 0.72 & 0.25  & 0.35$^{+0.14}_{-0.10}$ & 0.017$^{+0.008}_{-0.008}$ & 0.37 & 0.08 & 2100, 2800 & This paper\\
& \it{-3.72} & \it{5.55} & \it{13.18} & \it{38.8} & \it{1.0} & \it{-3.7} & \it{0.5} & 0.4 &\it{0.8} & \it{0.1} &\it{1} & \it{0.05} & \it{2800} & \it{G98}\\
& \it{-4.4} & \it{5.68} & \it{6.0} & \it{62} & \it{0.1} & \it{-3.9} & \it{0.74} & \it{0.22} & \it{0.30} & \it{0.03} & \it{0.34} & \nodata & \it{3200} & \it{C02}\\
BAT99-90 & -4.82 & 5.49 & 2.6 & 85 & 0.05 &-4.2 & 0.52 & 0.38 & 0.73$^{+0.30}_{-0.23}$ & 0.092$^{+0.040}_{-0.044}$ & 0.91 & 0.27 & 1600, 2400 & This paper \\ 
& \it{-4.06} &\it{5.13} & \it{6.46} & \it{43.6} & \it{1.0} & \it{-4.1} & \it{0.3} & \it{0.4} & \it{1.3} & \it{0.3} & \it{2.3}&  \it{0.28} & \it{2300} & \it{G98}\\ 
& \it{-4.8} &\it{5.44} & \it{3.4} & \it{72} & \it{0.1} & \it{-4.3} & \it{0.45} & \it{0.43} & \it{0.96} & \it{0.11} & {\it 1.20} & \nodata & \it{2600} & C02\\ \hline
Sanduleak~2 & -4.90 & 5.41 & 1.4 & 109 & 0.05 & -4.2 & 0.30 & 0.62 & 2.07$^{+1.05}_{-1.05}$ & 0.071$^{+0.013}_{-0.020}$ & 2.30 &  0.08 & 3800 & This paper \\
& \nodata & \nodata & \nodata & \nodata & \nodata & \nodata & \it{0.34} & \it{0.51} & \it{1.50} & \it{0.15} & \it{1.94} & \it{0.10} & \it{4500} & \it{K95}\\
& \it{-4.4} & \it{5.10} & \it{1.2} & \it{101} & \it{1.0} & \it{-4.4} & \it{0.1} & \it{0.4} & \it{4} & \it{0.5} & \it{9} & \nodata & \it{3600} & \it{G99} \\
& \it{-4.9} & \it{5.28} & \it{1.2} & \it{110} & \nodata & \nodata & \it{0.27} & \it{0.56} & \it{2.07} & \it{0.16} & \it{2.67} &\nodata & \it{2600} & \it{C00}\\
& \it{-4.85} & \it{5.20} & \nodata & \it{170} & \it{0.3} & \it{-4.3} & \it{0.3} & \it{0.55} & \it{1.83} & \it{0.15} &\it{2.33} & \nodata & \it{3300} & \it{T15}\\
LMC195-1 & -4.93 & 5.41 & 1.4 & 111 & 0.05 & -4.2 & 0.30 & 0.62 & 2.07$^{+1.05}_{-1.05}$ & 0.071$^{+0.013}_{-0.020}$ & 2.30 & 0.05 & 3800 & This paper\\
\enddata
\tablenotetext{a}{R$_{2/3}$ and $T_{\rm eff}$ are the radius and temperature at an optical depth of 2/3.}
\tablenotetext{b}{The extreme uncertainties on X(He) are $\pm0.06$ for BAT99-61, $\pm0.07$ for BAT99-90, and $\pm^{0.15}_{0.08}$ for the WOs.}
\tablenotetext{c}{The extreme uncertainties on X(C) are $\pm0.07$ for BAT99-61, $\pm0.09$ for BAT99-90, and $\pm^{0.09}_{0.15}$ for the WOs.}
\tablenotetext{d}{Ratios are in terms of the mass fractions.}
\tablenotetext{e}{Determined using a foreground reddening component E(B-V)=0.08 plus the LMC component listed here.}
\tablenotetext{f}{We used a two-component velocity law, while the previous studies (other than A22) used a one-component law.}
\tablerefs{A22 - \cite{Aadland2022}; C00 - \cite{Crowther2000}; C02 -\cite{Crowther2002}; G98 - \cite{Grafener1998}; G99 - \cite{Grafener1999};   K95 - \cite{Kingsburgh1995}; T15 - \cite{Tramper2015}.}
\end{deluxetable}

\section{Results} \label{sec:results}
 
All of the physical parameters found in our study (including those from \citealt{Aadland2022}) are included in Table~\ref{tab:PP}. For comparison, we also include the results from previous studies, i.e., \citet{Grafener1998,Grafener1999, Kingsburgh1995,Crowther2000,Crowther2002} and \citet{Tramper2015}. Note that although there have been previous studies of the WO star Sanduleak~2, this is the first study of the other single LMC WO star, LMC195-1. 
In this section we now discuss to what extent these properties agree, and compare the results to evolutionary models.  
These properties will allow us to determine how chemically evolved the WOs are compared to the WCs. 

Additionally, a comparison of the physical properties of these stars with evolutionary models can help us answer an even more fundamental question: how do WRs form?  The formation of WRs might require a binary companion that stripped off the outermost layers of the star \citep{P1967}. The lack of evidence of a companion for these stars could simply mean that the two have since merged.  Or, as the ``Conti scenario'' \citep{Conti1976} offers, mass loss due to the stellar winds of the WR progenitors (O-type stars) may have been sufficient onto themselves to have done the stripping. We doubt that either of these mechanisms is responsible for the formation of {\it all} the WRs in the Universe.  Rather, here we address the specific question of whether or not a prior binary companion is {\it necessary} to explain the observed properties of the eight (apparently) single LMC WC and WO stars that we have observed here.

\subsection{Comparison to Previous Work}

All six of our WCs, and one of our WOs (Sanduleak~2), have been previously analyzed.  Here we compare our results to those of previous studies.
As a reminder, modeling only really determines $\dot{M}/\sqrt{f}$, not $\dot{M}$ itself. The actual mass-loss rates are thus dependent upon what we assume for $f$. (The value adopted for $f$ also affects the quality of the fits; early spectroscopic evidence for clumping in WR stars came about by the poor agreement of unclumped models with the red side of the wings in many of the emission line profiles, as mentioned above; see \citealt{Hillier1991b,Hillier2000}.)  Thus to compare our mass-loss rates with those of others, it is fairer to use $\dot{M}/\sqrt{f}$; it is for that reason that we tabulate those values in Table~\ref{tab:PP}. 

The two newly modeled WC stars have very similar properties to the four WC stars modeled by \citet{Aadland2022}.  Recall that our goal in including these two stars was to increase our sample size and make sure that our results were representative of the WC class itself.  However, it should be remembered that all six of our WCs have the identical excitation type, WC4.  It would be interesting to see what differences, if any, are found with stars of lower excitation, although nearly all WCs (23 total, including binaries) in the LMC are of this subclass except one WC5 and one WC6 binary (see, e.g., \citealt{2018ApJ...863..181N}). On average, the subtypes of Galactic WCs are of later types;  \citet{2000MNRAS.315..407D} and \citet{Sander2012} performed analysis of several such examples. 

\subsubsection{Comparison of Results with Previous Modeling of the WC Stars}

The WC stars have been previously analyzed by \citet{Grafener1998} and \citet{Crowther2002}.  We made comparisons for the first four in \citet{Aadland2022}, and we find the same trends here. As a reminder, the \citet{Grafener1998} modeling used the Potsdam Wolf-Rayet modeling code PoWR \citep{Grafener2002} which, at the time, was not fully blanketed, while \citet{Crowther2002} used an earlier version of the \cmfgen code that we used here. 

In all cases, the \citet{Grafener1998} luminosities are 0.2-0.3~dex lower than our results. The \citet{Grafener1998} modeling did not include clumping (i.e., equivalent to a filling factor $f=1$).  Comparing their values for the mass-loss rates to ours (corrected for the clumping factor) shows excellent agreement, with differences of order 0.1~dex. (Our actual mass loss rates are thus $1/\sqrt{0.05}=4.5\times$ lower.) There is strong disagreement with the \citet{Grafener1998} abundances: their carbon abundances are all 40-50\%, with oxygen mass fractions much higher than ours or other studies. These differences are likely due primarily to the limitations of WR modeling at the time (see below), but may also be partially attributable to the data they used, primarily the old SIT-Vidicon spectrophotometry of \citet{1987ApJS...65..459T} taken by one of the present authors.  Although the SIT-Vidicon data spectroscopy was a great advance over photographic spectra, a comparison to our modern data shows some saturation issues with the strongest lines.

The comparison with \citet{Crowther2002} shows better agreement with the stellar luminosities, $\leq$0.1~dex, although in most cases the luminosities of our stars are slightly higher.  The mass-loss rates, corrected for the different values of $f$ used in our study and theirs, also show good agreement, with differences of 0.2~dex or less.  (Note that the \citealt{Crowther2002} values for $\log{\dot{M}/\sqrt{f}}$ were inadvertently left out of Table 2 in \citealt{Aadland2022}; they are -4.4, -4.4, -4.0, and -4.0 for BAT99-8, BAT99-9, BAT99-11, and BAT99-52, respectively.)  In general, our mass-loss rates are a bit lower than theirs.  A comparison of the abundances show excellent agreement for helium and carbon, with the mass fractions in agreement by 5\% of the total.  Similarly the oxygen abundances also agree very well, with the largest disagreement found for BAT99-9 \citep{Aadland2022}, where we find an oxygen mass fraction of 4\% and \citet{Crowther2002} finds 10\%.  Generally the oxygen mass fractions agreed to far better than this, as can be seen in Table~\ref{tab:PP}, i.e., 2\% vs.\ 3\% for BAT99-61 and 9\% vs.\ 11\% for BAT99-90. In general, our values for the mass fraction of oxygen are slightly lower.

One reason for the enhanced oxygen abundance found by \cite{Grafener1998} is their adopted oxygen model atoms which have considerably fewer levels than used in this study. Their models also neglected clumping and iron blanketing. The latter probably explains the lower luminosities they derived  -- a boost in luminosities was derived with additional blanketing in CMFGEN models of WN and WC stars \cite[e.g.,][]{Hillier1998,Hillier1999}. At that time the abundances found by \cite{Grafener1998} were broadly consistent with other models. The  O (and C) model atoms adopted by \cite{Crowther2002} are similar to those adopted here. Their models also allowed for clumping, and included the influence of iron.

\subsubsection{Comparison of Results with Previous Modeling of the WO Star Sanduleak~2}

Four previous studies have determined the physical properties of the WO3 star Sanduleak~2 (Brey 93). \cite{Kingsburgh1995} used  recombination theory to determine the star's chemical abundances using UV data from the International Ultraviolet Explorer satellite and their own ground-based optical data. As noted earlier, this method works best when the temperature and densities are constant, and when the lines are optically thin, simplifications which are not achieved in the accelerating winds where the WR lines are formed.   \citet{Grafener1999} performed analysis using PoWR of the same FOS data we use here, along with optical spectrophotometry from \citet{1987ApJS...65..459T}; as noted above, those optical data suffered from some saturation issues for the strongest lines.  \citet{Crowther2000} carried out their own detailed analysis with \cmfgen using FUV data from the Far Ultraviolet Spectroscopic Explorer (FUSE), the same FOS UV data used here, and their own optical data. More recently, \cite{Tramper2015} conducted an analysis with \cmfgen of their own optical and NIR spectra using {\sc cmfgen}.

The results of all of these studies are included in Table~\ref{tab:PP} for comparison.  The \citet{Grafener1999} modeling is an extension of the \citet{Grafener1998} study, and suffers from the same limitations as discussed above.  Their abundances are at variation with all other studies. 
As for the \citet{Crowther2000} and \cite{Tramper2015} studies, neither were able to obtain consistent fits with the O\,{\sc vi} $\lambda\lambda$3811,34 doublet which distinguishes WOs from WCs.
In particular, \citet{Crowther2000} found that the O\,{\sc vi} $\lambda \lambda$3811,34 doublet and C\,{\sc iv} $\lambda 7700$ indicated a very high temperature (170,000 K) while the FUV resonance doublet O\,{\sc vi} $\lambda \lambda$1032,38 suggested a much lower temperature (120,000 K). (A principle advantage of their use of FUSE data was to obtain this line. Here we obtained our own COS data for the same purpose.)  This meant that for their abundance determination, Crowther et al.\ (2000) had to assume an intermediate temperature (150,000 K).   (The \citealt{Grafener1999} study found a much lower temperature, 101,000 K.) \citet{Tramper2015} was unable to fit the O\,{\sc vi} $\lambda\lambda$3811,34 line, attributing the problem to over-population of the line's upper level by X-ray excitation.  They did not attempt to utilize the publicly available FUV or UV data. They found carbon and oxygen abundances that were intermediate between the \citet{Kingsburgh1995} and \citet{Crowther2000} studies.

During our modeling, a single set of parameters worked well for the vast majority of the lines, unlike the problem encountered by \citet{Crowther2002}.  The effective temperature of our Sanduleak~2 model is 109,000 K, which is in excellent agreement with the compromise temperature of 110,000~K adopted by \citet{Crowther2000}.

A comparison of these earlier results with ours is instructive.  Our luminosity is a bit higher (0.1-0.2~dex) than most previous studies.  Our mass-loss rate agrees well.  All but the \citet{Grafener1999} study are in surprisingly good agreement with the mass fraction of helium, about 30\%, including the recombination study by \citet{Kingsburgh1995}.  

With the exception of \citet{Grafener1999}, all studies of Sanduleak~2 have found similar values for the carbon mass fraction.  The values range from 51\% \citep{Kingsburgh1995} to 62\% (our study), with the other two \cmfgen studies showing 56\% \citep{Crowther2000} and 55\% \citep{Tramper2015}. Unlike our WC4 stars, carbon, rather than helium, is the dominant constituent (by mass) in WO stars. Our oxygen mass fraction is the lowest, 7\%, with the \citet{Crowther2000} and \citet{Tramper2015} studies indicating a value about twice as high, 15-20\%.
In fact, our oxygen abundances for the WOs are  about the same as what we found for some of the WCs.

Both \citet{Crowther2000} and \citet{Tramper2015} used \cmfgen to model their spectra.  We are using a version that has been updated in multiple ways, and our methodology has also led to the first success at obtaining good fits to both the O\,{\sc vi} $\lambda \lambda$1032,38 resonance doublet and the O\,{\sc vi} $\lambda \lambda$3811,34 doublet simultaneously. That said, can we rule out this higher oxygen abundance?  To answer this, we ran another model for Sanduleak~2 with enhanced oxygen: X(He) = 0.31, X(C)= 0.52, X(O)=0.15. 

\begin{figure}[ht!]
\plotone{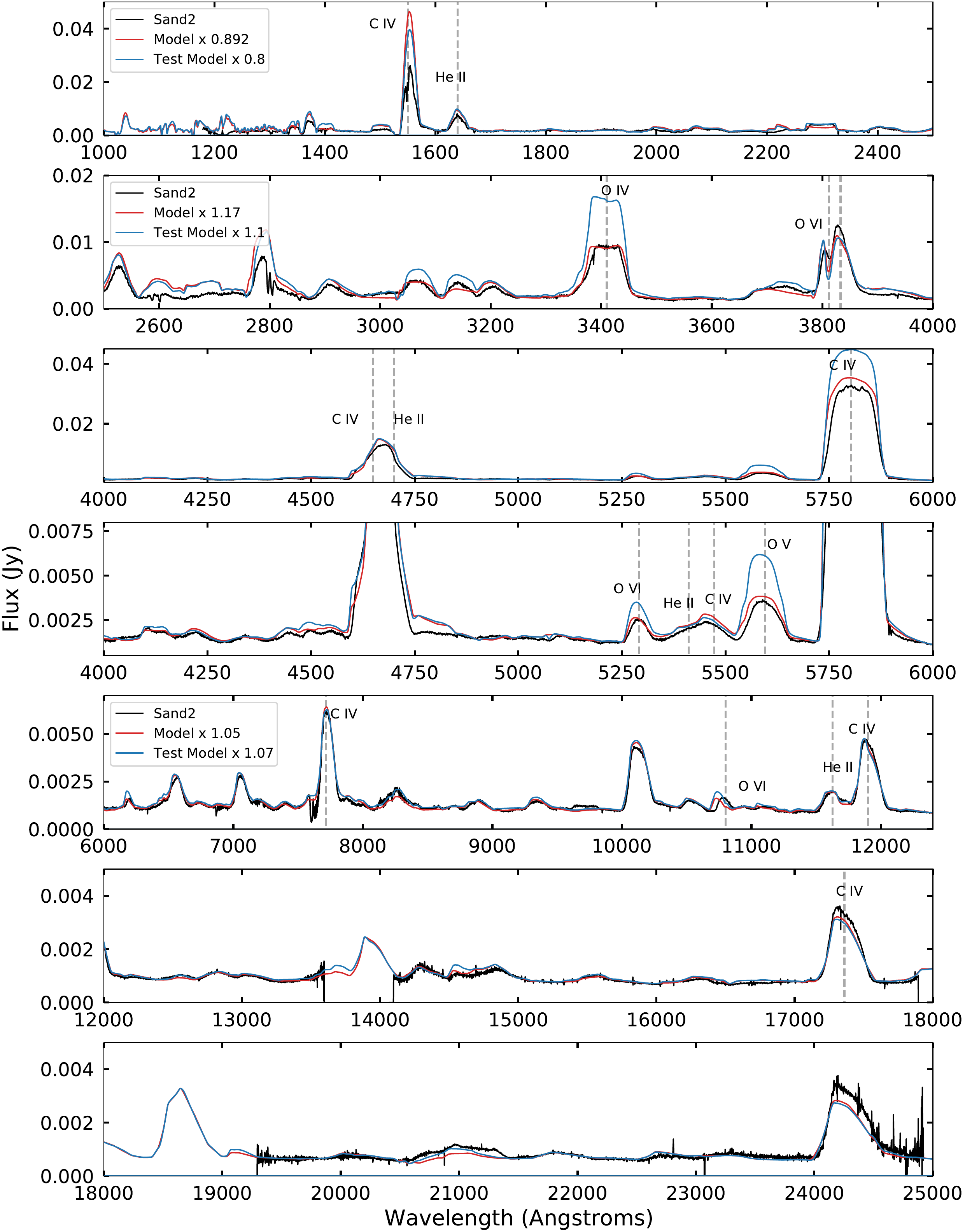}
\caption{Test to see if a modern model based on older parameters can be made to fit Sanduleak~2. The observed spectrum of Sanduleak~2 (black) is shown with the best fit model (red) and the test model (blue).  The test model was constructed  with chemical abundances similar to those of previous studies of Sanduleak~2 as described in the text.  Both models were scaled to the stellar continuum.  The labeled emission lines are the primary ones used for modeling the spectrum.
\label{Fig-Sand2-test}}
\end{figure}

We show the comparison between the enhanced oxygen model and our adopted one in Figure~\ref{Fig-Sand2-test}.  In the model with the higher oxygen abundance, almost all of the oxygen lines including O\,{\sc iv} $\lambda$3400, O\,{\sc vi} $\lambda$5291, and O\,{\sc v} $\lambda$5596 have much stronger fluxes than the observed spectrum.
(These are the primarily reference lines we used in our work.) The C\,{\sc iv} $\lambda$5801,12 doublet is also too strong by about 20\%.  Changing the oxygen by a factor of two, and the helium and carbon slightly, greatly impacts the fit of the oxygen lines.  Our original model with the smaller oxygen abundance is clearly a better fit to the spectrum, and we can rule out the higher oxygen abundance found by \citet{Crowther2002} and \citet{Tramper2015}.

Another note is how little the O\,{\sc vi} $\lambda\lambda$3811,34 doublet has changed in strength, emphasizing our earlier point that this high-excitation line is not impacted by the oxygen abundance.  {\it Our analysis thus shows that the strong O\,{\sc vi} $\lambda\lambda$3811,34 doublet in WOs is not due to them having higher oxygen abundance than WCs, but rather to their having higher effective temperatures and lower wind densities.}

\subsection{Comparison of the Chemical Abundances of the WCs with that of the WOs}

The  key to understanding the relation between the WCs and the WOs is found in a comparison of their surface chemical abundances.  If the WOs are more evolved than the WCs, than we expect them to have less helium, with a larger fraction of the mass in carbon and oxygen. A comparison of the oxygen abundance itself will show if the strong O\,{\sc vi} $\lambda \lambda$3811,34 doublet is due to a higher oxygen content or due to other factors, such as higher  effective temperature and lower wind density. However, as found in the previous section, the strength of this doublet is highly insensitive to the actual oxygen abundance, indicating that the latter is probably responsible.

 First, let us note that the two newly analyzed WCs, BAT99-61 and BAT99-90, show chemical abundances very similar to those of the four WCs previously analyzed by us in \citet{Aadland2022}.  The (C+O)/He ratios of the six WC stars vary from 0.37 (BAT99-61) to 1.09 (BAT99-52).  The individual helium mass fractions range from 47\% to 72\%, while the carbon mass fractions range from 25\% to 42\%.  In all cases, the oxygen abundance is a minor contributor to the mass, ranging from 3\% to 9\%. 
 
 We had previously found that one of our WC stars, BAT99-9, still shows traces of nitrogen, with mass fraction of 0.1\% \citep{Hillier2021}.  When helium burning begins, the $^{14}$N in the core (produced by the hydrogen-burning CNO cycle) will quickly be converted to $^{22}$Ne; mixing and mass-loss (primarily mass-loss) will then remove the nitrogen from the surface that gives WN-type WRs their distinctive spectra.  BAT99-9 is the first WC star found to show any nitrogen, and we suggest that it has only recently entered the WC phase (see further discussion in \citealt{Hillier2021}). Here we find that it is one of the least chemically evolved of the WCs based on its relatively high helium and low carbon abundances.  \citet{Hillier2021} also notes that similarly BAT99-11 seemed to show signs of nitrogen as well, although we did not attempt to model this.  BAT99-11 is even less chemically evolved. (We do not see nitrogen in BAT99-61 with a similar ratio, but suspect that higher signal-to-noise UV and optical data might help.)

  As for the WOs, most previous studies of Sanduleak~2 agree with ours that the helium abundance is about 30\%; we find the same value for LMC195-1.  This is significantly lower than the range of helium abundances found for the WCs. This by itself suggests that the WO stars are more chemically evolved than the WCs.  As for the oxygen abundances, the WCs show a range in oxygen abundances from 2$\pm$1\% to 9$^{+4}_{-2}$\%.  The two WOs have an oxygen abundance of 7$^{+1}_{-2}$\%.  Thus, the WOs are richer in oxygen than the least evolved
WCs, but have about the same oxygen abundance (to within the uncertainties) as the most oxygen-rich WCs.

\subsection{Comparison to Evolutionary Tracks}
\label{Sec-evol}

Evolutionary models provide the framework for interpreting our results. Previous analyses of WRs have compared luminosity and temperature to evolutionary models, but have mostly eschewed comparisons of the chemical abundances, with the notable exception of the study by \citet{Grafener1998}. The temperature is particularly hard to relate between the results of atmospheric analysis and that of evolutionary models due to the atmospheric extent of WRs (sometimes referred to as ``inflation" in the WR literature).  Here we compare our derived abundances, luminosities, and mass-loss rates to the evolutionary models, and evaluate what the models tell us about the lifetimes of these stars.

We compared our results to modern-day evolutionary models: the Geneva single-star evolutionary models \citep{Ekstrom2012, GenevaLMC} and the ``Binary Population and Spectral Synthesis" (BPASS) models \citep{Eldridge2017}.

The Geneva single-star models include the effects of rotation and
many improvements over the older versions. Here we make comparisons with both the models computed with an initial metallicity $Z=0.006$ characteristic of the LMC \citep{GenevaLMC}, and ones computed with $Z=0.014$ \citep{Ekstrom2012}.  The reason for including the latter is that we found better agreement with the properties of the WCs analyzed in \citet{Aadland2022}.  Both metallicities predict surface compositions in accord with what we observed (see Figures 8-9 in \citealt{Aadland2022} and Section~\ref{Sec-abunds} below), but the single-star $Z=0.006$ models do not produce WCs with luminosities as low as what we find (see e.g., Figure 10 in \citealt{Aadland2022} and Section~\ref{Sec-lums} below).  This may be indicative that the mass-loss rates used during the main-sequence phase are too low in the $Z=0.006$ models, or may be telling us that binary evolutionary is necessary to produce WC stars when starting with an initial metallicity as low as that found in the LMC.

To see if our results are consistent with binary evolution, we turn to the BPASS models \citep{Eldridge2017}.  For these we adopt the $Z=0.006$ models and restrict our study to those with WC/WO lifetimes greater than 0.05 Myr (i.e., 50,000 years).  The BPASS models do not include the effects of rotation.

Note that we recognize the WC/WO phase in the evolutionary models using the criteria suggested by \citet{Georgy2012}, with a surface mass fraction of hydrogen $X(H)<0.3$, $\log{T_{\rm eff}}>4.0$, and that $X(C)>X(N)$.  We add the additional constraint that $\log L/L_\odot>4.9$, as binary stripping can, in theory, produce hot stars with the surface compositions of WRs, but whose luminosities are too low to develop the optically thick winds that produce the characteristic WR emission lines \citep{2020MNRAS.491.4406S,Shenar2020}.  In extracting the relevant BPASS models, we made use of the python package {\it hoki} \citep{hoki2020}.

\subsubsection{Chemical Abundances}
\label{Sec-abunds}

 The surface chemical abundances indicate that the WOs are indeed more chemically evolved than the WCs, such that the WOs have a higher carbon abundance and a lower helium abundance.  The oxygen abundance remains unchanged between the WCs and the WOs.  Do the evolutionary models replicate these abundances?
  
  As helium burning proceeds, we expect the helium mass fraction will decrease as the carbon and oxygen mass fractions increase.  Figure~\ref{Fig-COHe-He} shows that the abundances of all eight stars tightly follow the predicted relationship: we see here stellar evolution in action. The (C+O)/He values are significantly larger for the two WO stars compared to the WCs.   Thus, we have answered the question: yes, the WOs are more highly evolved than the WCs.  
  
  The agreement between the various evolutionary models in that figure is slightly deceptive, as we expect that during the helium burning phase the sum of the mass fractions of helium, carbon, and oxygen to equal unity, and hence there should be a very tight theoretical relation between the (C+O)/He ratio and He mass fraction.   But the figure serves to easly demonstrate that the WOs are more chemically evolved, and that among the WCs we studied, BAT99-61 is the least evolved.  Figure~\ref{Fig-HeC} (top) compares our carbon mass fraction as a function of helium  to those predicted by the evolutionary tracks.  As we found in \citet{Aadland2022}, there is excellent agreement for the WC stars with the predictions of the Geneva single-star models, and for a subset of the binary models. As discussed above, the C/He ratios are actually better determined than the individual values, and Figure~\ref{Fig-HeC} (bottom) shows the C/He ratio plotted as a function of helium abundance.  
  
 We find that the two WO stars stand out as having higher than expected carbon. {\it We note that there should be nothing surprising here: both the \citet{Crowther2000} and \citet{Tramper2015} found very similar carbon abundances and C/He ratios (see Table~\ref{tab:PP}).}  What is new here is the realization that these values are strongly out of accord with the predictions of stellar evolution models.
  
  What, then, about the oxygen abundances?  As shown above, the (C+O)/He ratios agree perfectly with the predictions of the models.  Thus, since the carbon abundance is high, indeed the oxygen value is lower than expected for the WOs as shown in Figure~\ref{Fig-HeO}.
  
  We show this again from a slightly different perspective in Figure~\ref{Fig-ca}, where we plot the ratios of the C/He mass fraction
  against that of O/He.  The WOs do not match the predictions of the models.
  
  One thing that is striking in all of these is that the single-star Geneva models do an excellent job of predicting the abundances for the WCs. A subset of the binary BPASS models do as well, but they also allow for much lower carbon abundances and higher oxygen abundances than we observe in any of our stars.

We will return to a discussion of what the carbon and oxygen discrepancies may mean in Section~\ref{sec:PM}.

\subsubsection{Luminosities}
\label{Sec-lums}

The six WC stars have luminosities that span the range from $\log L/L_\odot=5.48-5.86$.  The two WOs have luminosities just slightly smaller,
with $\log L/L_\odot=5.41$.

In Figure~\ref{Fig-L-abunds} we show the luminosities as a function of the
chemical abundances. As we did in \citet{Aadland2022}, we find that the LMC metallicity Geneva models do not predict WCs/WOs with luminosities as low as what we observe.  If we instead rely upon the $Z=0.014$ Geneva models (solid curves), we find that the WCs span a range of initial masses from 32$M_\odot$ to 85$M_\odot$.  The two WOs, on the other hand, both cluster near the 32$M_\odot$. The BPASS models have a wide range of luminosities that include the range we find, as well as predicting WCs at higher luminosities than we have observed.

Figure~\ref{Fig-logL-m} shows the luminosity plotted as a function of the
{\it current} mass.  According to $Z=0.014$ (solar metallicity) Geneva tracks, the BAT99-61 should have a mass in the range of 18 to 28 M$_\odot$ and BAT99-90 would have a mass close to 10 M$_\odot$.  With BAT99-61 being close to the Geneva single-star models that have initial masses of 60 M$_\odot$ and 85 M$_\odot$, this means that BAT99-61 would have lost over half its mass before and during the WC phase.  BAT99-91, which falls between the Geneva single-star initial mass 32 M$_\odot$ and 40 M$_\odot$ solar metallicity models, would have lost 20 to 30 M$_\odot$. The two WOs would have masses around 10 $M_\odot$, meaning they would have lost around 20 M$_\odot$, or 60\% of their initial masses!

\subsubsection{Mass-loss Rates}

How well do our measured mass-loss rates compare to what is assumed in the evolutionary models?  For mass-loss during the WR phase, the Geneva  and BPASS  models both adopt the \citet{2000A&A...360..227N} prescription, where the rates are a strong function of stellar luminosity and abundances. 

When we compare our measured rates to those adopted by the models as a function of luminosity (Figure~\ref{fig-L-massloss}, top) we find reasonable agreement.  However, our mass-loss rates are lower than those expected when we compare the rates as a function of surface abundances (bottom two panels). This is not surprising, as we adopted a filling factor of 0.05 (Table~\ref{tab:PP}) in order to obtain good fits.  In general, this will lead to lower mass-loss rates than those used by \citet{2000A&A...360..227N} in deriving their prescription and suggests that perhaps this issue should be revisited.

\subsubsection{Lifetimes}

We do not know the ages of our stars to compare with the evolutionary models, but we can use the physical parameters we
derived to see how much longer the stars have before they undergo core collapse.  Such a comparison also potentially reveals conflicts with the properties we have measured and the evolutionary models.  For instance, do the models predict that the range of chemical abundances we measure will occur over a sufficiently long time frame that is consistent with us observing them (i.e., 0.5~Myr vs 0.5 yr)?

In Figure~\ref{Fig-t-CHe-OHe} we plot the chemical abundances as a function of time prior to core-collapse for the models and indicate the abundances we find. If we use the strict ``evolutionary" definition of a WC star, we see that the WC/WO phase starts about 300,000 yr before core-collapse.  The WCs in our sample would be mostly ``mid-life" WCs with $\sim$150,000 yrs left.  However, as described earlier, two of the least chemically evolved WC stars in our sample, BAT99-9 and BAT99-11, also shows trace amounts of nitrogen \citep{Hillier2021}, suggesting that they have only recently become WCs. As mentioned above, we expect that the surface nitrogen content of WNs will be ``quickly" removed by mass-loss and mixing. As an example, the surface content
of nitrogen of the Geneva 60$M_\odot$ model (Z=0.014) drops from its 0.8\% CNO equilibrium value to 0.1\% (what we
observe in BAT99-9) in just 40,000 years.  So, we conclude that either the mixing or mass-loss is less efficient at removing the surface nitrogen, or that the threshold
between the WN and WC stage defined as usual (mass fraction of carbon greater than
mass fraction of nitrogen) is too conservative.  We believe it is the latter,
as the peak nitrogen content in WNs can not be greater than the initial
metallicity (carbon and oxygen are converted to nitrogen as part of the CNO cycle), but the actual carbon content of WCs is much, much higher than that, at least 25\%.  As we see
in Figure~\ref{Fig-t-CHe-OHe} raising the surface C/He content from 0 to 40\% takes about 150,000 years.  The C/N ratio is BAT99-9 is 250, suggesting that using a ratio of 1 to define the transition makes the expected lifetime of the WC stage longer than it is.  It would be a useful exercise to run a series of \cmfgen models to see at what C/N ratio a WR begins to look more like a WC than a WN type.

\section{Discussion} \label{sec:PM}

\begin{figure}[ht!]
\plotone{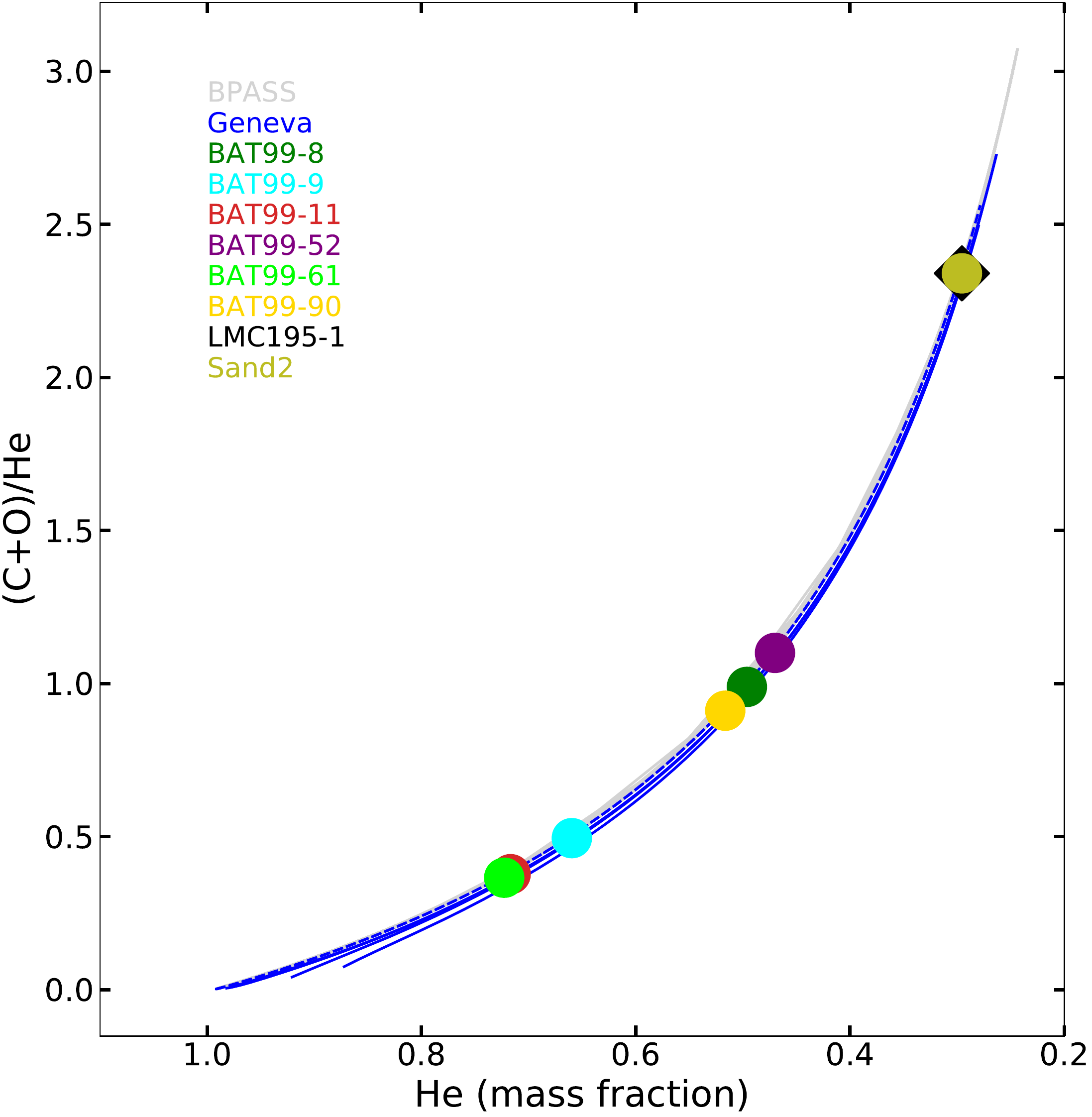}
\caption{The (C+O)/He mass ratio as a function of the He mass fraction.  As helium burning proceeds, helium will be converted into carbon and oxygen. The eight stars in our sample are plotted relative to the predictions for the surface composition from the BPASS and Geneva models. The dashed blue lines are using the \citet{GenevaLMC} $Z=0.006$ models, while the solid blue lines are from the \citet{Ekstrom2012} $Z=0.014$ models. The BPASS is shown by the grey lines.
The two points on the right at the identical location represent the two WO stars, Sanduleak~2 and LMC195-1; the other points are the WC stars analyzed here and in \citet{Aadland2022}.
\label{Fig-COHe-He}}
\end{figure}

The motivating question for this work has been whether the WO-type WRs are more evolved than WC-type WRs. If so, they would likely represent the final stage in evolution of the most massive stars prior to undergoing core-collapse and the subsequent supernova events.  
As helium burning proceeds, the helium mass fraction will decrease as the carbon and oxygen mass fractions increase.  As we showed above (Figure~\ref{Fig-COHe-He}), the abundances of all eight stars tightly follow the relationship predicted by evolutionary theory. The fact that the (C+O)/He ratios are higher for the WOs than for the WCs agrees qualitatively  with the findings of all previous studies, and thus, we have answered the question.  Indeed, WOs are more highly evolved than the WCs.  Ironically, however, the strong O\,{\sc vi} $\lambda \lambda$3811,34 that
caused the WOs to be recognized as a separate subclass is not strong due to higher oxygen content, and in fact the behavior of this line is quite insensitive to the oxygen abundance.  

Our success in fitting the O\,{\sc vi} $\lambda \lambda$ 3811,34 is due to an improved treatment of clumping, as described in the Appendix.  What actually causes this line to be stronger in WO stars than in WC4s?  The effective temperatures are higher, the wind densities are lower, and the average molecular weight is higher in WO stars.  The
higher effective temperature and lower wind density are related, in that a lower wind density will lead to a higher effective temperature if all other parameters are equal.  We experimented with a WC4 model and found that by lowering the mass-loss rate (in order to lower the wind density) also resulted in a higher effective temperature, and produced a stronger O\,{\sc vi} $\lambda \lambda$ 3811,34 doublet, and overall increased the ionization, consistent with that seen in a WO star.  As expected, this model differs from the spectra of the WO stars modeled in this paper: the lines are narrower than in the WOs,  and some line ratios also differ because of the different abundances. Further discussion can be found in the Appendix.

Another question we had wanted to address is whether our analysis would allow us to distinguish the most likely formation mechanism for Wolf-Rayet stars.  None of the eight stars in our sample show current evidence of binarity. As discussed earlier, this was a requirement for inclusion in our study.  However, this does not mean that binarity was not essential for their formation, as such companions could have merged, or be sufficiently low in mass and luminosity to be undetected.  While we cannot answer this question definitively, we can shed some light on the matter.  Figure~\ref{Fig-COHe-logL} shows the (C+O)/He ratios as a function of luminosity for the eight stars in our sample. The single-star Geneva models computed for LMC-like metallicity, $Z=0.006$, do not produce WRs with as low luminosities as what we observe.  In order to match their luminosities, we must refer to the solar-metallicty $Z=0.014$ tracks.  
This may suggest that
the adopted mass-loss rates during main-sequence evolution may be too low in the $Z=0.006$ Geneva models.  Using the \citet{2001A&A...369..574V} relation that mass-loss rates will scale as $Z^{0.7}$ for massive stars on the main-sequence, we expect that the LMC OB stars will have mass-loss rates about half as much as their solar-neighborhood counterparts.  However, due to uncertainties in the treatment of clumping, and other factors, the ``true" mass loss rates for OB stars are uncertain at a factor of two or worse \citep{2008A&ARv..16..209P,2020Galax...8...60H}, and the scaling with metallicity remains an area of active research (e.g., \citealt{2022arXiv220207811M,2022arXiv220208735G}).  Thus, the fact that we had to use the $Z=0.014$ single-star Geneva models, rather than the $Z=0.006$ versions, does not rule out single-star evolution given the uncertainties in the actual mass-loss rates.

\begin{figure}
\epsscale{0.7}
\plotone{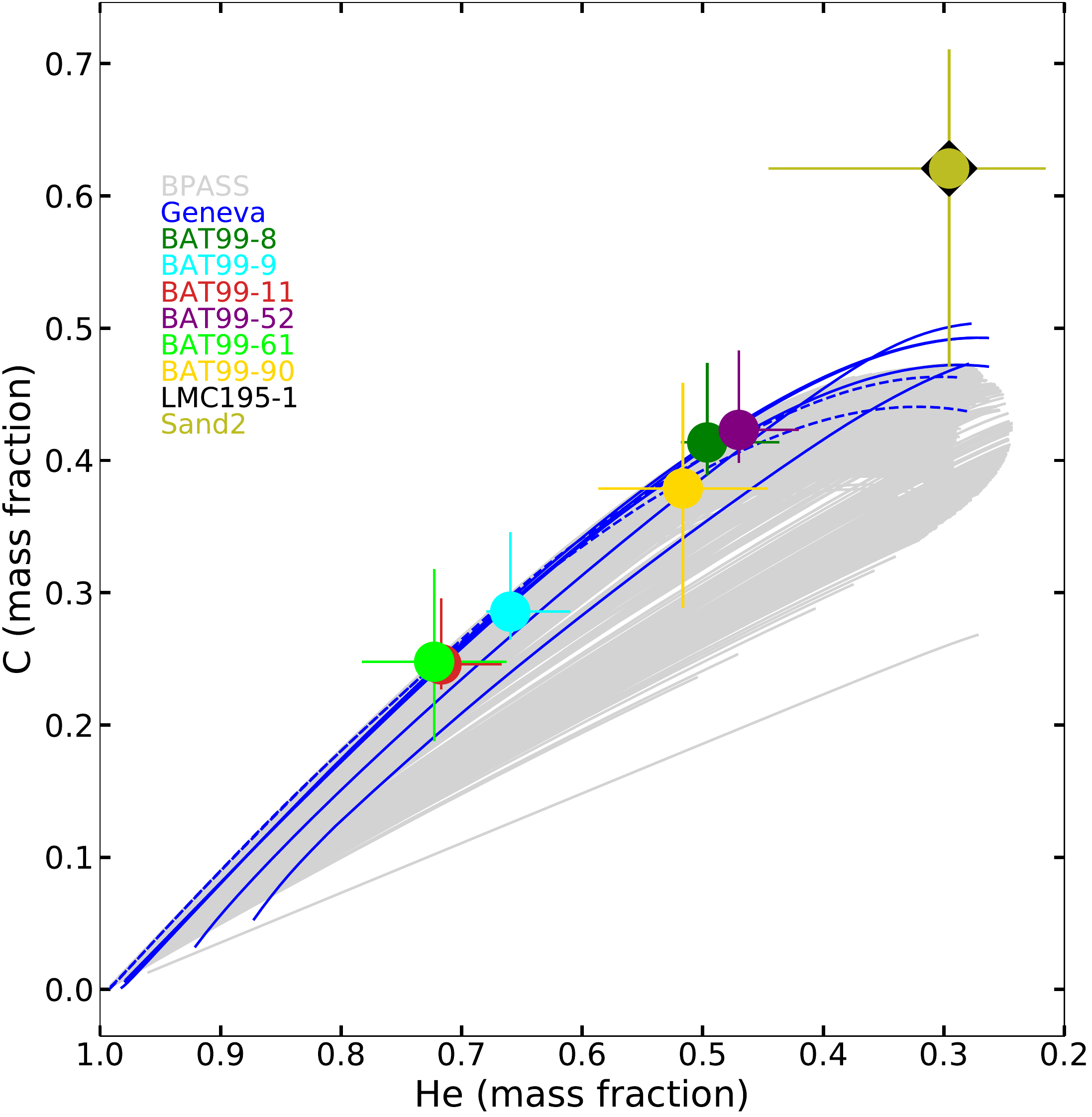}
\plotone{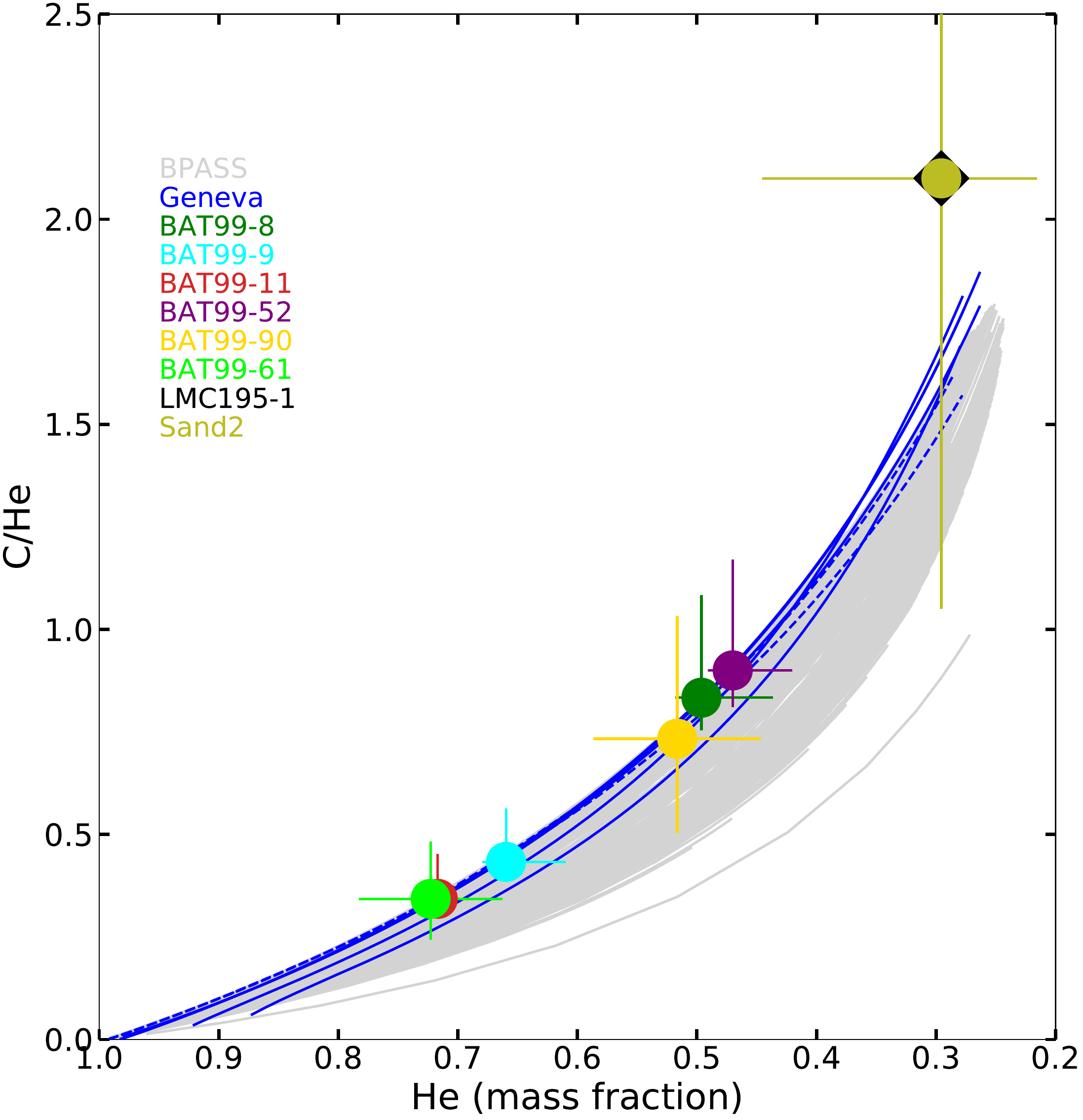}
\caption{Comparison of carbon abundance with the evolutionary models. The BPASS models (gray), Geneva solar models (solid blue), and Geneva LMC models (dashed blue) are used to show the predictions of the surface abundance of carbon (top) or the carbon to helium ratio (bottom) as a function of helium.  The six BAT99 stars are the WC stars; Sanduleak~2 and LMC195-1 are the two WOs. The carbon abundances of the WC stars agree with with the predictions of the models. However, the WO stars have a carbon abundance that is higher than any of the evolutionary models. 
\label{Fig-HeC}}
\end{figure}

\begin{figure}
\plotone{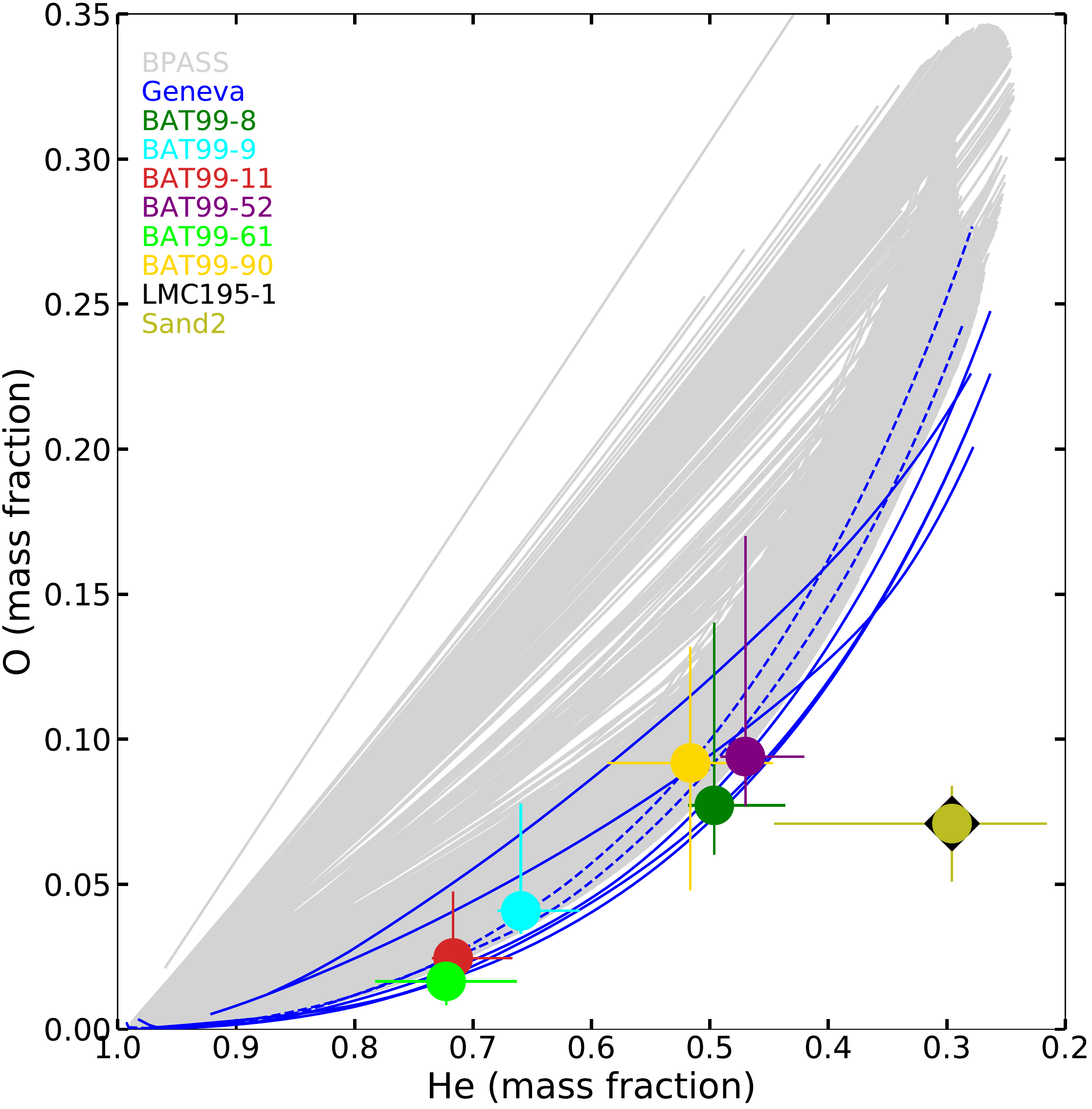}
\caption{Comparison of the oxygen abundance with the evolutionary models. The symbols are the same as in Figure~\ref{Fig-HeC}. This plot shows that the WC oxygen abundances agrees well with the Geneva single-star models and a narrow subset of the BPASS binary models.  However, the two WO stars have much lower oxygen abundances than predicted for their helium content. 
\label{Fig-HeO}}
\end{figure}

\begin{figure}[ht!]
\plotone{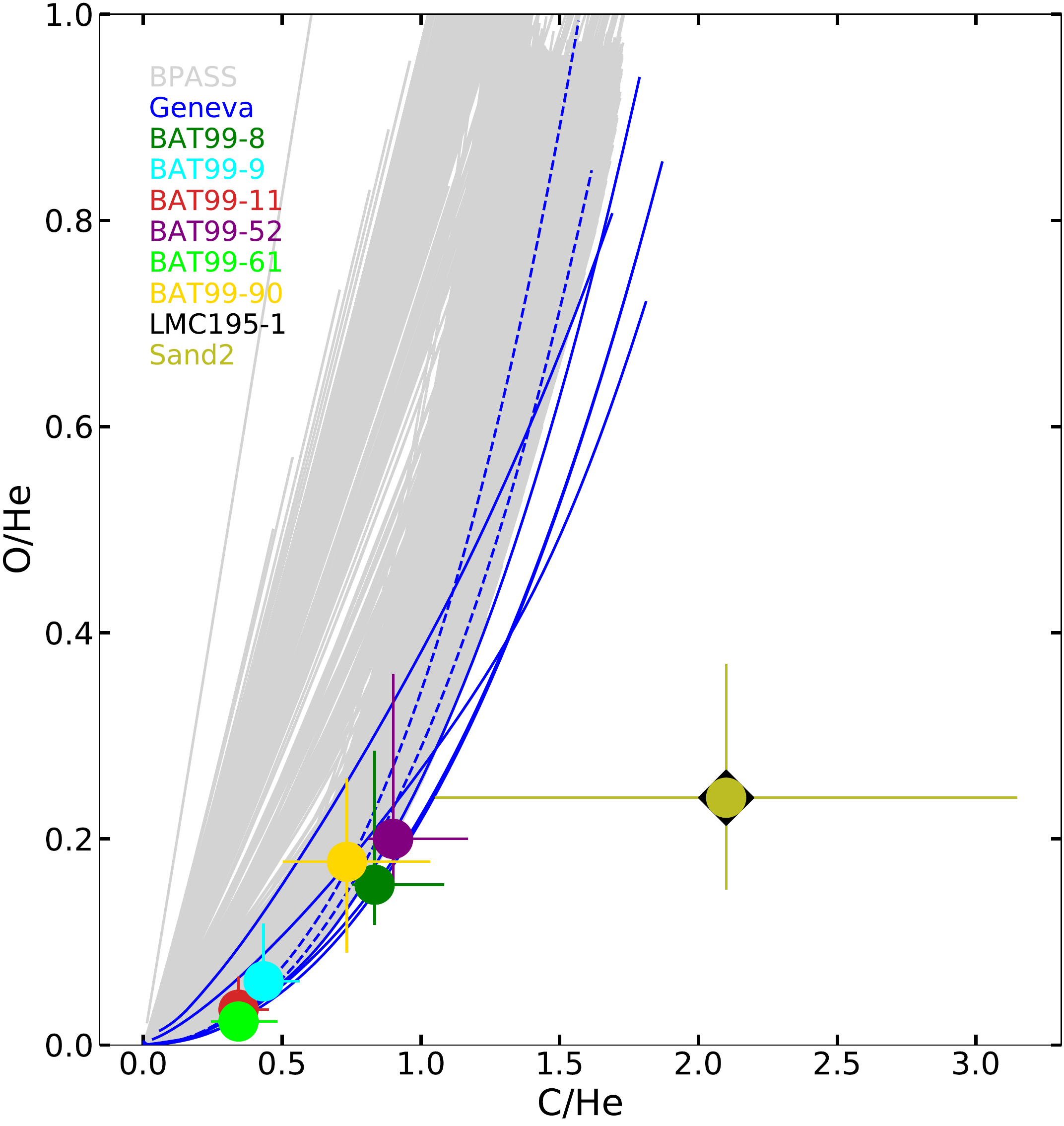}
    \caption{Comparison of the oxygen to helium ratio as a function of the carbon to helium ratio with the predictions of the evolutionary models.  The symbols are the same as in Figure~\ref{Fig-HeC}.  As we've seen in the previous two figures, the single-star evolution models, and a narrow subset of the binary models, do an excellent job of matching the surface abundances of the WC stars (BAT9-8, 9, 11, 52, 61, and 90).  The two WO stars (Sanduleak 2 and LMC195-1), however, have C/He ratios too high to match any of the predictions.  
\label{Fig-ca}}
\end{figure}

\begin{figure}[ht!]
\plotone{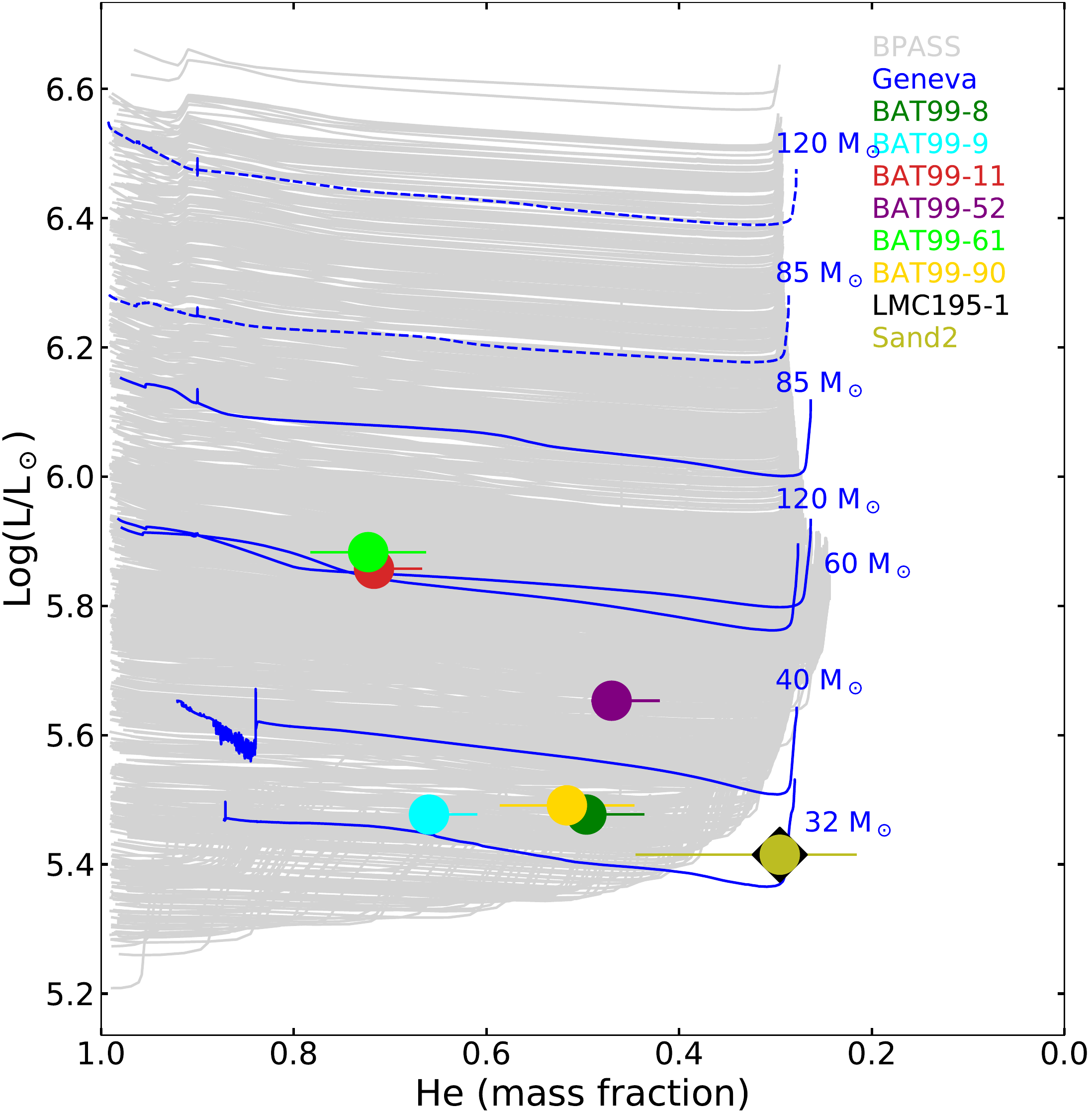}
\plottwo{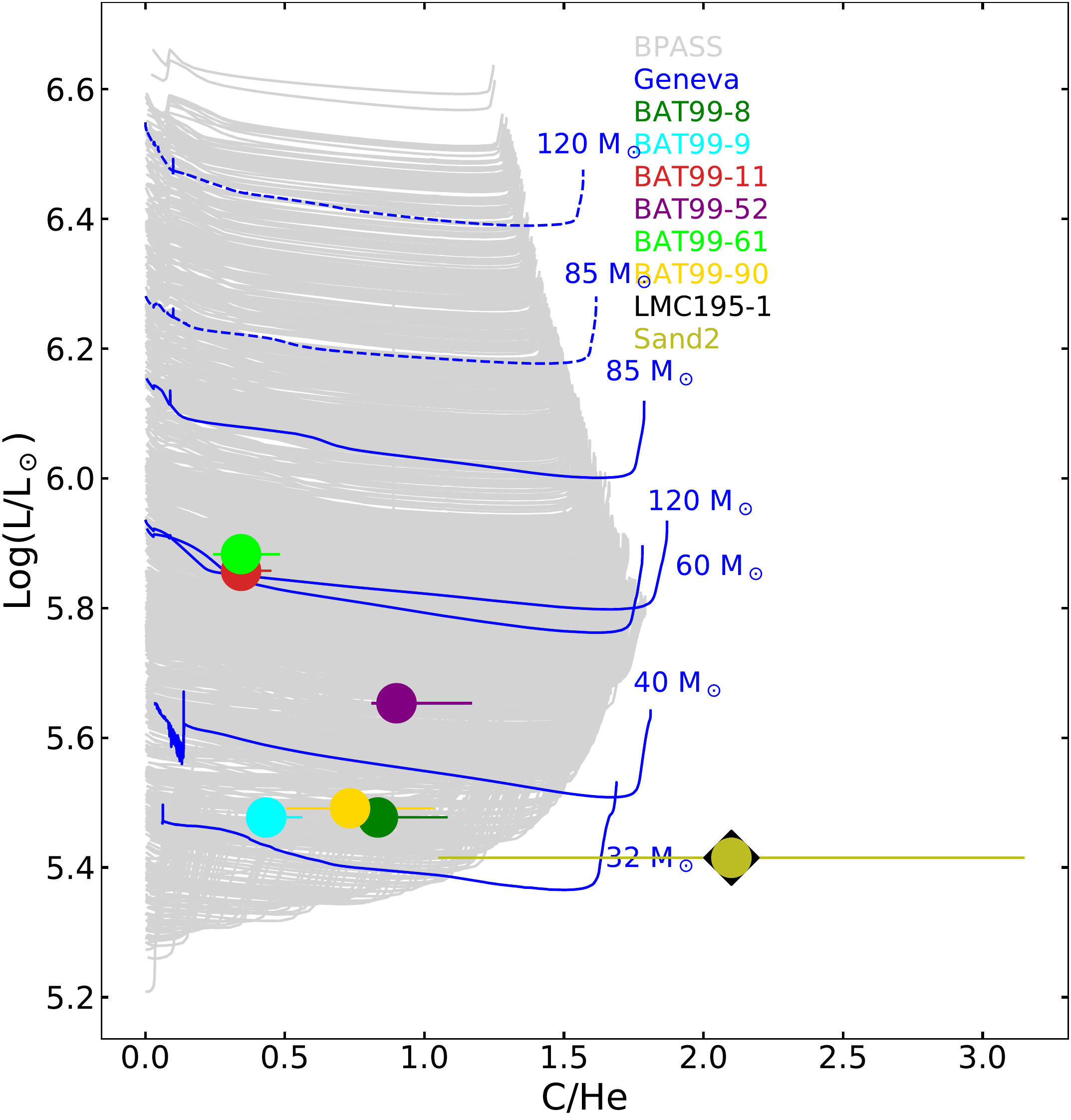}{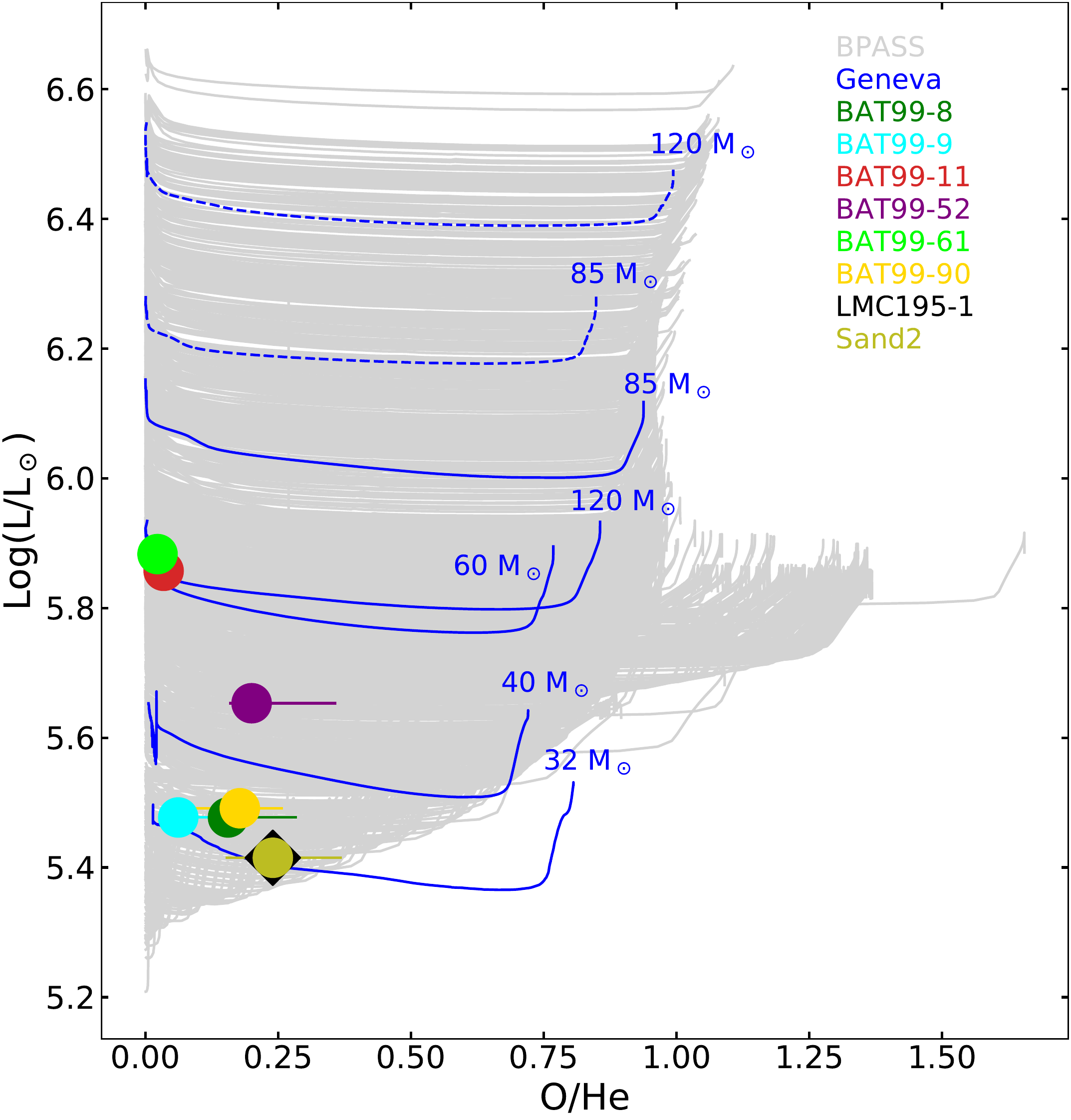}
\caption{Comparison of the luminosities with surface chemical abundances. As before, the predictions of the BPASS binary models are shown in grey, while the blue lines show the single-star Geneva models for solar (solid) and LMC (dashed) metallicities. 
\label{Fig-L-abunds}}
\end{figure}

\begin{figure}[ht!]
\plotone{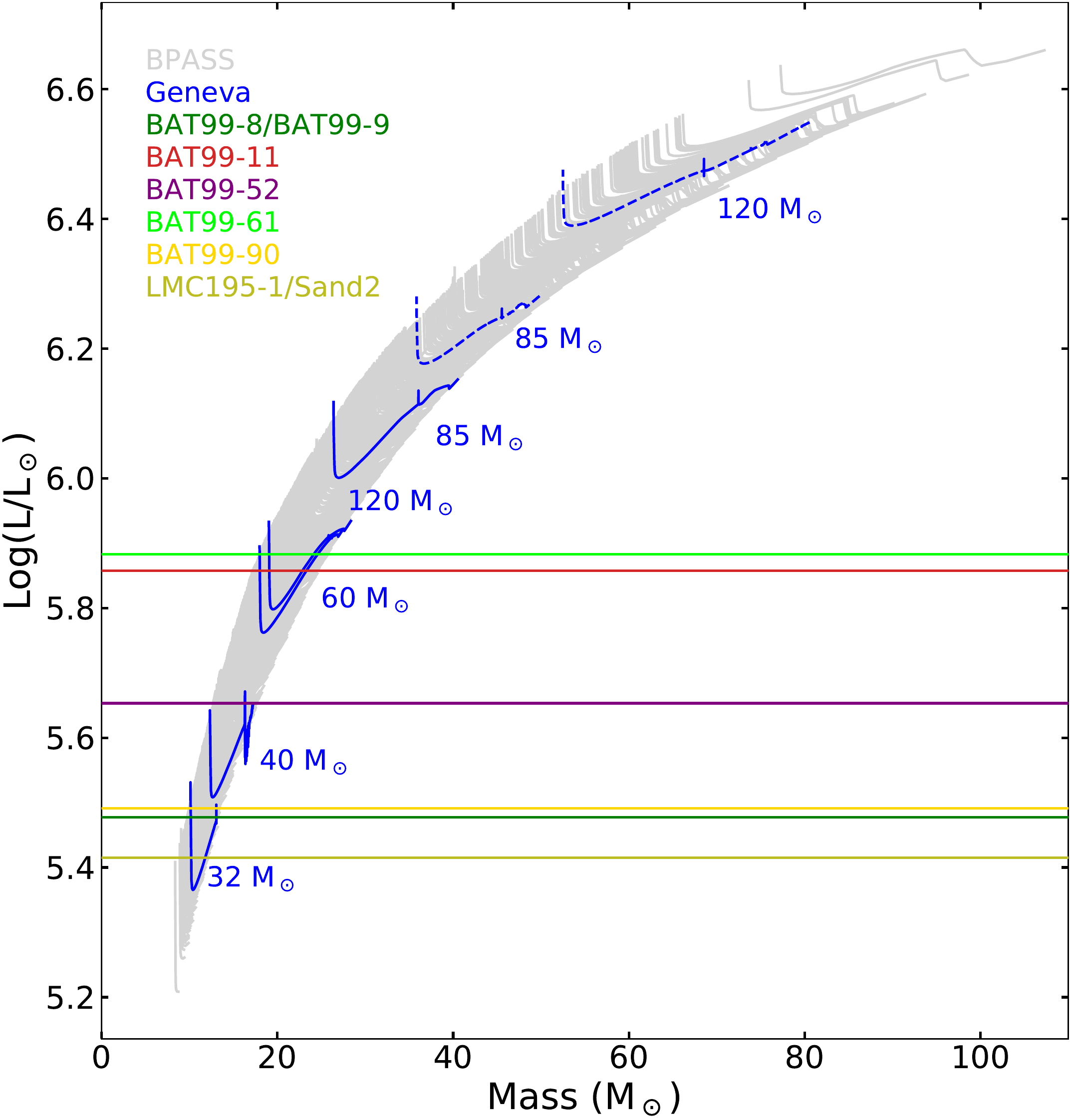}
\caption{Theoretical relationship between the WR luminosity and current mass for WC/WO stars. As before, the BPASS models are shown in gray, the Geneva solar metallicity models with solid blue lines, and the Geneva LMC metallicity as dashed blue lines. The luminosites of the stars in our sample are indicated by horizontal lines.  
\label{Fig-logL-m}}
\end{figure}

\begin{figure}[ht!]
\plotone{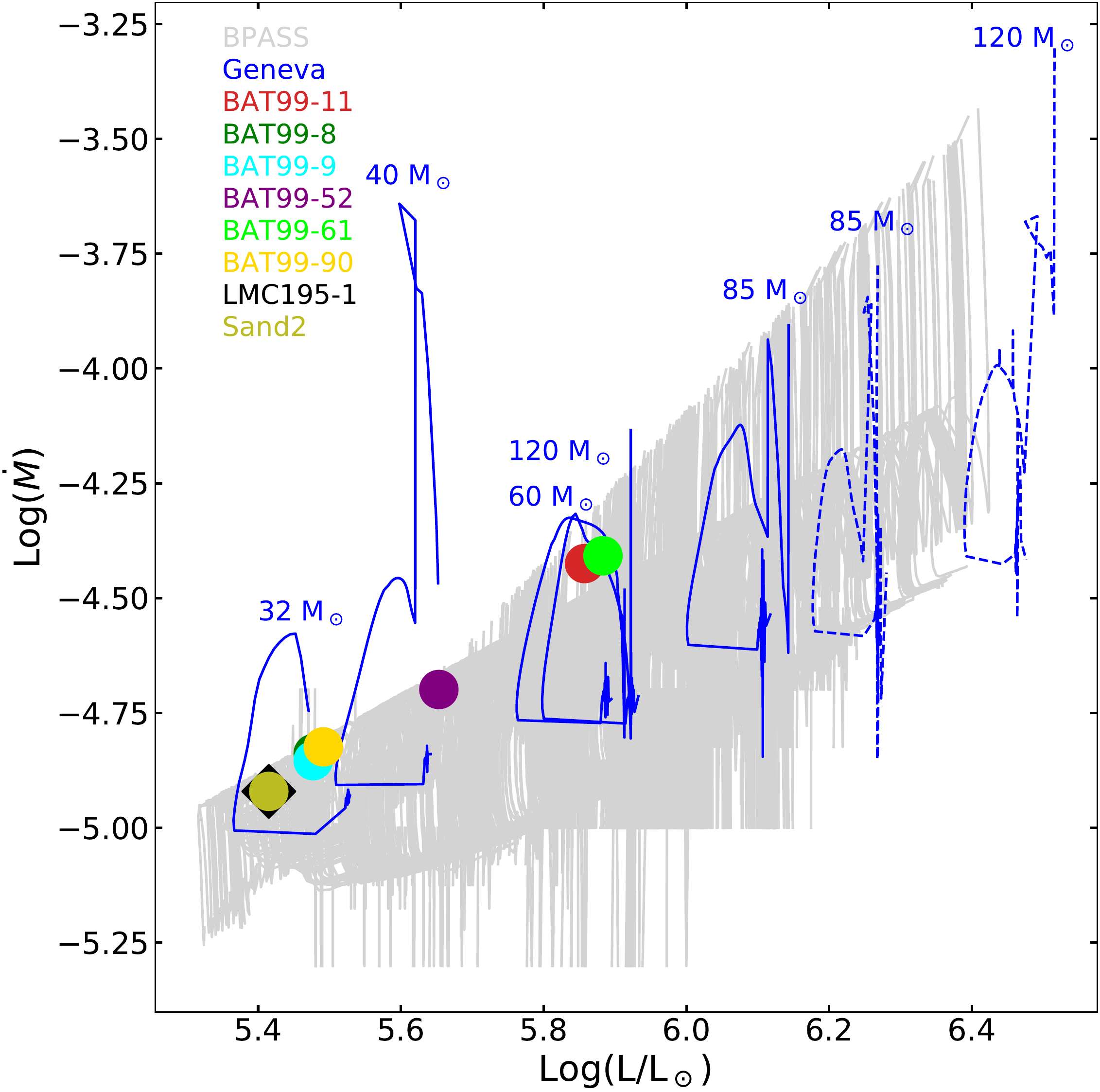}
\plottwo{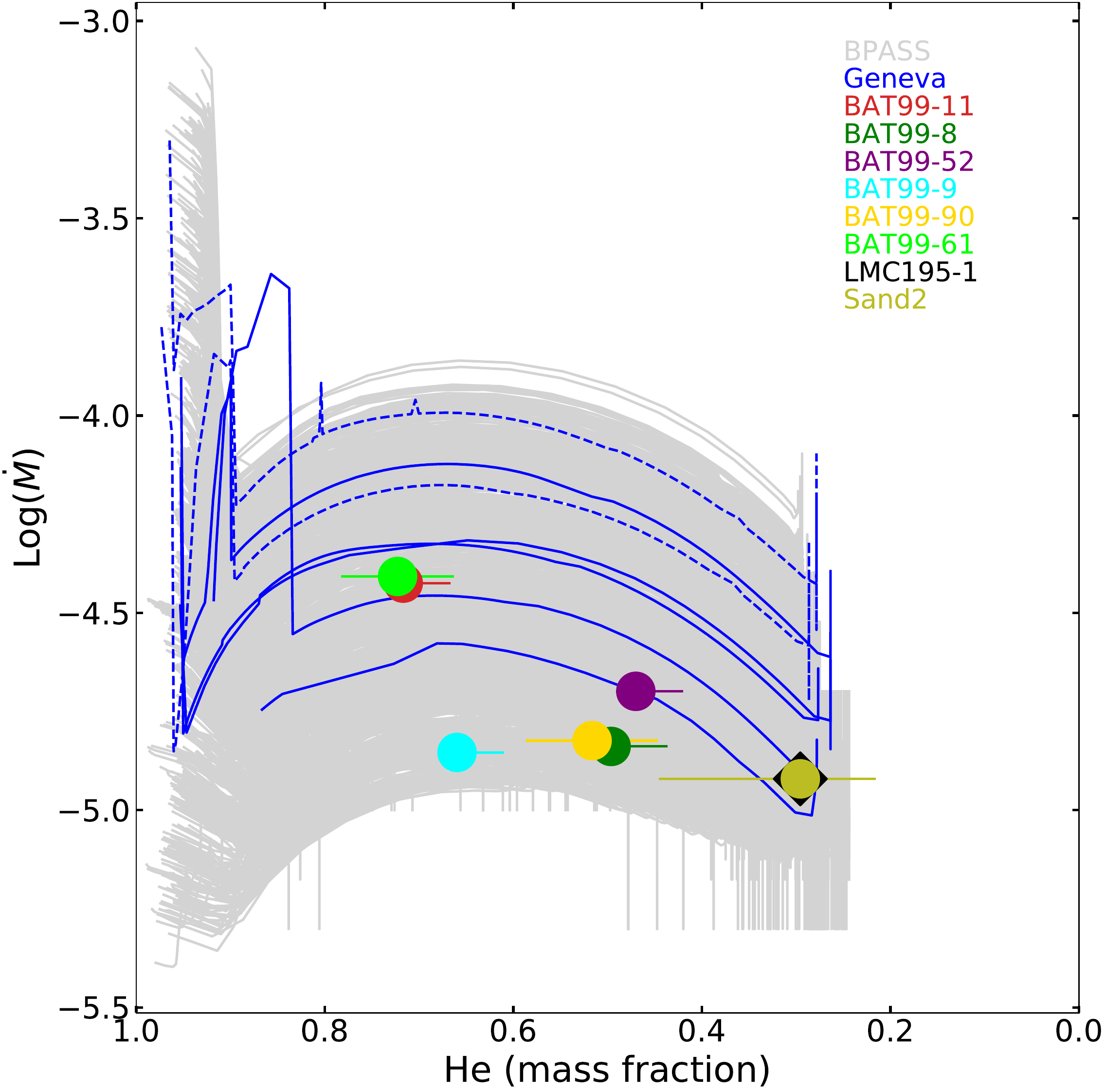}{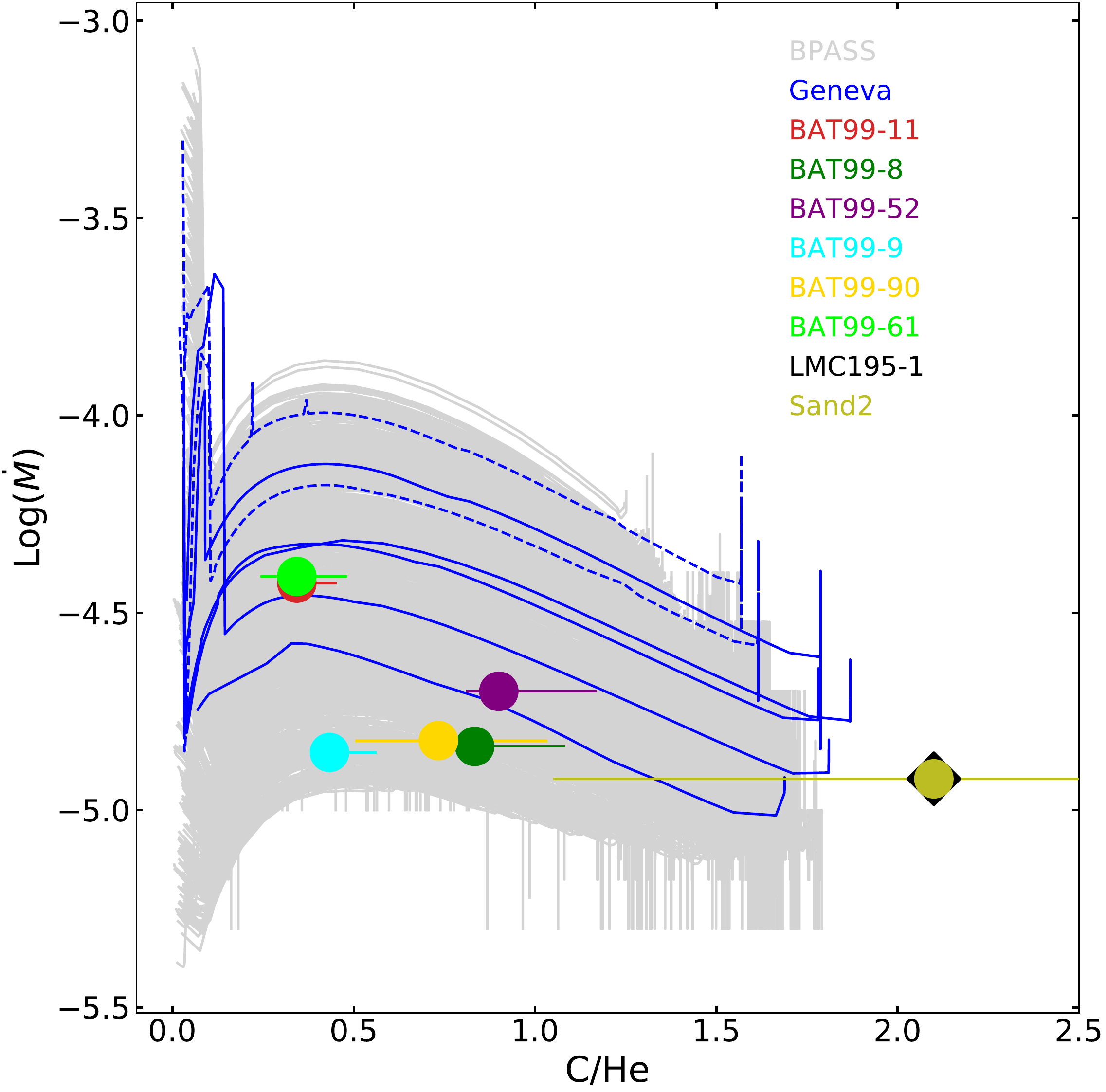}
\caption{Comparison of the measured mass-loss rates with those adopted in the evolutionary models.  The symbols are the same as in the previous plot.  Here we compare the mass-loss rates as a function of luminosity (top), helium abundance (lower left), and C/He ratio (lower right). Both the Geneva single-star models (blue lines) and BPASS models (grey) adopt the mass-loss prescription given by \citet{2000A&A...360..227N}, where the rates depend upon the luminosity and chemical abundances.  Our mass-loss ratios are derived using a filling factor $f$ of 0.05, and so are expected to be lower than the data used to derive their relations.  
\label{fig-L-massloss}}
\end{figure}

\begin{figure}[ht!]
\plotone{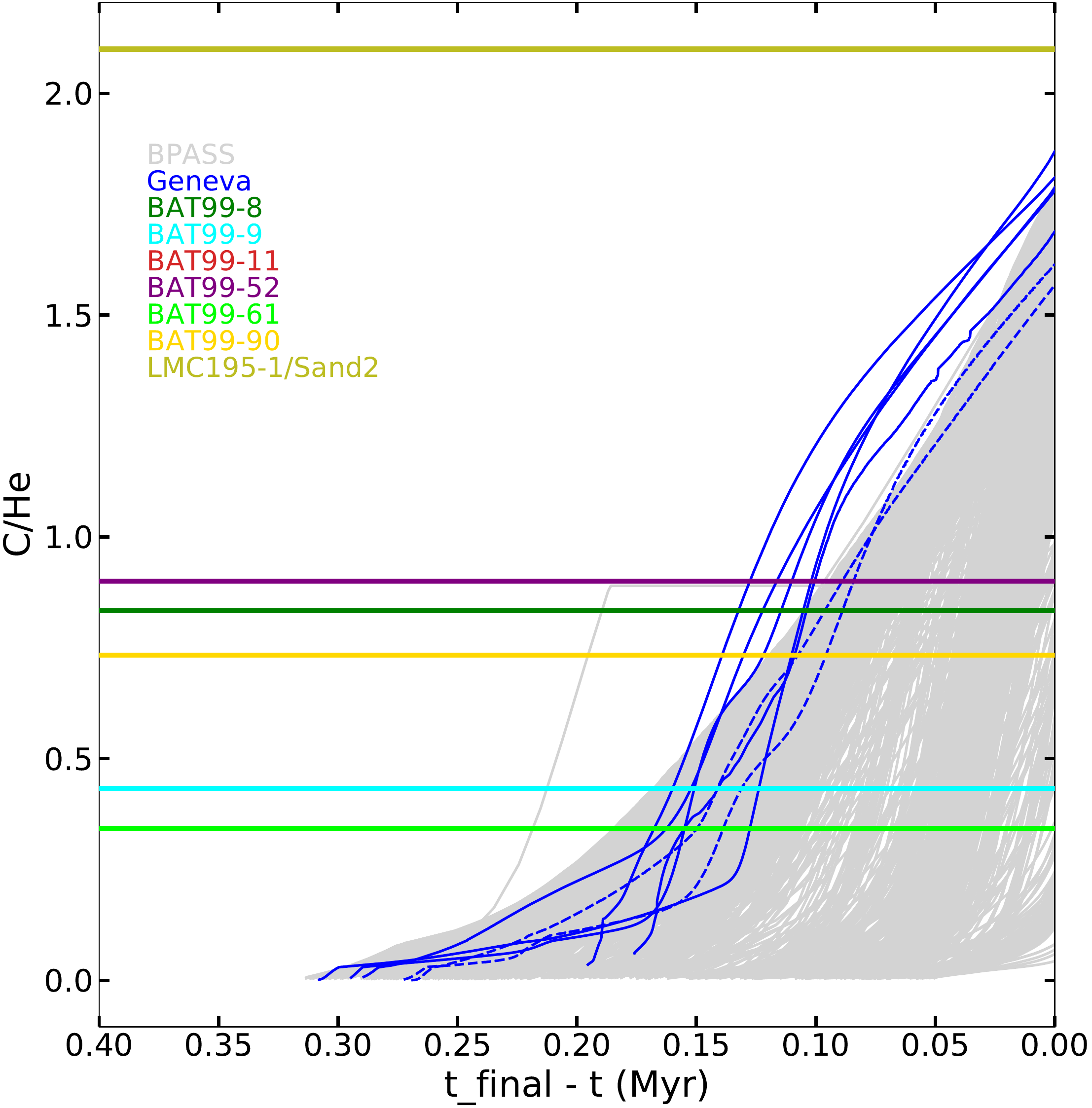}
\caption{Surface chemical abundance as a function of time before core-collapse.  
The measured abundances of our stars are indicated as horizontal lines. Note that the line for BAT99-11 is hidden beneath the line for BAT99-61 since they have very similar C/He values. We see that the WC stage lasts about 300,000 years using the strict X(C)$>$X(N) definition. If instead BAT99-9 and BAT99-11 are newly formed WCs (as argued in \citet{Hillier2021} and in the text) then the lifetime is shorter, perhaps 200,000 years.
\label{Fig-t-CHe-OHe}}
\end{figure}

\begin{figure}[ht!]
\plotone{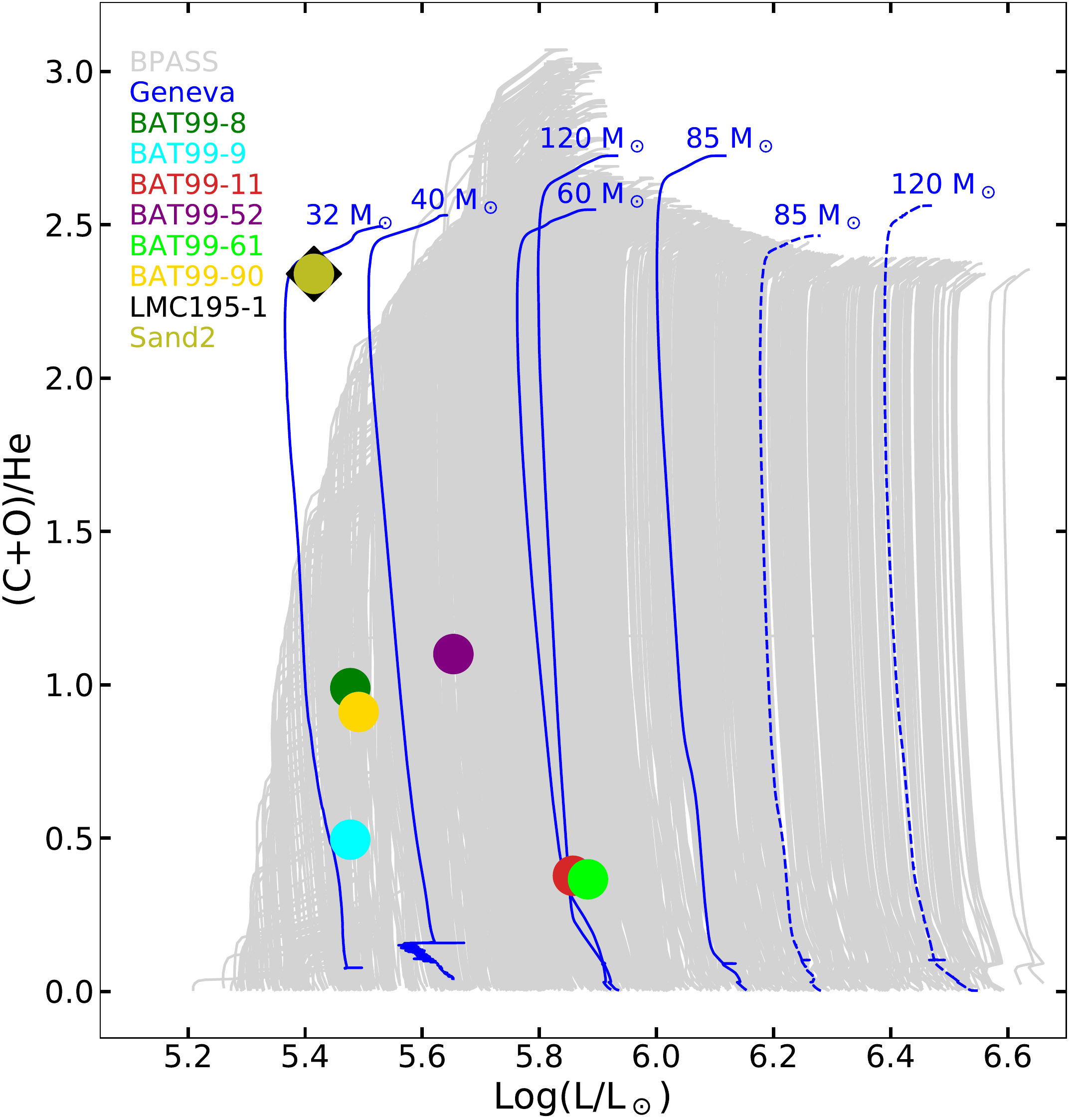}
\caption{The (C+O)/He ratio as a function of luminosity. The symbols are the same as in previous plots, and the initial mass of the Geneva models (blue lines) are marked.  There is good agreement between the models and the observations.
\label{Fig-COHe-logL}}
\end{figure}

Alternatively, it may be that binary stripping is needed to produce WCs and WOs at LMC-like metallicities. The LMC-metallicity BPASS models are able to produce WC/WO stars with luminosities as low as we find (see Figure~\ref{Fig-L-abunds}).  However, one concern with the binary explanation is that the BPASS models predict a much wider
range of properties than are actually observed.  For instance, consider the predicted values for carbon or oxygen as a function of helium (Figures~\ref{Fig-HeC},
\ref{Fig-HeO}, and \ref{Fig-ca}).  The single-star evolutionary models and a subset of the binary models do an excellent job of predicting the surface abundances of the WC stars. But the binary models show a huge range in abundances that are unpopulated.  We are still investigating what distinguishes the binary models that work from those that do not, but it appears that a revision in the nuclear reaction rates described below may eliminate many of the spurious parts of predicted parameter space.

Our most surprising finding is that the WOs have higher carbon abundances than expected from either the BPASS or Geneva models, as shown previously in Figure~\ref{Fig-HeC}. In many ways this should not have come as a shock: the high carbon mass fraction we measure for the WOs, 62\%, is similar to what has been found both by \citet{Crowther2000} and \citet{Tramper2015} in their analysis of one of our WO stars, Sanduleak~2.  \citet{2019A&A...621A..92S} finds similar values for the three Galactic WO stars they
recently studied, which have carbon mass fractions of 54-62\%.  However, previous studies have not made the comparisons to the evolutionary models that demonstrate that these values were in conflict with theory. At the same time, the oxygen abundance we find is not much higher (if at all) than what is found in the WC stars, as shown earlier in Figure~\ref{Fig-HeO}.

As a reminder, there are two competing nuclear reactions taking place during He-burning: the triple-$\alpha$ reaction $^4$He+$^4$He+$^4$He$\rightarrow^{12}$C increases the carbon content but leaves the oxygen abundance unaltered. The other reaction,  $^{12}$C+$^4$He$\rightarrow^{16}$O, increases the oxygen content at the expense of carbon and helium. Thus, both reactions decrease the helium abundance, but one increases the carbon abundance, and the other decreases it. 

The reaction rate of the $^{12}$C+$^4$He$\rightarrow^{16}$O reaction rate has long been recognized as being highly uncertain. \citet{1973ARA&A..11...73A} emphasized this in his landmark Annual Reviews paper, while \citet{2001ApJ...558..903I} note that despite significant progress in this area, that the rate of this reaction is still poorly known. The same point was also made by \citet{1993PhR...227...65W}. We know that helium burning increases the sum of carbon and oxygen at the expense of helium, but in what relative fraction? As N. Langer (2012)\footnote{https://astro.uni-bonn.de/~nlanger/siu\_web/nucscript/Nucleo.pdf} put it, it is not even clear if the final product of He-burning is carbon or oxygen!

Various studies show very different behavior of the reaction rate of as a function of the temperature; see discussion and references in \citet{2021FrASS...8...53E}. A comparison of various rates are shown in Figure 1 of \citet{2004ApJ...611..452E}.  The published BPASS models use the NACRE values from \citet{1999NuPhA.656....3A}; the Geneva models have adopted the newer values by \citet{2002ApJ...567..643K}.  

\begin{figure}
\plotone{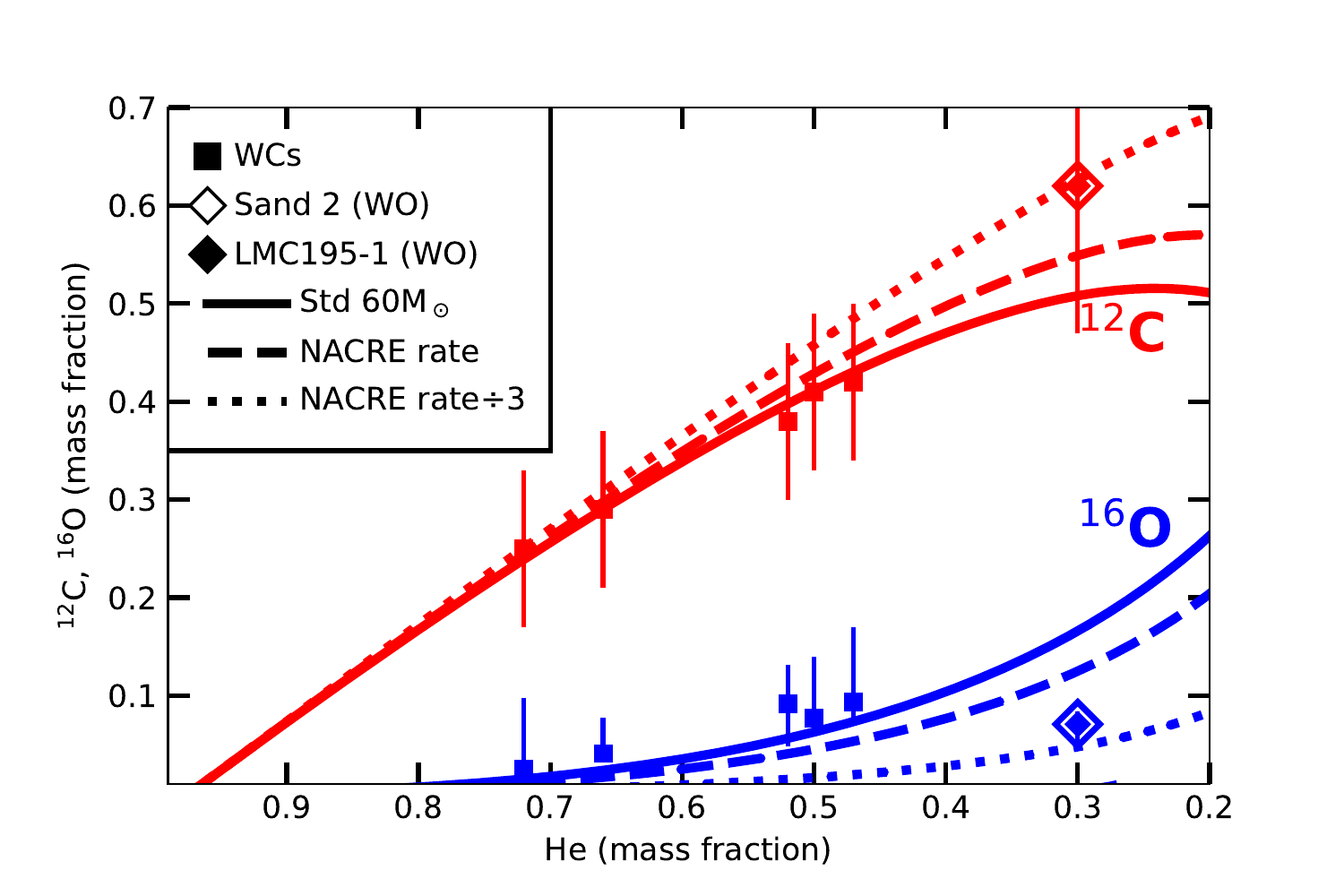}
    \caption{The effect of changing the $^{12}$C+$^4$He$\rightarrow^{16}$O reaction rate.  The solid line curves show the expected evolution of carbon (red) and oxygen (blue) for the standard Geneva evolutionary model with 60$M_\odot$. This was computed with the  \citet{2002ApJ...567..643K} reaction rate for $^{12}$C+$^4$He$\rightarrow^{16}$O.  The dashed curves show a more approximate model computed for illustrative purposes with the NACRE \citet{1999NuPhA.656....3A} reaction rate; this simple approximation does not include the effects of convection. The dotted line shows the same approximate model but now computed with the NACRE reaction rate reduced by a factor of 3.  The latter reproduces the measured abundances for the two WO stars, and provides a reasonable match to the abundances of the WCs.}
    \label{fig:Georges}
\end{figure}

Our colleague Georges Meynet was kind enough to make a preliminary exploration of what the effect would be of decreasing the reaction rate. (As emphasized by
\citealt{2021FrASS...8...53E} it is not necessarily true that decreasing a reaction rate will have the effect one expects on the chemical yields; see also discussion in \citet{2004cmpe.conf...31L}.) The results are shown in Figure~\ref{fig:Georges}. The solid curves provide a reference for what a Geneva 60$M_\odot$ solar metallicity model would predict for the mass fractions of $^{12}$C (red) and $^{16}$O (blue) as a function of the $^4$He using the \citet{2002ApJ...567..643K} reaction rate.  For comparison, the dashed curves show the results of a simplified stellar model with no convection and adopting the slightly older NACRE \citep{1999NuPhA.656....3A} reaction rate.
    It differs from what we see with the full model, but not dramatically--in other words, it is a reasonable approximation to what a full model would predict.  The dotted curves then show what would happen with this approximate model if the NACRE reaction rate was decreased by a factor of 3.  The expected abundances for the WCs would remain largely unchanged, as not enough carbon has built up for the $^{12}$C+$^4$He$\rightarrow^{16}$O reaction to be significant. (In other words, the triple-$\alpha$ reaction still dominates.) However, at later stages, a three-fold reduction in the NACRE reaction rate for $^{12}$C+$^4$He$\rightarrow^{16}$O would better reproduce the high carbon and low oxygen mass ratios we  measure for the WOs.  Of course, this approach is very approximate: any change in the reaction rate would likely be quite temperature dependent.
    
\begin{figure}
\plotone{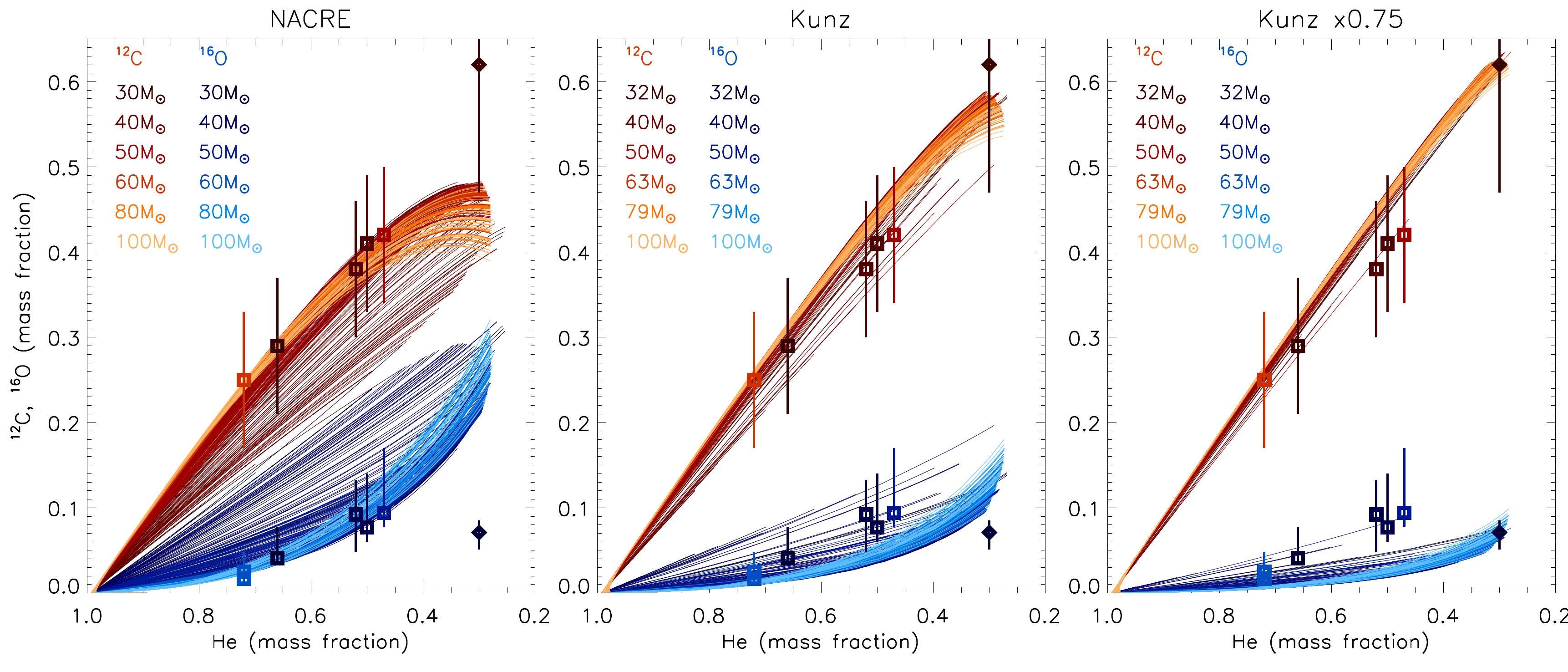}
\caption{\label{fig:JJE}Effect of changing the $^{12}$C+$^4$He$\rightarrow^{16}$O reaction rate on the BPASS evolution models. The first panel shows results for the BPASS v2.2 evolution models. The middle panel shows newly calculated models with the \citet{2002ApJ...567..643K} nuclear reaction rate. The right panel showing the models with this rate changed by a factor of 0.75. The red tracks show the carbon abundance while the blue tracks so the oxygen abundances for the models. The darker the shade of colour the lower the initial mass of the stellar evolution model.}

\end{figure}

Encouraged by this result, J.J.E. ran a new series of BPASS models to better understand the impact of changing the $^{12}$C+$^4$He$\rightarrow^{16}$O reaction rate. The results are dramatic, and shown in Figure~\ref{fig:JJE}.  This figure illustrates the expected mass ratios from the BPASS models using the NACRE rates, as well as newly computed binary models with the \citet{2002ApJ...567..643K} rates and the effects of decreasing the \citet{2002ApJ...567..643K} reaction rate by scaling it by a factor of 0.75.  We see the trend is similar to that as shown in Figure~\ref{fig:Georges}, in that lowering the nuclear reaction rate raises the carbon abundance at the expense of the oxygen abundance. More importantly, we also see that the range of possible parameter space covered by the models decreases as the nuclear reaction rate is decreased. These results suggest that the NACRE rate should not be used with stellar models and the \citet{2002ApJ...567..643K} rate is more accurate.

This figure suggests that when the uncertainties are taken into account, a reduction by a factor of 0.75 in the
\citet{2002ApJ...567..643K} rate would be enough to reproduce the carbon abundances of both the WCs and the WOs.   However, reproducing the oxygen abundances with a single numerical factor seems more problematical. If our oxygen abundance determinations are as exact as we think, then
these models suggest that a further reduction in the rate may be needed for the BPASS models to reach the oxygen abundance of the WOs. That said, it is unlikely that any correction to the nuclear reaction rate is a simple multiplicative scaling, but is likely to be temperature dependent, and the evolution in the core temperatures must
be also considered.   It is unlikely that the nuclear reaction rate varies from star to star, but one thing the binary evolution tracks show is that the conditions in the core, the density and temperature, can vary significantly between different stellar models. The BPASS models do not include the effect of rotation which could further change the core conditions during helium burning that could also make it more likely stars reach these very high carbon abundances at low helium abundances.

Further detailed analysis is beyond the scope of this study but it is clear the derived abundances of these WC and WO stars are a sensitive test of both the nuclear reaction rates and the conditions in the stellar cores during core-helium burning.

We conclude that WO stars are more chemically evolved than the WCs, but that their carbon content is higher, and their oxygen content lower, than modern evolutionary theory predicts given their low He content.  A decrease in the \citet{2002ApJ...567..643K} reaction rate of $^{12}$C+$^4$He$\rightarrow^{16}$O by 25-50\% in the relevant temperature range may be enough to explain this; it may also be that binary evolution is needed to fully explain the abundances of the WOs.

Replacing the NACRE \citep{1999NuPhA.656....3A} reaction rate for $^{12}$C+$^4$He$\rightarrow^{16}$O with that of \citet{2002ApJ...567..643K} in the BPASS models may eliminate the problem alluded to earlier, namely that the binary models predict a large range of abundances that are not observed. The reason is that the NACRE rates are higher so the final CO abundances are very sensitive to the burning conditions during helium burning of these massive stars. If the weaker \citet{2002ApJ...567..643K} rates are used, then the CO abundance becomes less sensitive to the conditions in the core during helium burning. 

     Andreas Sander (private communication, 2022) was kind enough to call our attention to one further implication: lowering the reaction rate for $^{12}$C+$^4$He$\rightarrow^{16}$O may also solve the conflict between evolutionary theory and the masses of some black holes. Analyses of gravitational wave events by LIGO/VIRGO have revealed several black holes with significantly higher masses than we expect for stellar remnants.  The most massive of these is the 85$M_\odot$ black hole detected from GW 190521 \citep{2020ApJ...900L..13A}, which presents a challenge for stellar theorists, as stars have been thought to be unlikely to produce black holes with masses between 50 and 130$M_\odot$ (see, e.g., \citealt{2017ApJ...836..244W,2021MNRAS.504..146V}). This is known as the black hole upper mass gap. Instead, stars in this mass range should blow themselves apart due to an instability resulting from electron-positron pair production \citep{1964ApJS....9..201F}. However, \citet{2019ApJ...887...53F} and \citet{2021MNRAS.501.4514C} have demonstrated that if the reaction rate for $^{12}$C+$^4$He$\rightarrow^{16}$O is lowered from what is usually adopted, this affects the resulting CO core masses and thus the (pulsational) pair instability, possibly even eliminating the black hole upper mass gap. Although other solutions have been proposed that would shrink the black hole upper mass gap (see, e.g., \citealt{2021MNRAS.504..146V}), our results provide support for the lower reaction rate solution.

\section{Summary} 
\label{sec:conclusions}

What is the evolutionary status of WO-type WRs?  Are these stars truly the ``last hurrah" of massive star evolution before core-collapse?  Spectroscopically these stars seem like they are higher excitation versions of WCs, and are distinguished by their strong O\,{\sc vi} $\lambda\lambda$3811,34 doublet. 

In order to answer this, we have obtained high quality UV, optical, and NIR spectra of six WCs and two WOs in the LMC,
and analyzed them with the best available tools.  We have compared the resulting properties to modern single-star and binary massive star evolution models.

Here is what we learned:

\begin{enumerate}
    \item The WO stars are not significantly higher in oxygen abundance compared to the WC stars.  In fact, the strength of the O\,{\sc vi} $\lambda \lambda$3811,34 line is relatively insensitive to the oxygen abundance in the wind temperature/density regime of these stars. Rather, its much larger strength in WO stars is due to their higher effective temperatures and lower wind densities. Previous difficulties in fitting this line were due to issues related to clumping, as described in more detail in the Appendix.
    \item The WO stars are, however, more chemically evolved than the WCs.  The carbon mass fraction is higher, and the helium mass fraction lower, in the WOs than in the WCs.
 \item Two of our least chemically evolved WC stars, BAT99-9 and BAT99-11, also show trace amounts of nitrogen \citep{Hillier2021}, suggesting that they transitioned from a WN stage to a WC stage recently.  If so, the WC/WO lifetime may be more like 200,000 years rather than 300,000 years.  Given their rarity, the WO stars likely have a 
 shorter lifetime than the WC phase, and perhaps significantly so, but improvements are in the evolutionary models are needed to be certain.
 \item The evolutionary models adopt the mass-loss formalism of \citet{2000A&A...360..227N} for the WC phase. We find relatively good agreement between our measured mass-loss rates with a function of luminosity, but relatively poor agreement when the rates are compared to the measured abundances. Since our modeling required using a volume filling factor of 0.05 to obtain a good fits to most of the lines, our derived mass-loss rates should be somewhat lower than that found in the earlier studies on which the \citet{2000A&A...360..227N} relations are based.   We suggest that it may be time to re-examine this commonly adopted relation.
    \item Neither the single-star nor binary models can reproduce the chemical abundances of the WO stars: they are much richer in carbon than they are expected to be. It has long been recognized that the $^{12}$C+$^4$He$\rightarrow^{16}$O reaction rate is highly uncertain, and show that a decrease of 25-50\% in the \citet{2002ApJ...567..643K} rate would be enough to remove the discrepancy. Lowering the reaction rate from usual values would also provide relief to the conflict between the large masses of black holes found by gravitational wave detections and theoretical expectations. 
    \item Our study sheds new light on the successes and problems with both the single-star and binary models in their ability to predict the properties of WC/WO stars. The single-star models with a metallicity appropriate to the LMC do not produce WCs/WOs with luminosities as low as we observed; the binary models do.  The higher metallicity single-star models do reproduce their luminosities, and, further, they reproduce the surface abundances of the WC stars very well.  This either suggests that the mass-loss rates adopted for LMC metallicities are too low, or that binary evolution is needed to produce WC/WO stars.  However, although a subset of the binary models predict surface abundances that match our results, they also predict a very wide range of properties that are not observed. This may, however, be solved by the BPASS models adopting the lower \citet{2002ApJ...567..643K} reaction rates, and such calculations are underway and will be presented elsewhere.

\end{enumerate}

\begin{acknowledgments}
Lowell Observatory and Northern Arizona University sit at the base of mountains sacred to tribes throughout the region. We honor their past, present, and future generations, who have lived here for millennia and will forever call this place home.

We are very grateful to Georges Meynet for his insight and calculations on the reaction rates, as well as other correspondence. Andreas Sander was also kind enough to send very useful comments on an early draft of the paper. An anonymous referee provided excellent comments, which helped us improve the manuscript. 

Our ground-based observing has been made possible by the Arizona and Carnegie Time Allocation Committees, and we are very appreciative of the excellent support we always have at Las Campanas Observatory.

This work has been supported primarily through the NASA ADAP grant 80NSSC18K0729, with the ground-based observing supported through the National Science Foundation grant AST-1612874.  Partial support was also provided by the Lowell Slipher Society funds.  We also acknowledge support for programs GO-5460, GO-5723, GO-12940, and GO-13781, as well as Archival Research programs AR-14568 and AR-16131, all of which were provided by NASA through a grant from the Space Telescope Science Institute, which is operated by the Association of Universities for Research in Astronomy, Incorporated.

\end{acknowledgments}

\facilities{Magellan:Baade (MagE optical spectrograph, FIRE near-IR spectrograph), HST(FOS, COS, STIS, WFC3)}

\clearpage

\clearpage
\clearpage

\clearpage

\clearpage

\clearpage
\clearpage

\clearpage

\clearpage

\clearpage

\clearpage

\clearpage

\clearpage

\clearpage

\clearpage

\clearpage

\clearpage
 
 \appendix
 
\section{On the cause of the spectral differences between WC4 and WO Stars}

The fundamental cause of the differences between WC4 spectra and WO spectra must be primarily related to the evolution state of the star. WO stars have a higher C/He ratio than WC4 stars and are more evolved. In our models, the WO stars have higher effective temperatures than the WC4 stars, and this leads to a higher excitation spectrum. In large part, the higher effective temperature is due to the lower density wind which also helps produce a higher excitation spectrum.  As noted in earlier papers (e.g., \citealt{Crowther2002}) the lower metallicity of the LMC leads to lower mass-loss rates which in turn leads to lower wind densities and higher excitation spectra, and the observed shift to earlier spectral types that occurs in the LMC.



To help understand the cause of the differences between the WO and WC4 spectra, 
(although there is probably a continuous change from one class to the next)
we examined the influence on spectra of changes in the mass-loss rate for the best-fitting models for Sanduleak~2 and BAT99-90 (with all other parameters held fixed). 

Consider our final model for Sanduleak~2. If we decrease the mass-loss rate by a factor of 1.5 there is a fundamental change in the spectrum. The  \osixdoub\ doublet becomes the most prominent feature in the optical spectrum,  O\,{\sc v} and  O\,{\sc iv} lines weaken,  \civopt\ becomes very much weaker (the intensity relative to the continuumis reduced from $\sim$\,30 to $\sim$\,4), and the optical continuum flux drops by a factor of $\sim$1.7. A similar change in the strength of \civopt\ was found for LMC195-1, as discussed  in Section 3.3. The spectrum would still be classified as WO, but  the spectrum is very different from the WO stars modelled in this paper. Because of the reduction in wind density, the effective temperature (defined at a Rosseland optical depth of 2/3) increased from 108800 to 122400 K.

We can also address what happens if we increase \Mdot\ by a factor of 2 for the best-fitting Sanduleak 2 model. Such a change causes the effective temperature to decrease from 108800  to  82350 K.  In this case the resulting spectrum is similar to that of WC4  stars, although the lines are broader, and there is a difference in some of the He/C line ratios (due to the different C and He abundances in Sanduleak 2). The comparison reaffirmed our conclusion that there is no need for the O abundance in WO stars to be higher than in WC4 stars.

We also played the same game with our best-fit model for BAT99-90. Decreasing the mass-loss rate by a factor of 1.5 leads to a more  WO-like spectrum and a significant weakening of C\,{\sc iii} lines, although it would still probably be classified as WC4. Decreasing the mass-loss rate by a factor of 2, however, leads to a situation where the \civopt\ doublet has weakened considerably, and the model spectrum is of higher excitation than the WO stars modelled in this paper. These results illustrate the extreme sensitivity of some lines in WC stars to model parameters (which has previously been noted for C\,{\sc iii} $\lambda$ 5696 by \citep{H89_WC}).

Understanding the WO/WC sequence remains a challenge. Our 1D models, while  complex, are essentially homogenous models, and they are probably a poor representation of real WR winds. The spectral properties of
WC and WO stars are undoubtedly determined by  the  properties of the CO core, the closeness of the stars to the Eddington limit and the Fe opacity bump. As has  been noted by several authors \citep[e.g.][]{NL02_mdot,2013A&A...560A...6G,2020MNRAS.491.4406S}, the latter two place strong constraints on the inner wind structure at the sonic point, and hence on the mass-loss rate.

\section{Clumping and O\,{\sc vi} $\lambda\lambda 3811, 3834 $}

An extensive discussion concerning clumping and its possible
consequences was provided by \cite{Aadland2022}.
Here we concern ourselves with issues specifically
related to the models presented in this paper.
As noted in the text, we typically adopt a two parameter
clumping law --- the two parameters basically specify an onset velocity,
and the volume filling factor at infinity. However there is no reason
to expect that this law is valid at all radii. For example, we might expect that
the velocity dependence of clumping may be different depending
on whether the wind is optically thick or thin. Further clumping may not only
be induced by instabilities in radiation driving; clumping is also expected to
occur in plasmas that are dominated by radiation pressure \citep[e.g.,][]{S01_edd_instab}.
In their work on WR hydrodynamics, \cite{2017A&A...603A..86S}  adopted a clumping law dependent
on optical depth. Unfortunately,  the richness of the line spectrum, 
the broadness of the lines, and line blending make it difficult to constrain the clumping and its variation
with radius \cite[e.g.,][]{Aadland2022}\footnote{
Progress towards realistic 3D radiation-hydrodynamic simulations of
Wolf-Rayet winds is being made by \cite{2022arXiv220301108M}.
Such simulations, while in their infancy, will provide important insights
into the velocity law and clumping in Wolf-Rayet winds.}.

While model predictions are relatively invariant to $\Mdot/\sqrt{f}$, radial variations in clumping
(and similarly the velocity law) can influence the relative strength of emission lines. For
example, a enhancement in the clumping factor in the outer wind increases the strength of
 \ion{C}{3}\ emission lines (e.g., $\lambda 2296$), and will have little influence
on the strength of emission lines from higher ionization stages. However clumping variations in
the inner wind can also alter line strengths, and such variations may be important
for explaining the strength of \osixdoub\ in both WC and WO stars
\citep{GH05_WC_wind,Aadland2022}.  However, as is well known,
such variations will also alter the continuous energy distribution.

\begin{figure}  
\includegraphics[width=1.0\linewidth, angle=0]{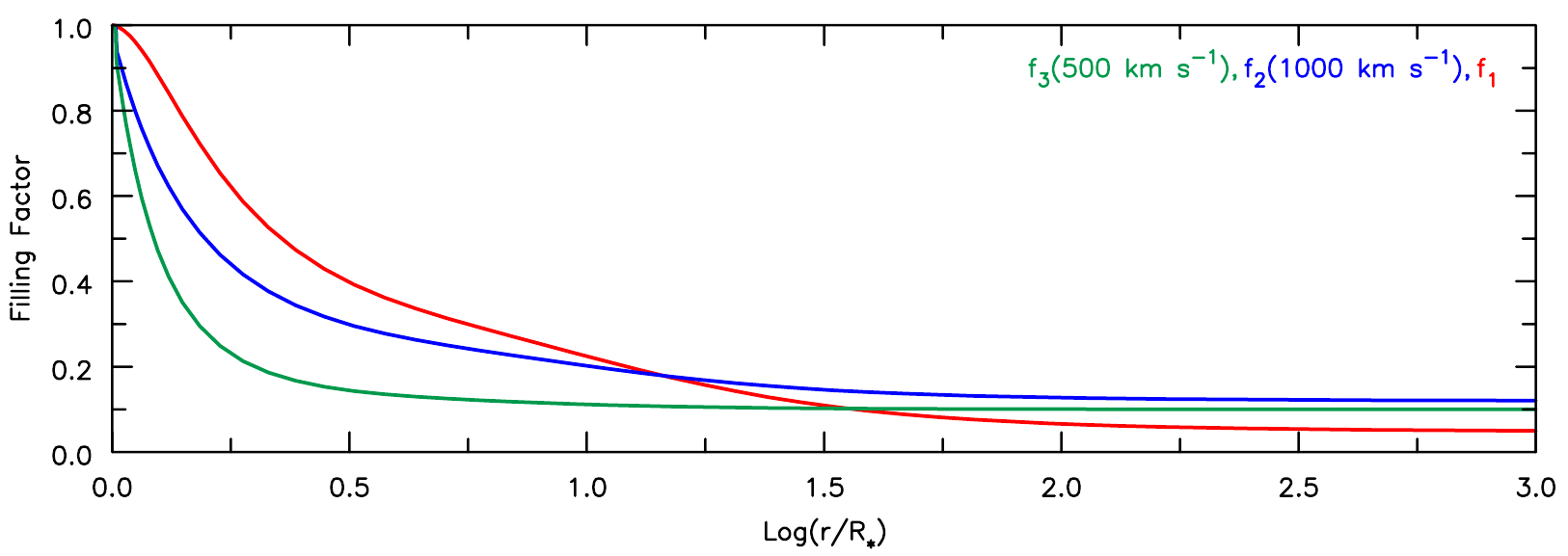}
\includegraphics[width=1.0\linewidth, angle=0]{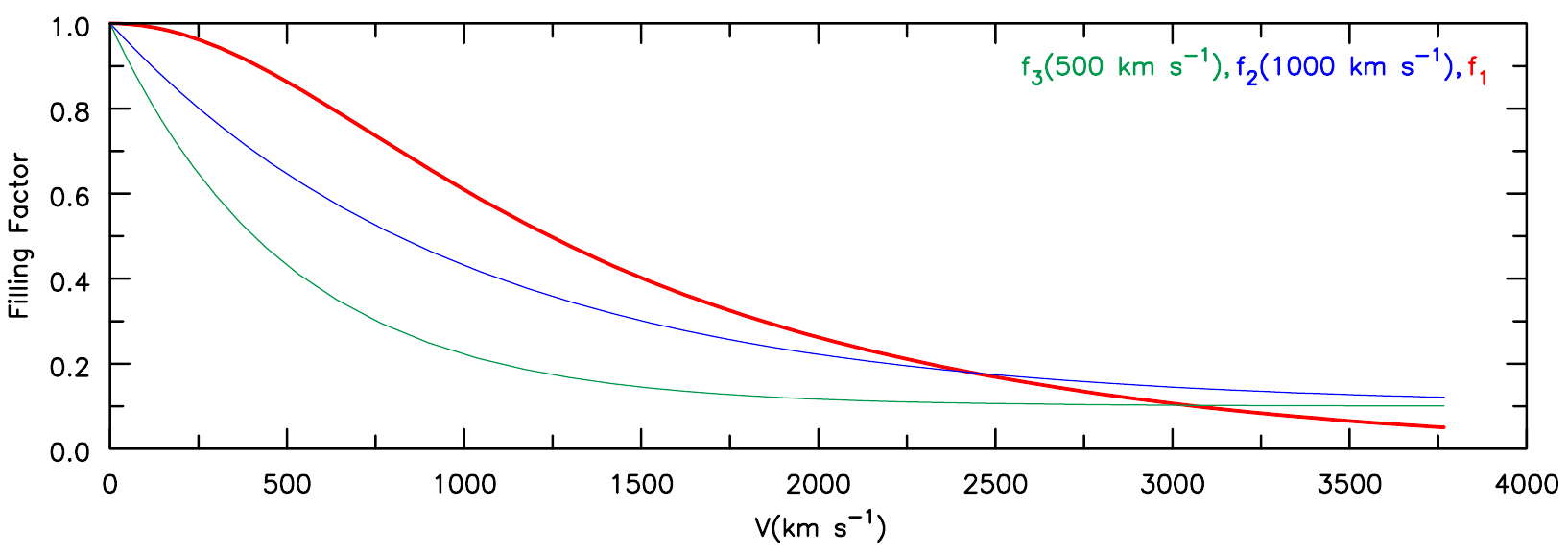}
\caption{Illustration of the variation of the volume factor with radius (top) and
velocity (bottom). Two of the curves are labeled by their ``switch on" velocity
which were 500 and 1000 km s$^{-1}$. The third curve belongs to that used
in final model for Sanduleak 2.}
\label{fig_clump}
\end{figure}

In Figure~\ref{fig_clump}, we show three different clumping laws that influence the
strength of \osixdoub\ and the  \civres\ resonance transitions.
Generally models with these three clumping laws will yield similar spectra although
slight changes in \Mdot\ or $L$ might be needed to yield best agreement with observation. 
In the model with the volume-filling factor illustrated by the red curve, a Rosseland optical depth of two-thirds
occurs at a velocity of $\sim$1400 km s$^{-1}$, or a $\log{(r/R_*)}$ value of 0.16 ($r=1.44R_*$). 
The red curve gives the best fit to the
O\,{\sc vi} doublet but yields a worse fit to \civres. 

Between the final clumping law, and that computed using a switch velocity
of 1000 km s$^{-1}$, there is a 10\% (relative) change in the continuum flux over the wavelength range from
1000 to 10,000\,\AA. This change, unfortunately, will be masked by
uncertainties in the reddening correction and flux calibration. Over a wider
band (e.g., out to  10\,\mum) the change is more significant, $-$30\% to 40\%.
The changes were similar when compared to a model using the switch
velocity of 500 km s$^{-1}$  over 1000\,\AA\ to 10,000\,\AA, but much larger at 10\,\mum.
The numerical values are illustrative only since the  models did not use identical parameters
(e.g., \Mdot, O abundance), but they do highlight the need for accurate continuum flux measurements,
especially beyond 1\,\mum\ where the effects of reddening are minimized.

\begin{figure}  
\includegraphics[width=1.0\linewidth, angle=0]{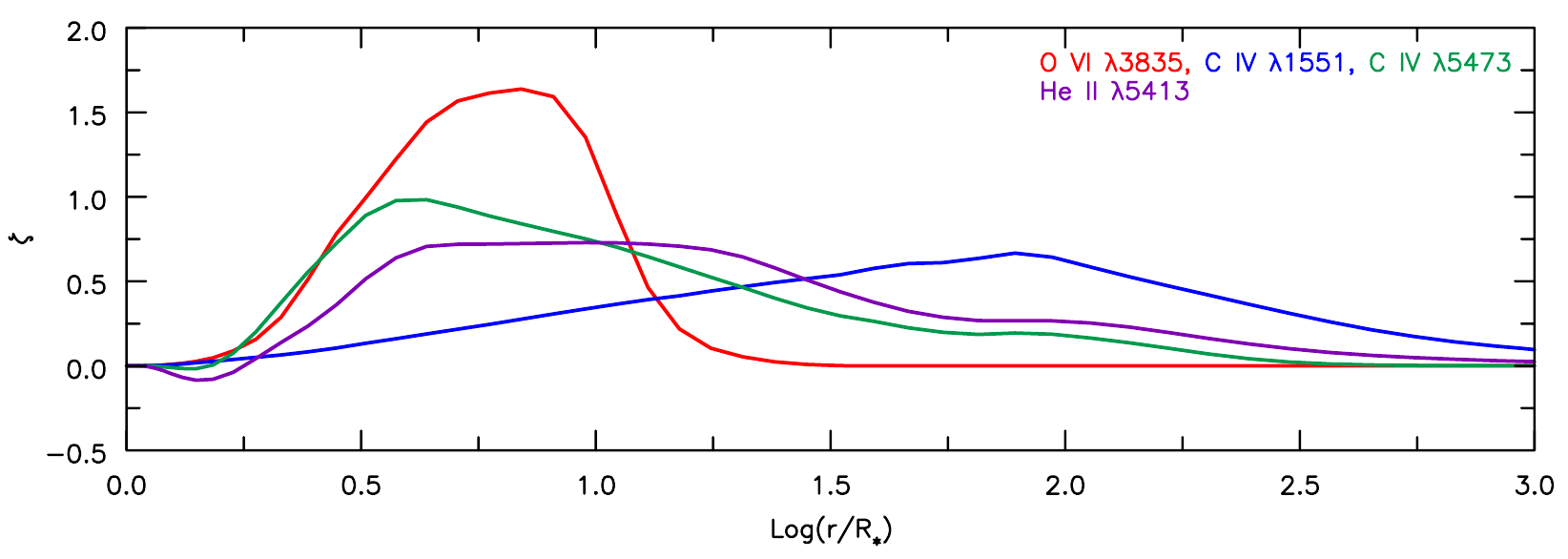}
\includegraphics[width=1.0\linewidth, angle=0]{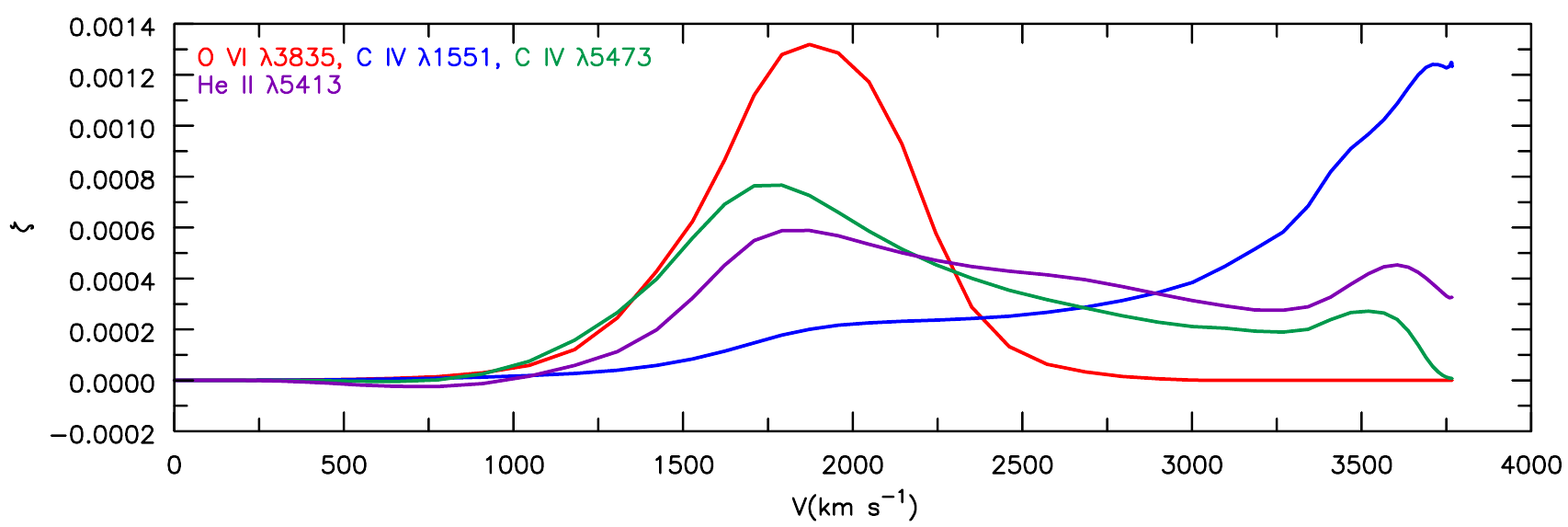}
\caption{Illustration of the variation of the  origin of \osixdoub, \civres,
 \ion{He}{2}\ $\lambda$5413, and  \ion{C}{4}\ $\lambda$5473. The quantity $\zeta$ is related to to the strength of the line, as defined in \citet{Hil87_B}. The area
 under each curve has been normalized to unity, with the area to the right of
 any location indicating the fraction of flux originating beyond that radius.
 As well as line optical depth effects, emission from the inner wind 
 (i.e., $V > 1000$ km s$^{-1}$) is limited by the continuum opacity. The rapid rise in the curve for 
 \civres\ at higher velocities is simply due to the near constancy of the velocity
 in the outer wind.}
\label{fig_orig}
\end{figure}

In Fig.~\ref{fig_orig}, we illustrate the origin of several emission lines using
the approach of \cite{Hil87_B}. This approach
uses the Sobolev approximation and ignores line overlap. As apparent
from the figure, the \osixdoub\ originates at lower velocities than the other lines, and its origin is
primarily constrained by the decreasing ionization state of oxygen as we move
out in the wind. Conversely, \civres\ originates over a broad swath of the wind. Also illustrated
is the origin of two of the key He and C abundances diagnostics (He\,{\sc ii} $ \lambda$5413,
and C\,{\sc iv} $\lambda$5473). These lines originate in roughly the same region of the wind,
and hence their ratio is relatively insensitive to clumping. 

\section{Radiative Driving.}
 
As noted in  \cite{Aadland2022}, {\sc cmfgen} computes the radiative force, and this can be compared with that
required to set the mass-loss rate and velocity law. For BAT99-61, the
model radiation force is typically within $\pm30$\% of that required to drive
the flow, and gives us confidence in both our model results and
the ability of WR stars to drive their winds by radiation pressure. This result
is better than shown for BAT99-9 in \cite{Aadland2022}, especially since the disagreement
is in both directions. This thus appears to be a case where we could get a hydrodynamical
mass-loss rate and velocity law solution that were broadly consistent with those derived
empirically.  Of course, there are still numerous issues to be resolved for a fully consistent
solution -- clumping is parameterized, and there are still issues with the accuracy of
the calculated radiative force. These arise, for example, form the adopted abundances, atomic
data, and form the adopted atomic models. In the inner wind, small errors can make 
a big difference in the solution due to the proximity of the star to the Eddington limit.

For BAT99-90 the situation is not quite as good, with larger discrepancies
at low velocities ($\sim$80~km s$^{-1}$).  This may be related to the use of a
smaller core radius in this model.  For the WO models the situation is worse, with the radiation force
needed typically underestimated by factors of 2 to 3. Understanding
the cause of this discrepancy will require further work. In \citet{Aadland2022},
we noted that while adding additional species/lines can make
large differences to the radiative force, the predicted spectrum was
insensitive to such changes (assuming a fixed mass-loss rate and
velocity law).

\begin{figure}  
\label{fig_hydro}
\includegraphics[width=1.0\linewidth, angle=0]{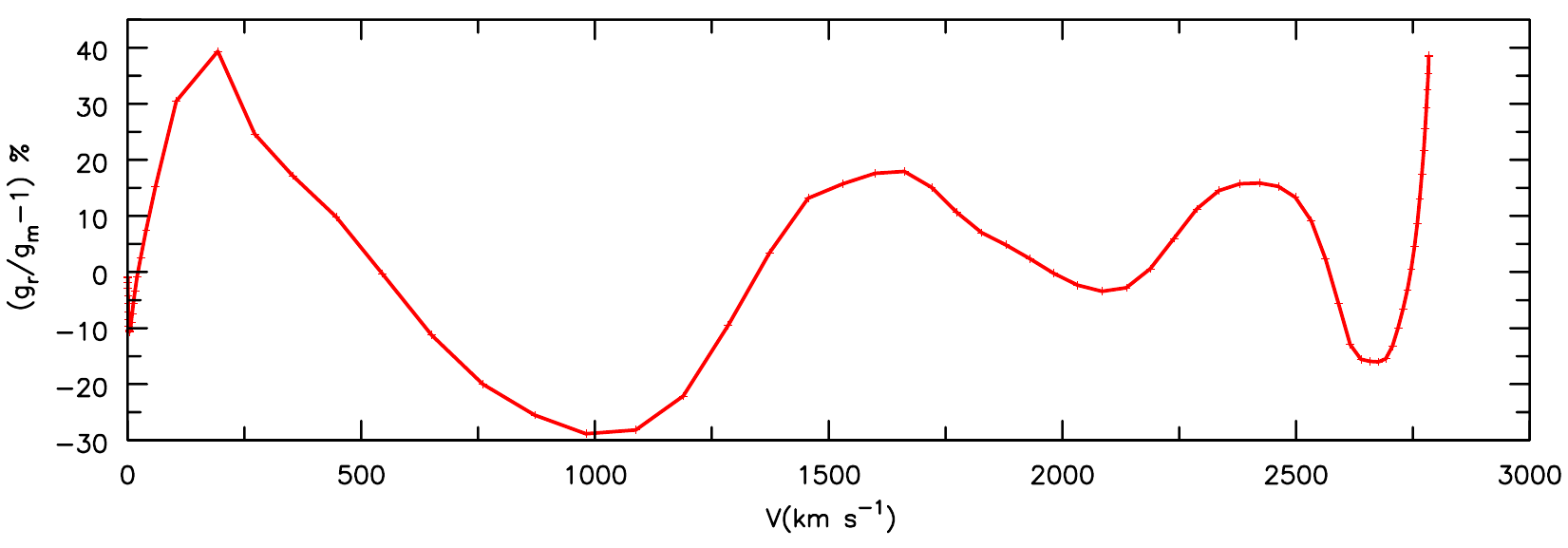}
\caption{Check on the consistency of the radiative force needed to drive the
mass-loss for the assumed clumping and velocity law for BAT99-61. $g_r$ refers to the radiative force 
required to achieve consistency, while $g_m$ refers to that computed from the model radiation field. No attempt
was made to revise the adopted velocity law (or clumping) to improve the
hydrodynamics.  This model had 19 chemical species (He, C, O, Ne, Na, Mg, Al, Si, P, S, Cl, Ar, K, Ca, Cr, Mn,
Ti, Fe, Ni),  4692 non-LTE atomic levels, 122 ionization states, and treated 646147 line transitions when computing
the atmospheric structure.}
\end{figure}

\clearpage

\bibliography{main}{}
\bibliographystyle{aasjournal}

\end{document}